\documentclass{aastex631}

\usepackage{amsmath}

\usepackage{commath}

\usepackage{ dsfont }
\usepackage{soul}

\received{September 24, 2021}
\revised{January 21, 2022}
\accepted{January 31, 2022}

\submitjournal{ApJ}

\shorttitle{MHD wave mode identification in sunspots}
\shortauthors{Albidah et al.}

\graphicspath{{./}{figures/}}

\usepackage{graphicx}
\usepackage{appendix}

\begin{document}

\title{Magnetohydrodynamic wave mode identification in circular and elliptical sunspot umbrae: evidence for high order modes}


\correspondingauthor{Abdulrahman Albidah }
\email{abalbidah1@sheffield.ac.uk, a.albedah@mu.edu.sa}

\author[0000-0001-7314-1347]{A. B. Albidah}
\affiliation{Plasma Dynamics Group, School of Mathematics and Statistics, The University of Sheffield, Hicks Building, Hounsfield Road, Sheffield, S3 7RH, UK}

\affiliation{Department of Mathematics, College of Science Al-Zulfi, Majmaah University, Al-Majmaah, 11952, Saudi Arabia}

\author[0000-0002-0893-7346]{V. Fedun}
\affiliation{Plasma Dynamics Group, Department of Automatic Control and Systems Engineering, The University of Sheffield, Mappin Street, Sheffield, S1 3JD, UK}

\author[0000-0003-2220-5042]{A. A. Aldhafeeri}
\affiliation{Mathematics and Statistic Department, Faculty of Science, King Faisal University, Al-Hassa, P.O. Box 400, Hofuf 31982, Saudi Arabia}

\author[0000-0002-3066-7653]{I. Ballai}
\affiliation{Plasma Dynamics Group, School of Mathematics and Statistics, The University of Sheffield, Hicks Building, Hounsfield Road, Sheffield, S3 7RH, UK}

\author[0000-0001-8161-4677]{W. Brevis}
\affiliation{School of Engineering, Pontificia Universidad Cat\'{o}lica de Chile, Chile}

\author[0000-0002-9155-8039]{D. B. Jess}
\affiliation{Astrophysics Research Centre, School of Mathematics and Physics, Queen’s University, Belfast, BT7 1NN, UK}
\affiliation{Department of Physics and Astronomy, California State University Northridge, Northridge, CA 91330, USA}

\author[0000-0001-7577-0913]{J. Higham}
\affiliation{School of Environmental Sciences, Department of Geography and Planning, University of Liverpool, Roxby Building, Liverpool, L69 7ZT, UK}

\author[0000-0002-5365-7546]{M. Stangalini}
\affiliation{ASI, Italian Space Agency, Via del Politecnico snc,
00133 Rome, Italy}

\author[0000-0001-5414-0197]{S. S. A. Silva}
\affiliation{Plasma Dynamics Group, Department of Automatic Control and Systems Engineering, The University of Sheffield, Mappin Street, Sheffield, S1 3JD, UK}

\author[0000-0002-9546-2368]{G. Verth}
\affiliation{Plasma Dynamics Group, School of Mathematics and Statistics, The University of Sheffield, Hicks Building, Hounsfield Road, Sheffield, S3 7RH, UK}

\begin{abstract}
In this paper we provide clear direct evidence of multiple concurrent higher order magnetohydrodynamic (MHD) modes in circular and elliptical sunspots by applying both Proper Orthogonal Decomposition (POD) and Dynamic Mode Decomposition (DMD) techniques on solar  observational data. These techniques are well documented and validated in the areas of fluid mechanics, hydraulics, and granular flows, yet are relatively new to the field of solar physics. While POD identifies modes based on orthogonality in space and it provides a clear ranking of modes in terms of their contribution to the variance of the signal, DMD resolves modes that are orthogonal in time. The clear presence of the fundamental slow sausage and kink body modes, as well as higher order slow sausage and kink body modes have been identified using POD and DMD analysis of the chromospheric H$\alpha$ line at 6562.808~{\AA} for both the circular and elliptical sunspots.  Additionally, to the various slow body modes, evidence for the presence of the fast surface kink mode was found in the circular sunspot. All the MHD modes patterns were cross-correlated with their theoretically predicted counterparts and we demonstrated that ellipticity cannot be neglected when interpreting MHD wave modes. The higher-order MHD wave modes are even more sensitive to irregularities in umbral cross-sectional shapes, hence this must be taken into account for more accurate modelling of the modes in sunspots and pores.
\end{abstract}

\section{Introduction} \label{sec:intro}
The qualitative and qualitative description of plasma dynamics in solar and space environment are one of the most challenging aspects of solar physics. The variety of plasma motions subject to restoring forces (e.g. pressure gradient, gravitational, Lorentz etc.) give rise to magnetohydrodynamic (MHD) waves and oscillations. In the absence of these restoring forces perturbations might evolve into laminar and turbulent flows, shocks, nonlinear patterns, etc. Waves have the unique property to carry energy and information about the medium in which they propagate, making them an ideal tool for plasma and field diagnostics. However, their true diagnostic potential can be put at use only if high resolution observations are available that could determine the accurate measurement of wave properties and their true nature.  
Sunspots are the most prominent manifestations of the emergence of magnetic field in the lower regions of the solar atmosphere and they are often the footpoints of active regions (ARs) that are able to considerably influence the space weather. Although, historically,  sunspots are the most studied features in the solar atmosphere, their dynamical properties are far from being understood. Thanks to modern observational capabilities (from visible to near infrared wavelengths) waves and oscillations in sunspots are observed from the photosphere to the corona. Over this height the properties of waves (amplitude, frequency, etc.) can change due to the intrinsic changes in the plasma environment and magnetic field, making their classification and study rather difficult. 

The presence of waves and oscillation of sunspots have been known since the pioneering work by \cite{beckers1969chromospheric}, who evidenced oscillatory behaviour in a sunspot by determining the observational parameters of umbral flashes. Shortly after, the three-minute oscillations in Doppler velocity in the umbral region have been identified by \cite{beckers1972oscillatory}. Later on, it has been shown that the most dominant oscillations in sunspots and pores have periods of 5 minutes at the photospheric heights, and 3 minutes at the chromospheric heights, while global oscillations of sunspots, as a whole, have periods that range from hours to days \citep[to name but a few]{nagashima2007observations,stangalini2011mhd,jess2012propagating,2015SSRv..190..103J, khomenko2015}. In contrast, \cite{stangalini2021novel} have shown that the dominant oscillations of a magnetic pore observed by means of The Interferometric BIdimensional Spectropolarimeter (IBIS) have periods of 3 minutes in the photosphere, instead of the expected 5 minutes period and this was the first time reporting the 3 min oscillations in a pore photosphere. Using high-resolution observations Solar Optical Telescope (SOT) on board Hinode, \cite{nagashima2007observations} investigated the spatial distribution of the power spectral density of the oscillatory signal in and around an AR. The G-band data showed that in the umbra the oscillatory power is suppressed in all frequency ranges. On the other hand, in CaII H intensity maps oscillations in the umbra, so-called umbral flashes, are clearly seen with the power peaking around 5.5mHz (3 minutes). The CaII H power distribution showed the enhanced elements with the spatial scale of the umbral flashes over most of the umbra, with a region with suppressed power at the center of the umbra. The relation between the 3-min and 5-min oscillation in sunspots has been studied by \cite{zhou2017relationship}, who showed that the running waves are propagating across the umbra–penumbra as 3-min oscillations when they are located at the umbra region, and 5-min oscillations in the penumbra region.

Magnetic structures observed in the solar atmosphere are perfect environments for the propagation of guided waves. Traditionally (considering only the pressure gradient and Lorentz force, as restoring forces), MHD waves propagating in plasmas are classified according to their relative propagation speed (slow and fast magnetoacoustic modes, Alfv\'en or intermediate modes), their radial structure (surface or body) and their number of nodes in the radial direction (fundamental or overtone). Within the high plasma-$\beta$ region, i.e. where sound speed, $c_S$, is greater than Alfv\'en speed, $v_A$, slow modes are propagating within the tube mainly along magnetic field lines with the local Alfv\'en speed. The fast modes are allowed to propagate in any direction. Along the direction of field lines the fast mode travels with the sound speed and in the direction perpendicular to the field lines it propagates with the phase speed $v_{ph}=\sqrt{c_S^2+v_A^2}$. In the region where the plasma-$\beta$ is low ($v_A > c_S$), slow mode propagates approximately with the local sound speed and fast mode propagates with the local Alfv\'en speed along the magnetic field lines. The angular dependence for the phase speeds of the slow and fast modes are identical in the high plasma-$\beta$ region. Slow waves are prohibited from travelling perpendicularly to the magnetic field lines for both, high and low plasma-$\beta$ regions. Surface waves propagate in a way that their maximum amplitude is attained on the boundary of the waveguide and they are evanescent inside and outside the magnetic flux tube. In contrast, body waves have an oscillatory pattern in the radial direction inside the waveguide and their lowest amplitude is on the boundary of the waveguide. In the external region body waves are also evanescent (the wave power is localised and it is concentrated within the waveguide). The fundamental modes have only one radial node occurring at the umbra/penumbra boundary, however, overtones have more than one radial node. MHD waves can also be classified according to their motion with respect to the longitudinal symmetry axis of the waveguide. While sausage modes propagate without perturbing the symmetry axis, kink modes perturb the axis in a back and forth motion. Finally, fluting modes have a complex way of perturbing the axis. Sausage and kink modes are continuously observed in solar magnetic structures (and their literature is vast), however, higher order modes, i.e. the fluting modes, were so far elusive and their existence was hypothetical.  

Recently, \citet{2017ApJ...842...59J} have detected slow body kink modes propagating along the azimuthal direction in a sunspot, by implementing a $k-\omega$ Fourier filter ($0.45 - 0.90$ arcsec$ ^{-1}$ and $5 - 6.3$ mHz) on  H$\alpha$ images acquired by the Hydrogen-Alpha Rapid Dynamics camera \citep[HARDcam;][]{2012ApJ...757..160J}. More recently, \citet{albidah2020RS} have applied the Proper Orthogonal Decomposition (POD, \citet{pearson1901liii}) and Dynamic Mode Decomposition (DMD, \citet{schmid2010dynamic}) techniques for the same set of observations as \citet{2017ApJ...842...59J}, to identify the fundamental slow body sausage and kink modes. 

\cite{keys2018photospheric} showed the separate existence of surface and body sausage modes in pores that have approximately elliptical cross-sectional shape. The authors have taken a one-dimensional cross-cut along the pores and assumed that for sausage surface modes the magnitude of the power along the time series has its maximum at the boundary and has its minimum at the center, and the opposite for the sausage body mode. However, it has been recently shown \citep{aldhafeeri2021magnetohydrodynamic} that the magnitude of the surface sausage mode has its maximum amplitude at the boundary along the minor axis, while it has its minimum amplitude at the boundary along the major axis. Therefore, the assumption of \cite{keys2018photospheric} may only work for a pore that has a circular cross-sectional shape. Our present study that involves the use of the POD/DMD techniques will address this issue and will show how reliable these methods are in the identification of modes in waveguides of different cross-sectional shapes.

The POD and DMD are methods which are commonly used in fluid mechanics and granular flows \citep{murray2007application,berry2017application,higham2017using,higham2018modification,higham2020using,higham2021modification}. The POD technique allows the determination of spatially orthogonal patterns from signals whilst the DMD technique allows the determination of temporally orthogonal patterns as the DMD provides a spatial pattern of the indicated modes with a pure frequency \citep[]{tu2014dynamic}. Both of these methods are separately very useful and when combined, following the method developed by \citet{higham2018implications}, can be used to elucidate temporally and spatially orthogonal structures from solar observations, see e.g. \citet{albidah2020RS}. For completeness, we provide a brief mathematical overview of the methods in Section \ref{sec:Method}. 

To a very large extent the traditional  analysis of oscillations in sunspots involves applying Fourier analysis to provide the power spectra, and that can be carried out by integration over a region of interest (ROI) or even on pixel by pixel basis. The assumption of a sinusoidal basis in the spatial domain can be taken as a disadvantage of using Fourier analysis since we are applying it in a cylindrical, or even an elliptical models. In the case of cylindrical waveguide the basis functions in the radial direction are Bessel functions, which are orthogonal to each other by definition, which give the application of POD the strength as the method looks at the orthogonality in space. Furthermore, POD and DMD have a further advantage over Fourier analysis as they cross-correlate individual pixels over the ROI, in the spatial and temporal domain, respectively. Moreover, the shape of the sunspot may be affected by the surrounding background and due to that, the shape may lose the property of orthogonality in the spatial domain, and hence the advantage of using POD will be lost as it will not work very well anymore. However, DMD can detect modes which are orthogonal in time. Therefore the best approach is to use the POD and DMD techniques in combination.

The wavelet time series analysis has also been widely used to study MHD wave modes and its properties in the sunspot umbra region. \cite{o2002oscillations} have applied the wavelet and Fourier analysis on an umbral regions of an observed sunspot using different spectral lines of the umbral region that covers the range of temperatures from the low chromosphere to corona to show the appearance of oscillations at all investigated temperatures, with frequencies in the range of 5.4 mHz to 8.9 mHz. \cite{christopoulou2003wavelet} have used this methodology for identification of the 3-min oscillations in the sunspot umbral region. By using radio (Nobeyama Observatory) and EUV (TRACE, SDO/AIA) observations, \cite{sych2012frequency} have applied the wavelet analysis to study the amplitude and frequency modulation of 3-min oscillations of microwave and extreme ultraviolet (EUV) emission generated at different heights of a sunspot atmosphere.

The existence of higher order modes was so far mostly predicted theoretically \citep[see e.g.][]{edwin1983wave} and the very few studies of these modes used indirect methods to show their existence. Using the observations obtained with the help of the Fast Imaging Solar Spectrograph installed at the 1.6 m Goode Solar Telescope (GST) \citet{Kang2019} suggested that the observed two-armed spiral wave patterns in pores could be explained in terms of a superposition of slow sausage body mode (corresponding to an azimuthal wavenumber $n=0$) and a flutting mode ($n=2$). However, correlation analysis between numerically simulated and observed modes to validate obtained results was not included in their study.

Our paper is organized as follows: a brief overview of the observations of sunspots we study is given in section \ref{sec:Observations}. Section \ref{sec:Method} contains a short, but necessary description of the POD and DMD decomposition techniques. The theoretical models that have been used in our analysis (the cylindrical model, elliptical model and the irregular shape model) are presented in Section \ref{sec:model}. The modes' identification methodology in the circular and the elliptical sunspots and the discussion on the nature of the modes are presented in Section \ref{sec:identification}. Finally, our the conclusions are presented in Section \ref{sec:Conclusion}.

\begin{figure*}[!t]
\begin{center}
 \begin{tabular}{cc}
 \includegraphics[scale=0.4]{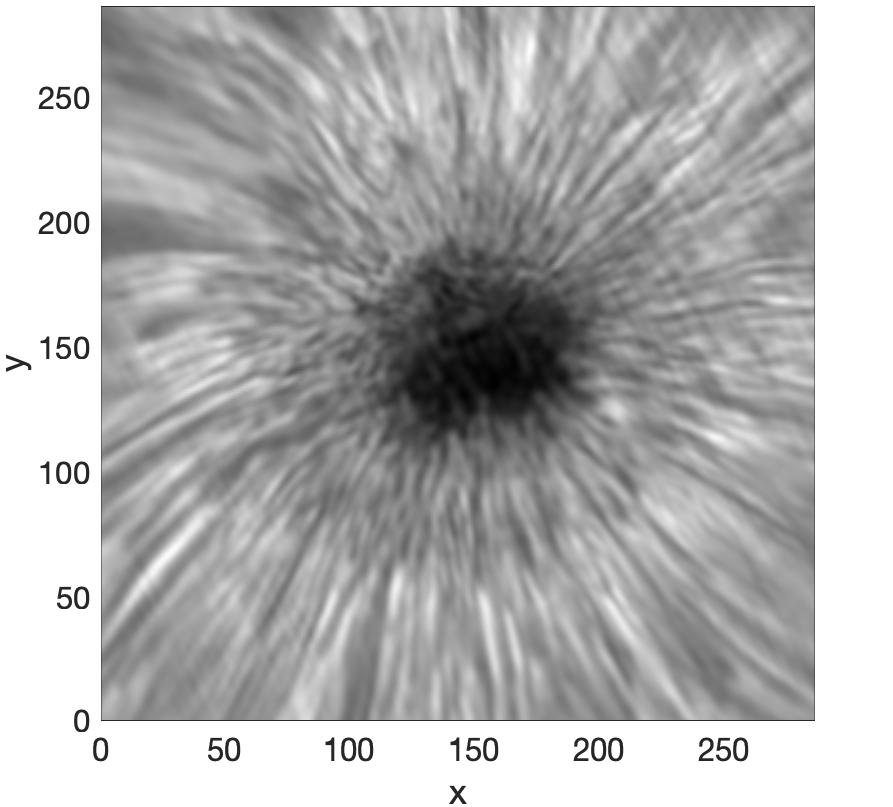}&
 \includegraphics[scale=0.4]{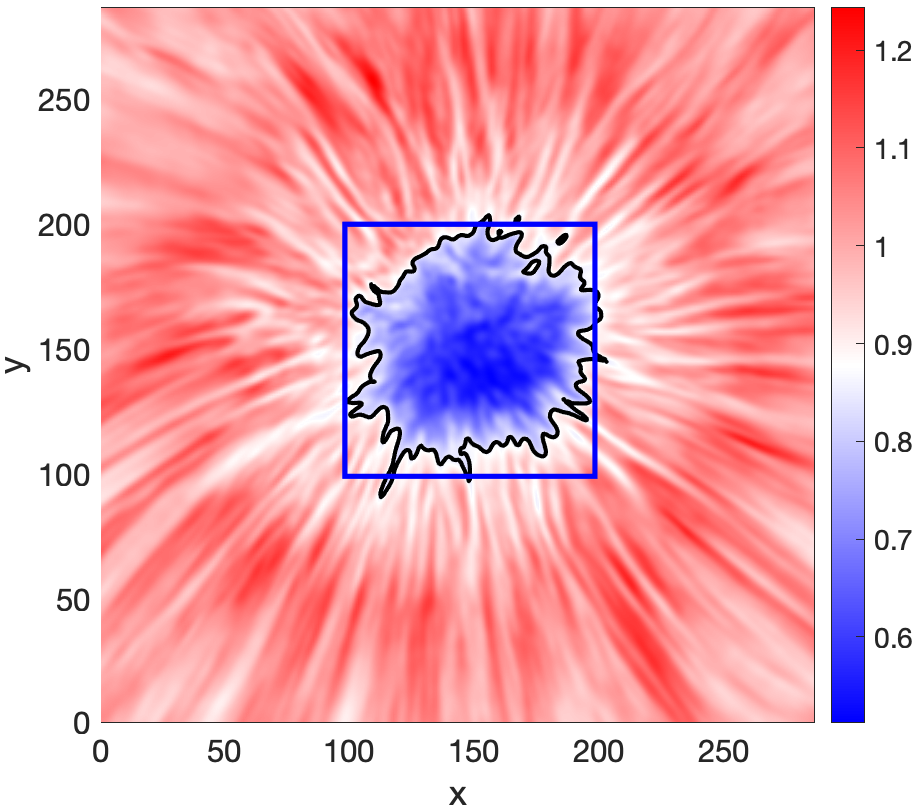}\\
 \includegraphics[scale=0.4]{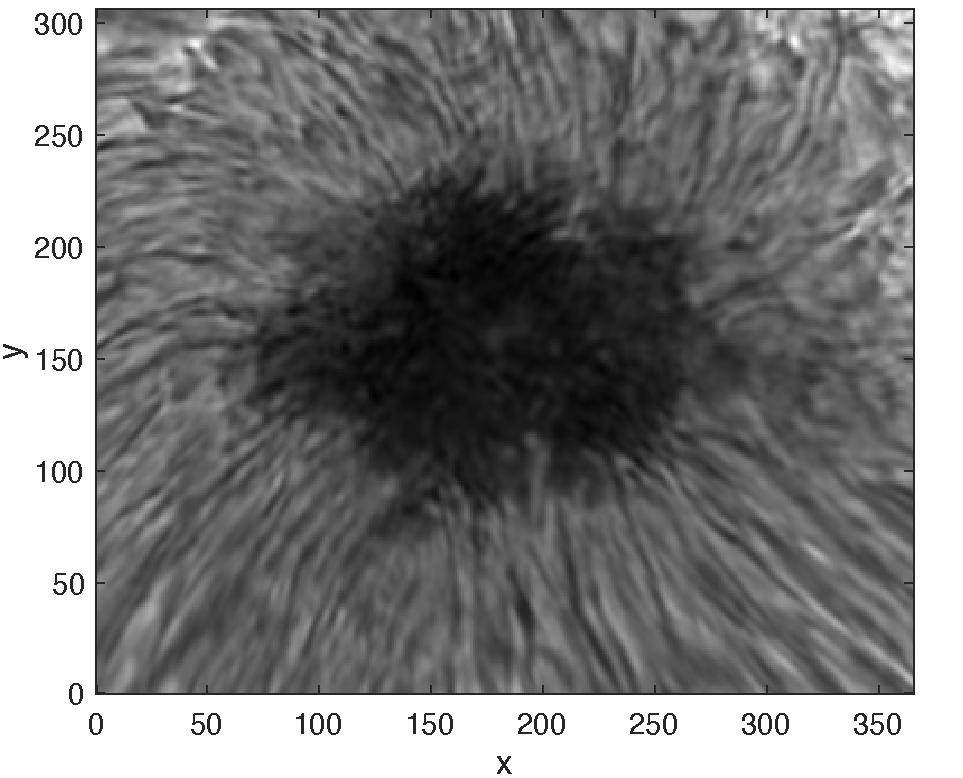}&
 \includegraphics[scale=0.4]{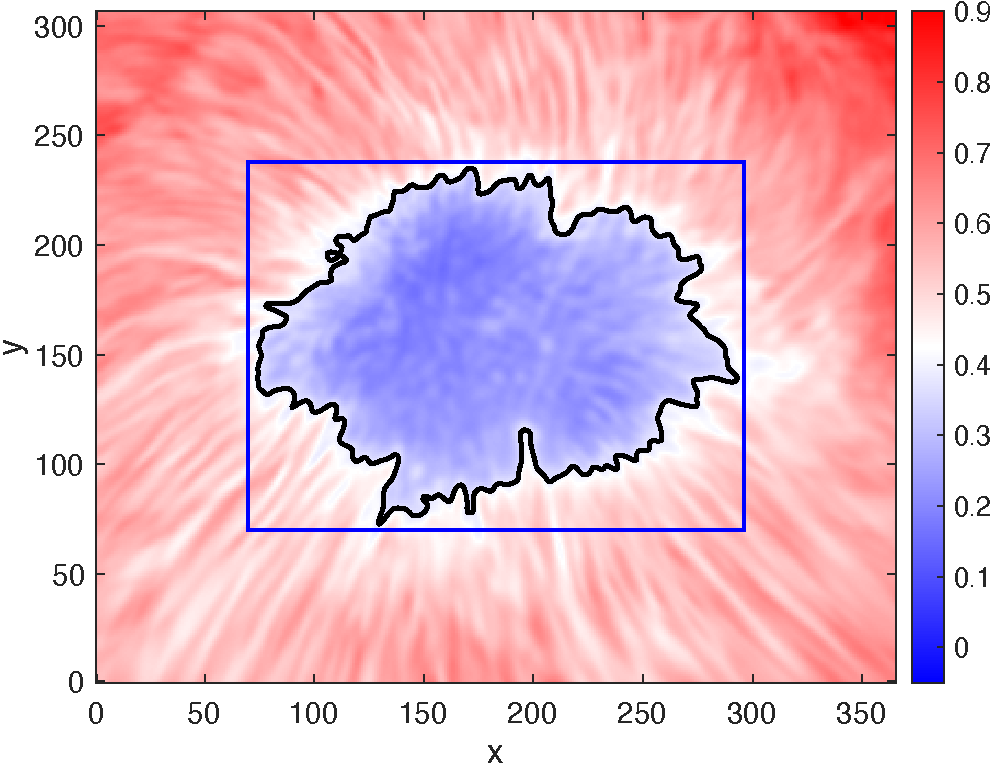}
\end{tabular}
\end{center}
\caption{The first column shows snapshots from the H$\alpha$ time series of the circular (upper panel) and elliptical (lower panel) sunspot. The second column displays the mean intensity of the time series of the circular (upper panel) and elliptical (lower panel) sunspot, the colourbar displays the magnitude of the mean time series, the solid black line shows umbra/penumbra boundary with intensity threshold level at 0.85 for the circular sunspot (and 0.4 for the elliptical sunspot) and the blue box shows the region where we apply our POD and DMD analysis. \label{fig:snapshot_circular&E}}
 \end{figure*}

\section{Observations} \label{sec:Observations}
The sunspot observations employed in the present study were both acquired using the HARDcam, which is an upgrade to the Rapid Oscillations in the Solar Atmosphere \citep[ROSA;][]{2010SoPh..261..363J} imaging system available as a common-user instrument at the National Solar Observatory's Dunn Solar Telescope (DST). Each observational image sequence was acquired through a narrowband 0.25~{\AA} (full-width at half-maximum) filter centered on the chromospheric H$\alpha$ absorption line at 6562.808~{\AA}. During each of the observing sequences, high-order adaptive optics \citep{2004SPIE.5490...34R} were employed, with the acquired images further improved through the application of speckle reconstruction algorithms \citep{2008A&A...488..375W}. Specific acquisition details are provided below.

\subsection*{10 December 2011}

The sunspot formed part of NOAA~11366, which was located at heliocentric coordinates ($356''$, $305''$), or N18W22 in the conventional heliographic coordinate system. A pixel size of $0{\,}.{\!\!}{''}138$ per pixel was chosen to provide a field-of-view size equal to $71''\times71''$. Images were acquired over the course of 75~minutes (16:10 -- 17:25~UT) at a cadence of 0.050~s. The dataset has previously been employed in a host of scientific studies \citep{2013ApJ...779..168J, 2016NatPh..12..179J, 2017ApJ...842...59J, 2015ApJ...812L..15K, albidah2020RS} due to the excellent seeing conditions and the highly circularly symmetric shape of the sunspot umbra. Following speckle techniques, the final cadence for the reconstructed images is 1.78~s. A sample H$\alpha$ image of the sunspot is displayed in the upper left panel of Figure \ref{fig:snapshot_circular&E}.

\subsection*{24 August 2014}

The sunspot formed part of NOAA~12146, which was located at heliocentric coordinates ($496''$, $66''$), or N10W32 in the conventional heliographic coordinate system. A diffraction-limited pixel size of $0{\,}.{\!\!}{''}108$ per pixel was chosen to provide a field-of-view size equal to $180''\times180''$, which is the maximum allowable by the DST optics. Images were acquired over the course of 120~minutes (13:56 -- 15:56~UT) at a cadence of 0.017~s. The dataset has previously been employed in a study that examined the presence of Alfv{\'{e}}n wave driven shocks in sunspot atmospheres \citep{2018NatPh..14..480G}. Following speckle techniques, the final cadence for the reconstructed images is 1.00~s. A sample H$\alpha$ image of the sunspot is displayed in the lower left panel of Figure \ref{fig:snapshot_circular&E}.

\section{POD and DMD Analysis of Observational Data} \label{sec:Method}
\noindent For both the POD and DMD we consider a sequence of ROI intensity snapshots of time domain of size $\text{T}$ and spatial domain of size $\text{X}\times \text{Y}$, with each snapshot regularly temporally spaced. Each snapshot is column vectorised such that $\text{N}=\text{XY}$ and a matrix $\mathbf{W}$ is created from them such that $\mathbf{W} \in \mathds{}^{\text{N}\times \text{T}}$. To apply the POD technique we use a Singular Value Decomposition (SVD):
\begin{equation}
    \mathbf{W}=\mathbf{\Phi}\mathbf{S}\mathbf{C}^{*}. \label{Eq:POD}
\end{equation}
\noindent The decomposition results in the spatial structure of each mode given in the form of columns of the matrix $\mathbf{\Phi}$, and their temporal evolution is given by the columns of the matrix $\mathbf{C}$, where $^{*}$ denotes a conjugate transpose operation. Modes are ranked according to their contribution to the total variance of the snapshot series. This contribution is given by the diagonal elements of matrix $\lambda$, by means of the vector $\lambda = \text{diag}(\mathbf{S})^{2}/(\text{N})$.
 
\noindent To perform DMD, analogously to POD, the snapshots are organized in columns, but in two matrices, $\mathbf{W}^{A}$ and $\mathbf{W}^{B}$, such that $\mathbf{W}^{B}$ is shifted by a snapshot of $\mathbf{W}^{A}$. Then, the matrix $\mathbf{W}^{A}$ is decomposed using the SVD:
\begin{equation}
  \mathbf{W}^{A}=\tilde{\mathbf{\Phi}}\tilde{\mathbf{S}}\tilde{\mathbf{C}^{*}}.
\end{equation}
Using this result and the matrix $\mathbf{W}^{B}$, the matrix 
\begin{equation}
    \mathbf{F}=\tilde{\mathbf{\Phi}}\mathbf{W}^{B}\tilde{\mathbf{C}}\tilde{\mathbf{S}}^{-1}, 
\end{equation}
\noindent is obtained. Using the matrix $\mathbf{F}$ we can calculate its complex eigenvalues, $\mu_{i}$, and eigenvectors, $z_{i}$, where $i=1...\tau$ and $\tau=T-1$. According to \citet{schmid2010dynamic}, to create a robust set of eigenvectors, a Vandermonde expansion of the eigenvalues can be employed as
\begin{equation}
    \mathbf{Q}_{i,j}=\mu^{j-1}_i,
\end{equation}
\noindent where $i=1...\tau$ and $j=1...\tau$. The spatial modes of DMD are calculated by:
\begin{equation}
    \mathbf{\Psi}=\mathbf{W}^A\mathbf{Q}^{*},
\end{equation}
\noindent and the frequencies associated to these modes can be determined using
\begin{equation}
    f_i=f_s arg(z_i)/2\pi,
\end{equation}
\noindent where $f_s$ is the snapshots-sampling frequency. Following the approach by \citet{higham2017using, higham2018implications}, modes are identified based on their contribution to the variance via POD. This step is followed by the calculation of a Fourier Power Spectra of the POD time coefficients associated with the dominant modes. These spatially important frequencies can then be used to identify temporally orthogonal modes determined by the DMD. 

\section{Models} \label{sec:model}
The behavior of MHD waves in sunspots approximated by magnetic configurations such as cylinder with circular or elliptical cross-section can be understood by analysing dispersion relations derived from the full set of MHD equations. More complicated cases (when the sunspot's cross-section shape is irregular) can be studied only numerically. In this study we used all these three types of models describing MHD wave propagation in magnetic flux tubes, and compared the theoretically predicted waves' signature with observed data using correlation techniques.

The first model is the cylindrical model with a cicular cross-section and the predictions of this model are compared with the modes that are observed from the circular sunspot (see Figures \ref{fig:fund_sausage_Circular}-\ref{fig:fluting(n=3)_circular}). The second model describes the possible wave modes in a waveguide with elliptical cross-section. The predictions of this model are compared with the observational data derived using the sunspot with approximately elliptical shape (see Figures \ref{fig: fun_saus_E}\ref{fig:m=3_E}). Finally, the third model assumes an irregular cross-sectional shape. The theoretical predictions were compared to both sunspot types, i.e. the circular  (see Figures \ref{fig:fund_sausage_Circular}-\ref{fig:fluting(n=3)_circular}) and elliptical (see Figures \ref{fig: fun_saus_E}-\ref{fig:m=3_E}).

\subsection{Cylindrical model}
Following the standard approach, we assume a magnetic flux tube with circular cross-section of radius $a$. The axis of the tube is oriented along the vertical $z$-axis. We will denote the quantities that
correspond to each of the internal and external regions with the indices $i$ and $e$, respectively. The plasma is permeated by a homogeneous magnetic field directed along the longitudinal symmetry axis ($B_i$ and $B_e$, respectively) and is characterised by constant plasma densities ($\rho_i$ and $\rho_e$) and kinetic pressures ($p_i$ and $p_e$). Assuming regular solution along the symmetry axis and a localized dynamics inside the flux tube (i.e. exponentially decaying solutions outside the magnetic cylinder), the dispersion relations of surface and body modes can be given as \citep[see][for more details]{edwin1983wave}:
 \begin{equation} \label{dispartion_cylinder_S}
   	\rho_i(k_z^2v_{A_i}^2-\omega^2)m_e\frac{K'_n(m_ea)}{K_n(m_ea)} = \rho_e(k_z^2v_{A_e}^2-\omega^2)m_i\frac{I'_n(m_ia)}{I_n(m_ia)},
   \end{equation} 
  \begin{equation} \label{dispartion_cylinder_B}
   	\rho_i(k_z^2v_{Ai}^2-\omega^2)m_e\frac{K'_n(m_ea)}{K_n(m_ea)} = \rho_e(k_z^2v_{Ae}^2-\omega^2)m_i\frac{J'_n(m_ia)}{J_n(m_ia)},
   \end{equation} 
where $k_z$ is the wavenumber in the vertical direction, $v_{Ai}=B_i/\sqrt{\mu_0\rho_i}$ and $v_{Ae}=B_e/\sqrt{\mu_0\rho_e}$ are the Alfv\'en speeds inside and outside the flux tube, $\omega$ is the frequency of waves, $I_n$, $J_n$ and $K_n$ are the Bessel functions of order $n$, $\mu_0$ is the permeability of free space and the {\it dash} denotes the derivative of the Bessel functions with respect to their argument. The magneto-acoustic parameters, $m_i$ and $m_e$, are defined as
\begin{equation} \label{Eq:1}
	m_{i}^2=\frac{(k_z^2c_{i}^2-\omega^2)(k_z^2v_{Ai}^2-\omega^2 )}{(c_{i}^2+v_{Ai}^2)(k_z^2 c_{Ti}^{2}-\omega^2)} \quad  \mathrm{and} \quad  m_{e}^{2}=\frac{(k_z^2c_{e}^2-\omega^2)(k_z^2v_{Ae}^2-\omega^2 )}{(c_{e}^2+v_{Ae}^2)(k_z^2 c_{Te}^2-\omega^2)}.
\end{equation}
The signs of these parameters determine the nature of MHD waves, e.g. Equation (\ref{dispartion_cylinder_S}) with $m_{i}^2>0$ corresponds to the dispersion relation of surface waves, Equation (\ref{dispartion_cylinder_B}) with $m_{i}^2<0$ describes body waves. In the case of both modes the condition $m_e^2<0$ ensures an exponentially decaying solution outside the magnetic flux tube. In the above expressions $c_{Si}=\sqrt{\gamma p_i/\rho_i}$ and $c_{Se}=\sqrt{\gamma p_e/\rho_e}$ are the adiabatic sound speeds, $\gamma$ is the ratio of specific heats and the quantities $c_{Ti}$ and $c_{Te}$ are the characteristic speeds of slow magnetoacoustic modes (tube) speeds in the two regions, defined as
\begin{equation} 
c_{Ti}^{2}=\frac{v_{Ai}^{2} c_{i}^2}{v_{Ai}^{2} + c_{i}^2}  	\quad  \mathrm{and} \quad c_{Te}^{2}=\frac{v_{Ae}^{2} c_{e}^2}{v_{Ae}^{2} + c_{e}^2}.
\end{equation}
The values of the parameter $n$ determine the symmetry of the mode with respect to the axis of the magnetic flux tube, that is $n=0$ corresponds to sausage modes, $n=1$ to kink modes and $n\geq 2$ to fluting modes.

\subsection{Elliptical model}
The model describing the wave propagation in a cylindrical magnetic flux tube \citep[][]{edwin1983wave} can be expanded to a more general case of a magnetic waveguide with an elliptical cross-section \citep[see ][for more details]{aldhafeeri2021magnetohydrodynamic}. In this configuration, dispersion equations for MHD surface and body waves can be represented as 

 \begin{equation} \label{dispartion_EE_S}
 \rho_{0e}(k_z^2v_{Ae}^2-\omega^2)\frac{\Xi_{m}^{'{E,O}}(\abs{\tilde {m}_i},s_0)}{\Xi_{m}^{E,O}(\abs{\tilde {m}_i},s_0)}= \rho_{0i}(k_z^2v_{Ai}^2-\omega^2)  \frac{\Psi_{m}^{'{E,O}}(\abs{\tilde {m}_{e}},s_0)}{\Psi_{m}^{E,O}( \abs{\tilde {m}_{e}},s_0)}, 
   \end{equation} 
 \begin{equation} \label{dispartion_EE_B}
 \rho_{0e}(k_z^2v_{Ae}^2-\omega^2) \frac{\Theta_{m}^{'{E,O}}(\tilde {m}_i,s_0)}{\Theta_{m}^{E,O}(\tilde {m}_i,s_0)}=  \rho_{0i}(k_z^2v_{Ai}^2-\omega^2) \frac{\Psi_{m}^{'{E,O}}(\abs{\tilde{m}_{e}},s_0)}{\Psi_{m}^{E,O}( \abs{\tilde {m}_{e}},s_0)}.
  \end{equation} 
Here the new magneto-acoustic parameters are
 \begin{equation} \label{Eq:2}
	\tilde{m}_{i}^2=-\frac{\sigma^{2}}{4}m_i^2,  \quad  \tilde{m}_{e}^2=-\frac{\sigma^{2}}{4}m_e^2,
\end{equation}
where $\sigma$ is the distance from the center of the ellipse and its focal points. Although the form of the dispersion relations (\ref{dispartion_EE_S}) and (\ref{dispartion_EE_B}) are rather similar to the case of a waveguide with circular cross-section, the functions involved in equations (\ref{dispartion_EE_S}) and (\ref{dispartion_EE_B}) are Mathieu functions, rather than Bessel functions. In the above dispersion relations $\Xi_{m}^{E,O}$, $\Theta_{m}^{E,O}$ and $\Psi_{m}^{E,O}$ denote the internal solution for body wave, the internal solution for surface wave and the external solution, respectively. The superscripts $E$ and $O$ denote the even and odd solutions and the prime denotes the derivative of the Mathieu function with respect to the confocal elliptic variable, $s$. 

The study by \cite{aldhafeeri2021magnetohydrodynamic} revealed that the cross-sectional shape introduces significant changes in the behaviour of waves, as this depends on the polarisation along the major or the minor axis of the ellipse. It was also found that higher order modes are strongly influenced by the change in the eccentricity of the waveguide.

\subsection{Irregular shape model}
\label{sec:irr}
In the reality, the cross-section of sunspots is far from being regular. The dispersion relations for regular circular and elliptical cross-sections reveal that these relations are sensitive to the transversal geometry of the waveguide \citep[][]{aldhafeeri2021magnetohydrodynamic}.
In order to determine the property of waves and their oscillatory patterns in waveguides with irregular cross-section a numerical approach was used to determine the eigenfunctions and the associated eigenvalues. For this problem we used a Cartesian coordinate system, assuming the photospheric level to be the $xy$-plane and the vertical direction to be along the vertical $z$ axis. The spatial structure of the eigenfunctions is physically constrained by the cross-sectional shape of the waveguide. The governing equation of the longitudinal velocity perturbation, $v_{z}$, was  derived and solved by using the observed cross-sectional shape, where the shape is obtained by taking the threshold level of the umbra and set $v_z=0$ at the umbra/penumbra boundary to be consistent with the observational data.  
 
By assuming linear MHD perturbations, the time-independent Helmholtz equation has been derived for the vertical component of velocity perturbation, $v_z$, of the form
\begin{eqnarray}
\frac{\partial^2 v_z}{\partial x^2}+\frac{\partial^2 v_z}{\partial y^2}-m_i^2 v_z =0, \label{a1} 
\end{eqnarray}
where $m_i^2$ is the eigenvalue defined by Equation (\ref{Eq:1}).
Equation (\ref{a1}) was solved by assuming Dirichlet-type boundary condition, i.e. at the boundary of the magnetic waveguide the $z$-component of the velocity perturbation vanishes. With this type of boundary condition the numerical solution describes only slow body modes, which constitutes a limitation of this model. 
 Body waves are guided waves whose longitudinal velocity amplitude is zero on the boundary of the waveguide, while taking their maximum value inside the waveguide. By choosing this type of boundary condition, we disregard those modes (known as surface modes), whose velocity amplitude is not zero on the boundary. Of course, the POD/DMD techniques can also recover surface waves, see section \ref{sec:surface}.

In order to apply the above approach to observations (line intensity), it is more convenient to write Eq. \ref{a1} in terms of density perturbation, $\rho$. The relationship between the density and longitudinal velocity component is 
\begin{eqnarray}
v_z=\frac{k_z c_i^2}{\omega \rho_0} \rho, \label{Eq:v_z} 
\end{eqnarray}
where $\rho_0$ is the unperturbed density that corresponds to the equilibrium state  \citep{aldhafeeri2021magnetohydrodynamic}. From Equation  (\ref{Eq:v_z}) it follows that the evolution of the density perturbation is governed by a similar Helmholtz equation. In general, all the dominant compressive variables are proportional to each other, therefore they can be assumed to be governed by a Helmholtz-type equation.


\section{MHD wave modes identification and discussion} \label{sec:identification}
The POD and DMD techniques were applied on the two data-sets associated with the sunspots, shown in Figure \ref{fig:snapshot_circular&E}. The oscillatory pattern of modes recovered with the help of these techniques is compared with the results drawn from theoretical models constructed for a waveguide with  cylindrical, elliptical, and the irregular cross-section. The comparison is quantified by means of a cross-correlation analysis \citep{correlation11,correlation22}, calculated on a pixel-by-pixel basis. The result of the correlation is a number between 1 and -1, where 1 means that the two pixels have a linear correlation while -1 denotes a linear anti-correlation. 

Furthermore, as POD provides information about the temporal evolution of the coefficients of the POD modes, we can determine the power spectrum density (PSD), which will show the most dominant frequencies of modes. Since DMD identifies modes in terms of their frequency, and by using the magnetoacoustic parameters $m_i$ and $\tilde{m}_{i}$ for the cylindrical and the elliptical models, respectively, the longitudinal wavenumber, $k_z$, was obtained by using Equation \ref{Eq:1} for the sunspot with a circular cross-sectional shape (see Table \ref{Table1}), and Equation \ref{Eq:2} for the sunspot with an elliptical cross-sectional shape (see Table \ref{Table2}). Here $\omega=2\pi f$ is the angular frequency, $f$ refers to frequency in Hz, $c_i=10$ km s$^{-1}$ is the assumed sound speed, $v_{A_i}=4c_i$ is the Alfv{\'e}n speed and $\sigma^{2}=0.4174$. With the help of these quantities the wavelength ($\lambda=2\pi/k_z$) of waves and the phase speed ($V_{ph}=f \lambda$) were calculated for the MHD modes identified by our analysis. 
The three-dimensional visualisations of the POD and DMD modes for the circular and the elliptical sunspots are provided in the Appendix (see  Figures \ref{fig:C_3D} and \ref{fig:E_3D}). The 3D surface were immersed in the volume rendering of the theoretical MHD wave model of the irregular cross-sectional shape.

\begin{figure*}[t!]
\centering
 \begin{tabular}{ccc}
 \includegraphics[width=40mm]{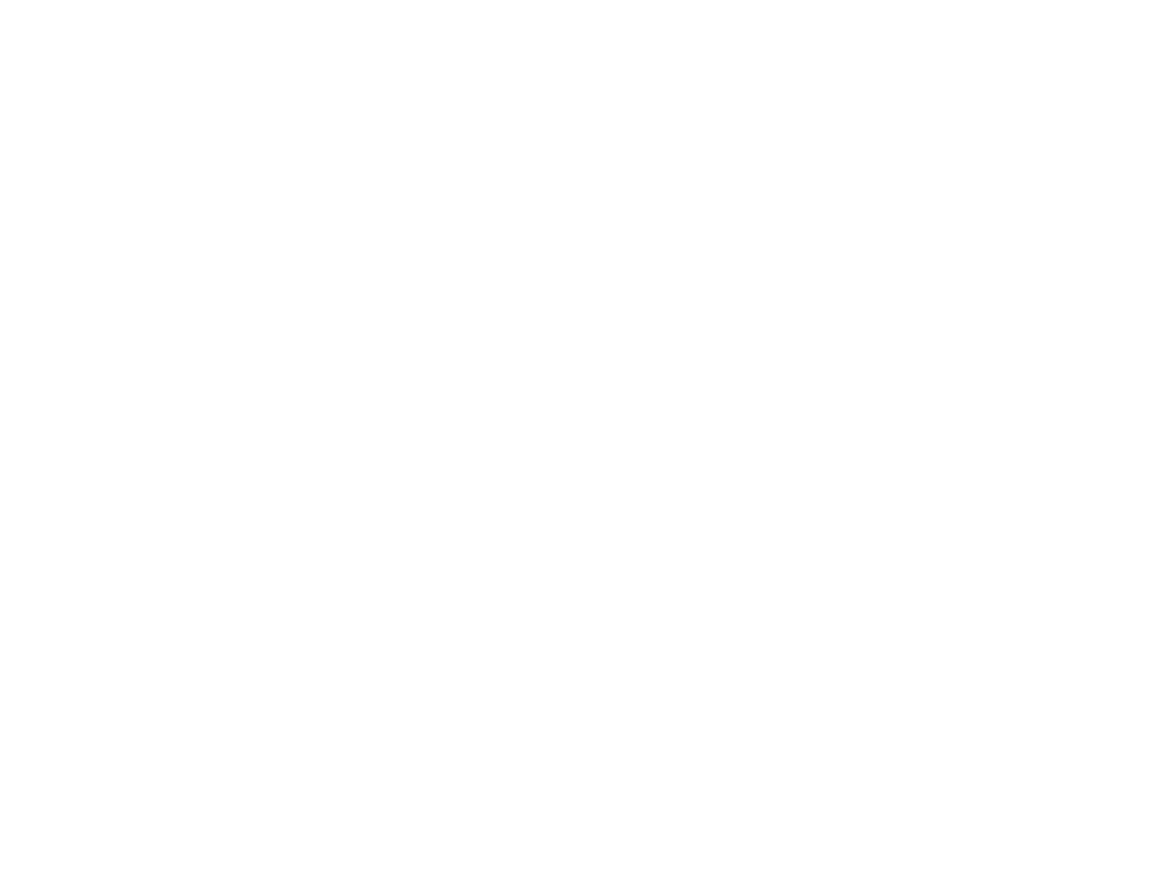}&
 \includegraphics[width=40mm]{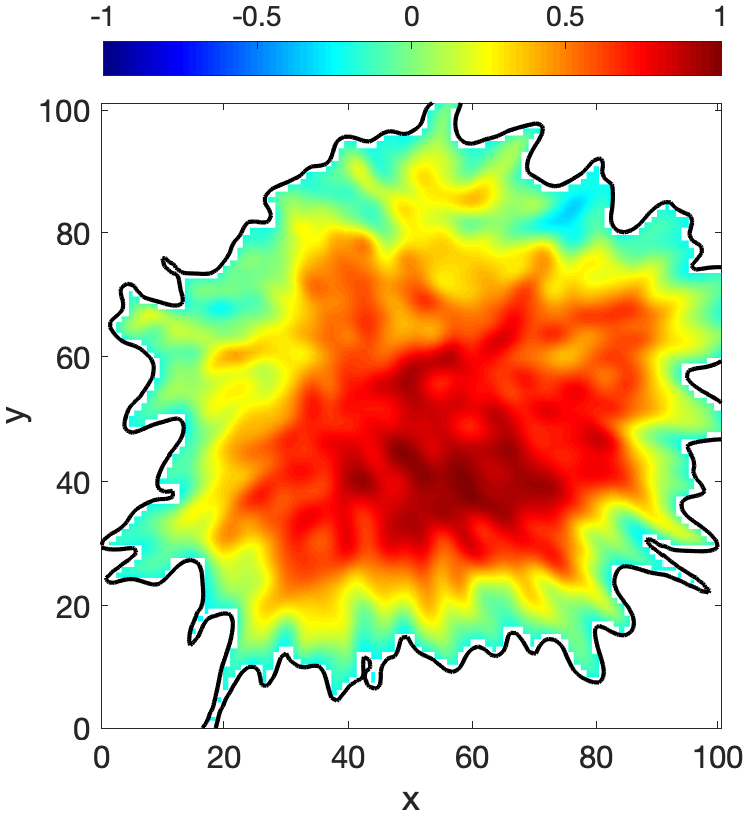}&
 \includegraphics[width=40mm]{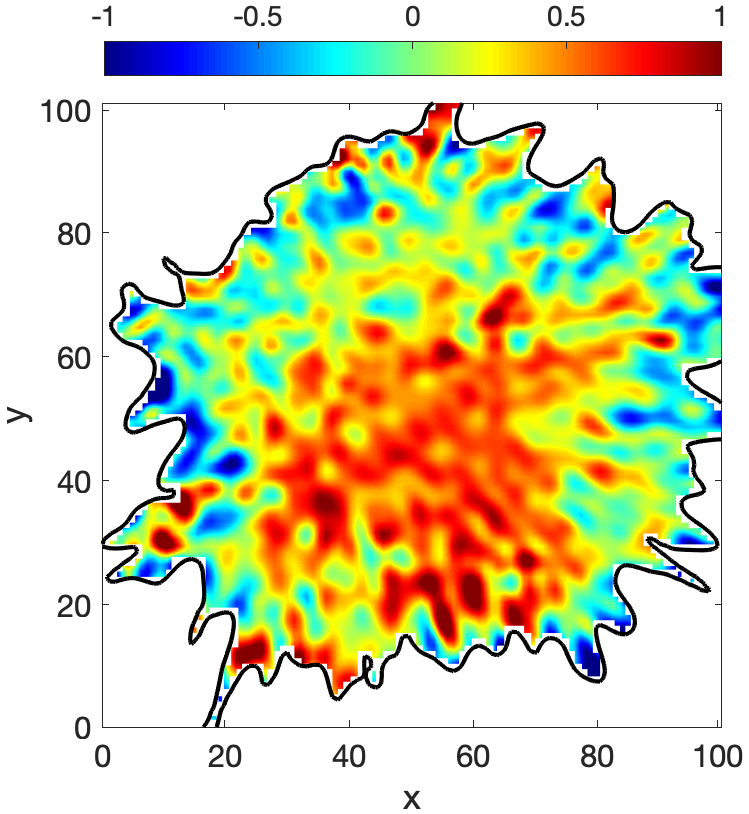}\\
 \includegraphics[width=40mm]{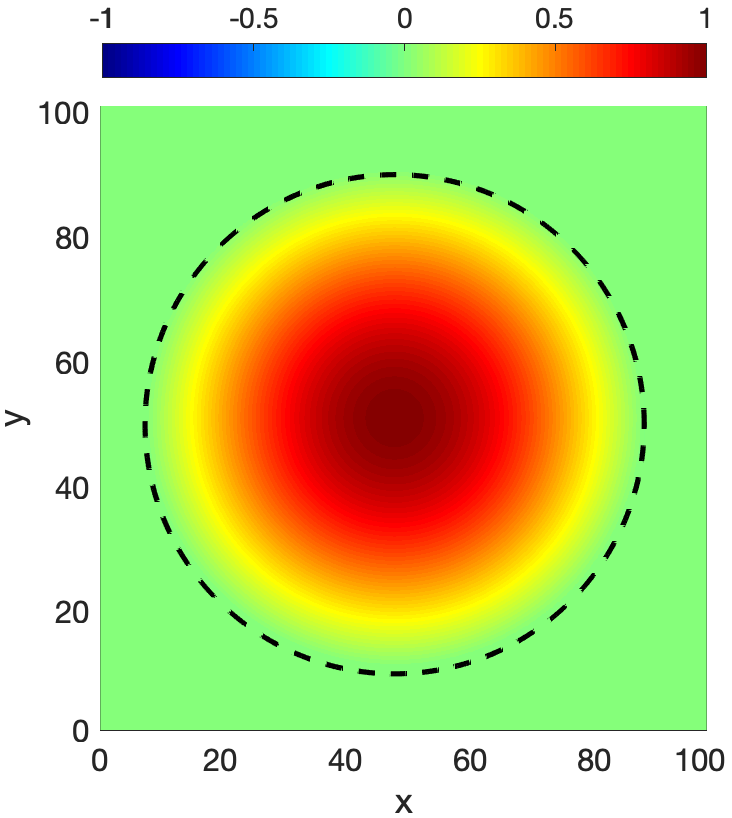}&
 \includegraphics[width=40mm]{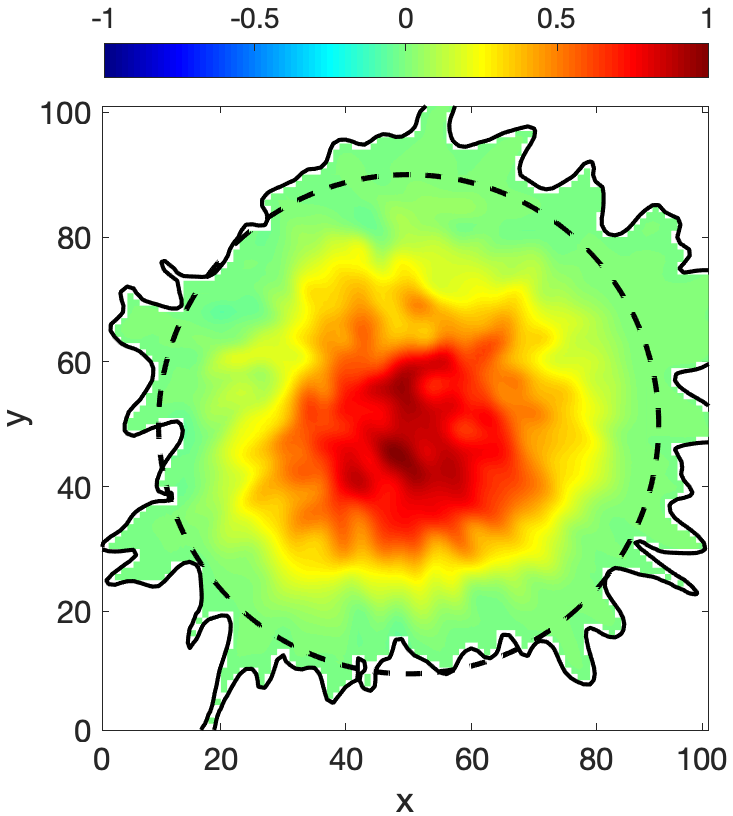}&
 \includegraphics[width=40mm]{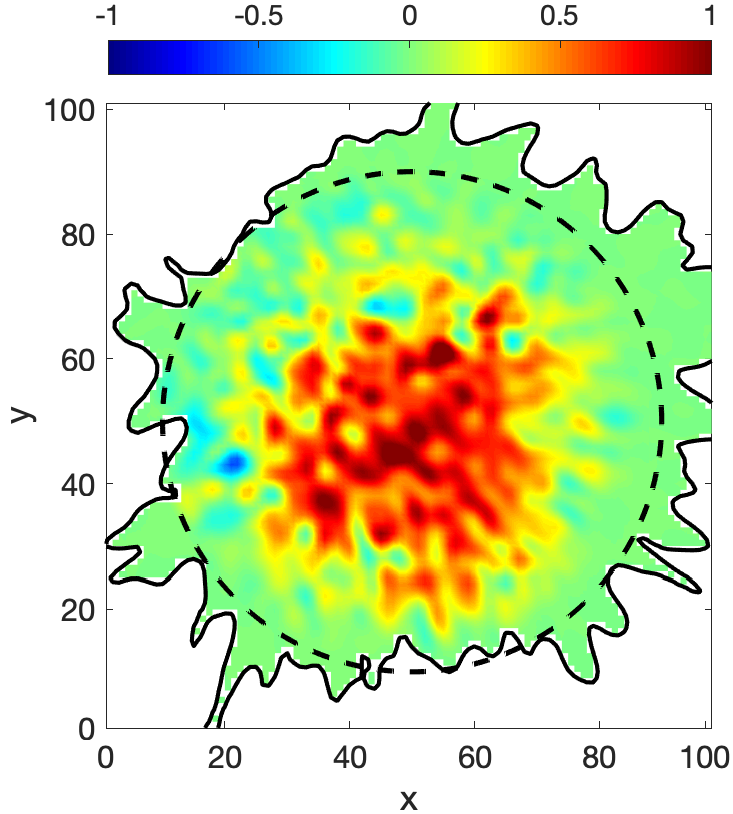}\\
 \includegraphics[width=40mm]{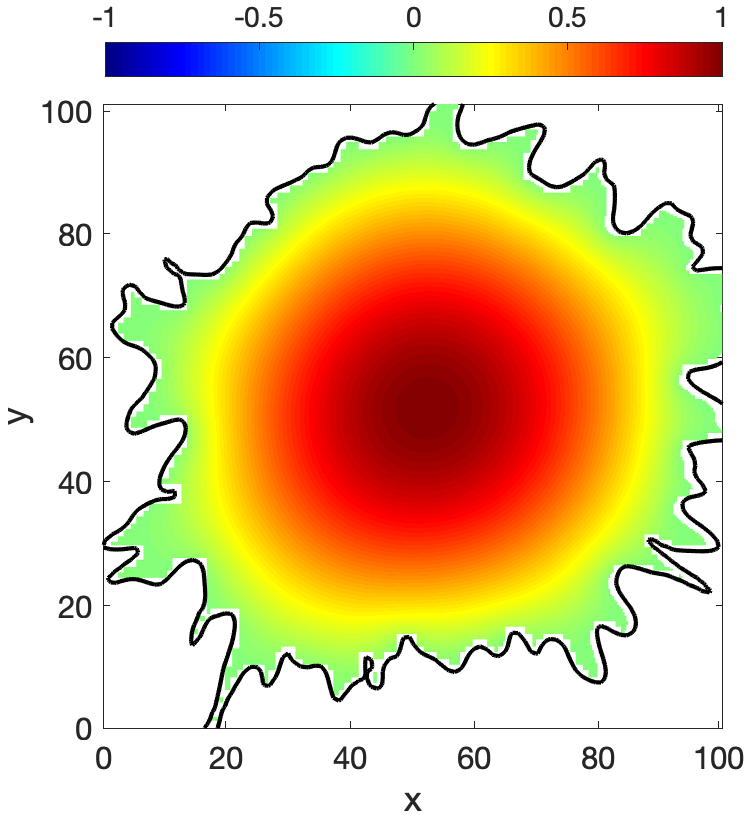}&
 \includegraphics[width=40mm]{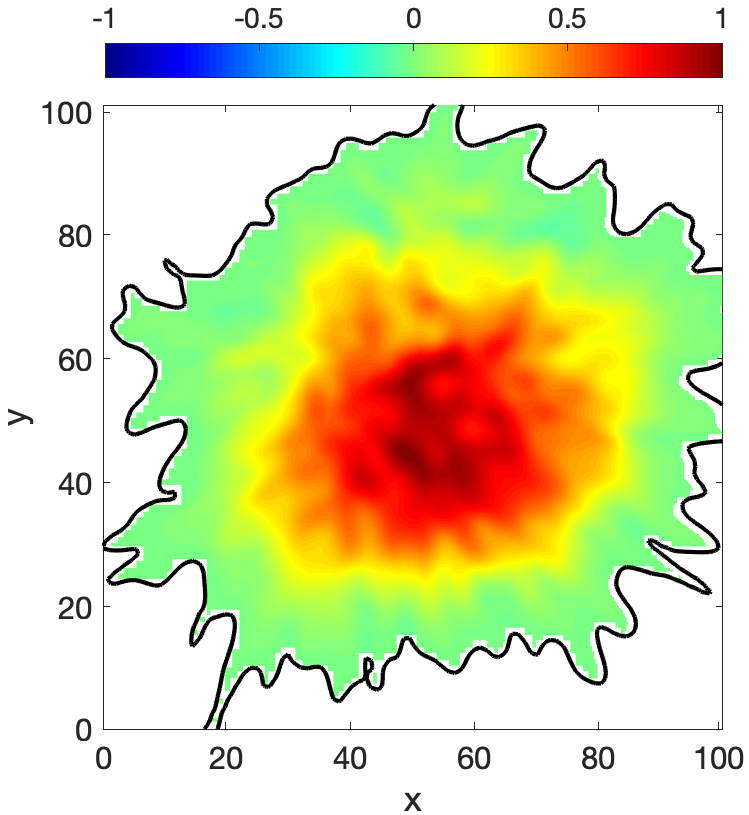}&
 \includegraphics[width=40mm]{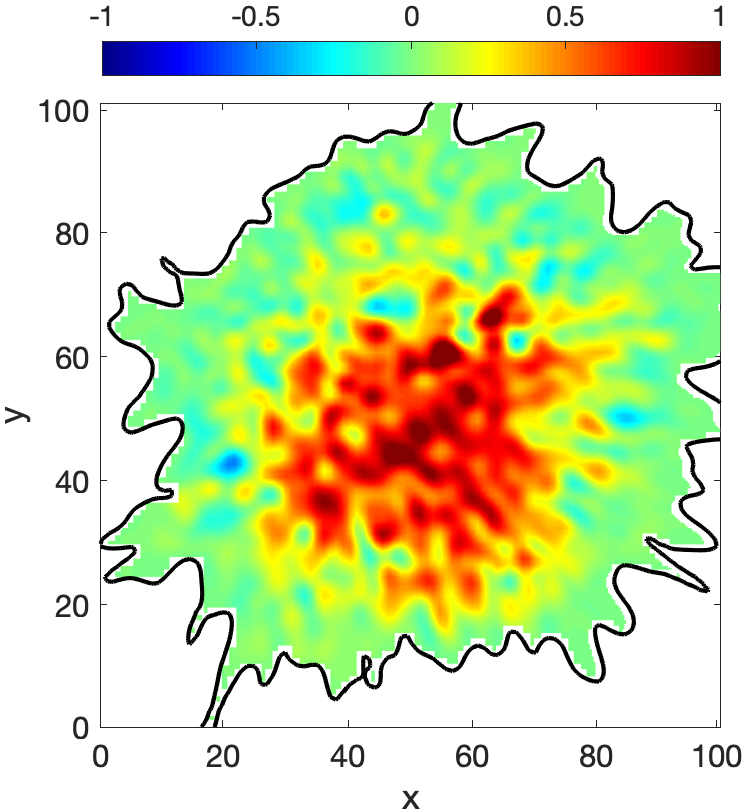}\\
\end{tabular} \label{fig:fund_Saus_Circular}
\caption{The first row displays the spatial structure of the modes that were detected from the observational data, the first POD mode (middle) and the DMD mode that corresponds to the frequency of 4.8 mHz (right) \cite{albidah2020RS}. In the first column we display the theoretical spatial structure of the  fundamental slow body sausage mode in the cylindrical magnetic flux tube model (middle) and the corresponding structure considering the realistic sunspot with irregular shape (bottom). The rest of the panels are showing the cross-correlation between theoretically constructed and observationally detected modes and the positive/negative numbers in the colourbar denote correlation/anti-correlation. The dashed circles show the boundary of the tube and the solid black line shows the umbra/penumbra boundary. The 3D visualisation of this mode can be found in Figure \ref{fig:C_3D} in the Appendix. The same configuration was used for Figures \ref{fig:fund_kink_Circular} to \ref{fig:fluting(n=3)_circular}.} \label{fig:fund_sausage_Circular}
\end{figure*}

\begin{figure*}[!t]
\centering
 \begin{tabular}{ccc}
 \includegraphics[width=40mm]{C_blank_white.png}&
 \includegraphics[width=40mm]{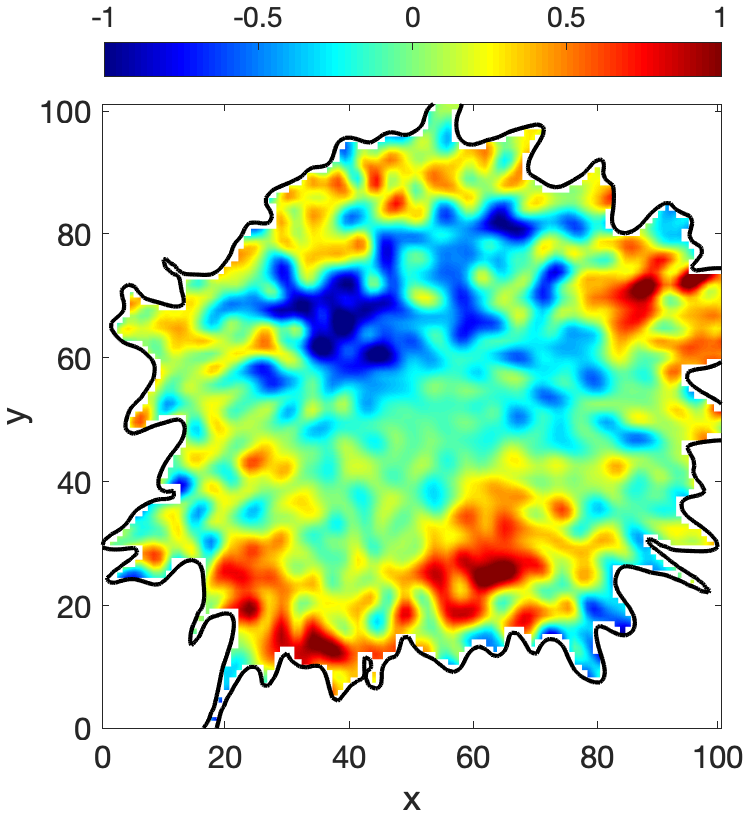}&
 \includegraphics[width=40mm]{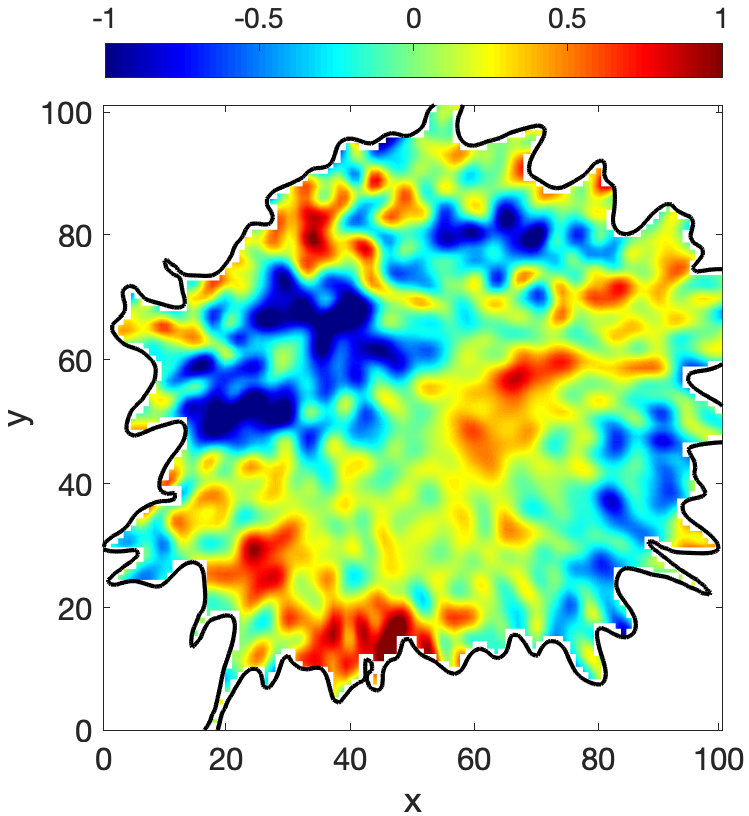}\\
 \includegraphics[width=40mm]{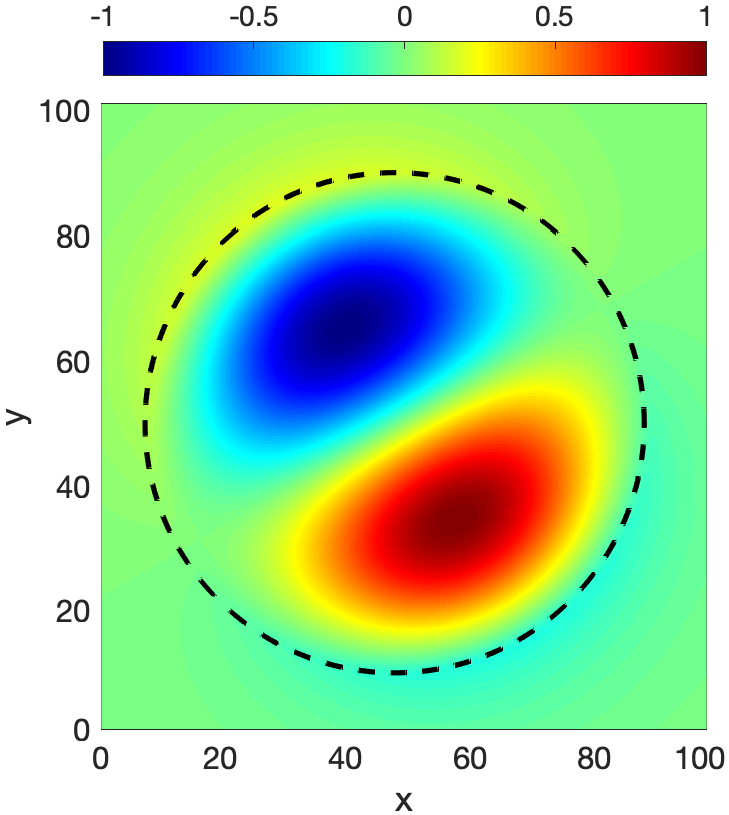}&
 \includegraphics[width=40mm]{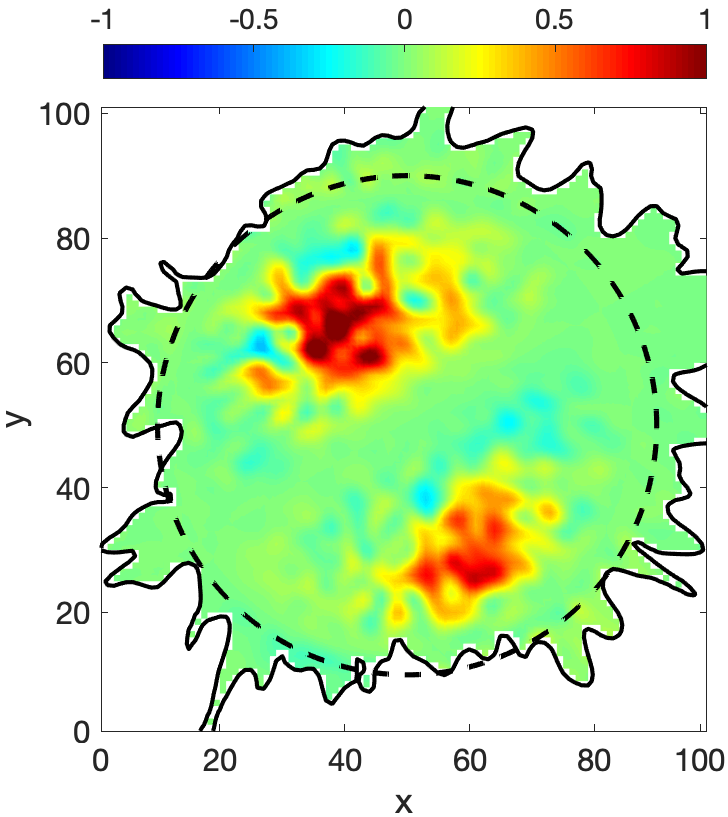}&
 \includegraphics[width=40mm]{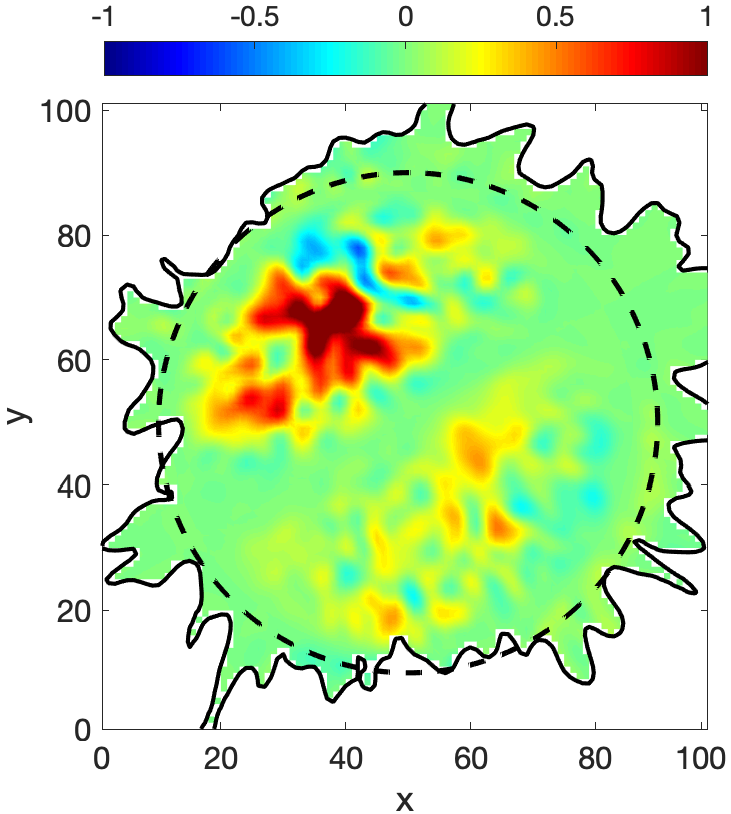}\\
 \includegraphics[width=40mm]{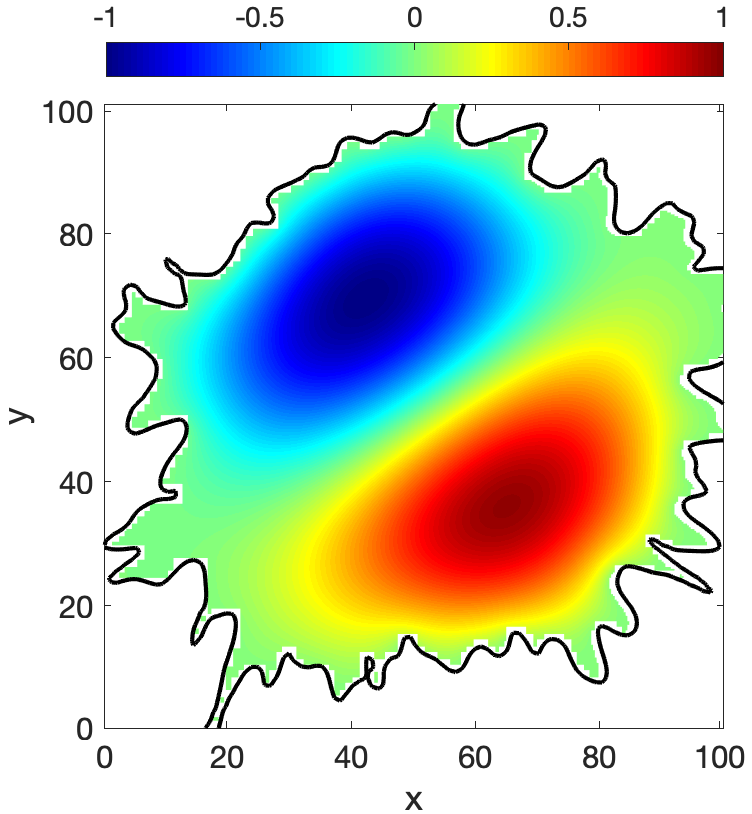}&
 \includegraphics[width=40mm]{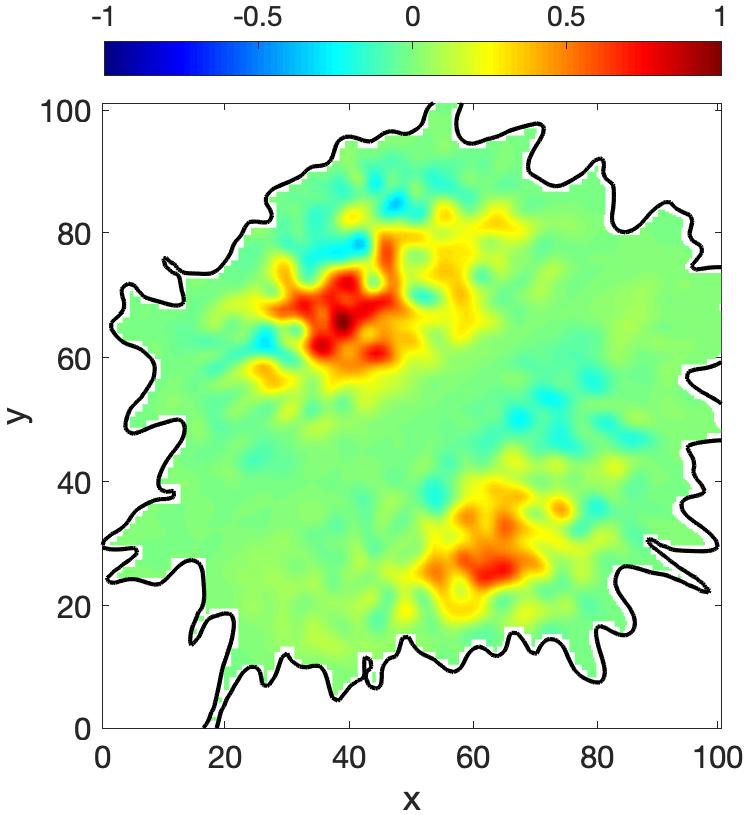}&
 \includegraphics[width=40mm]{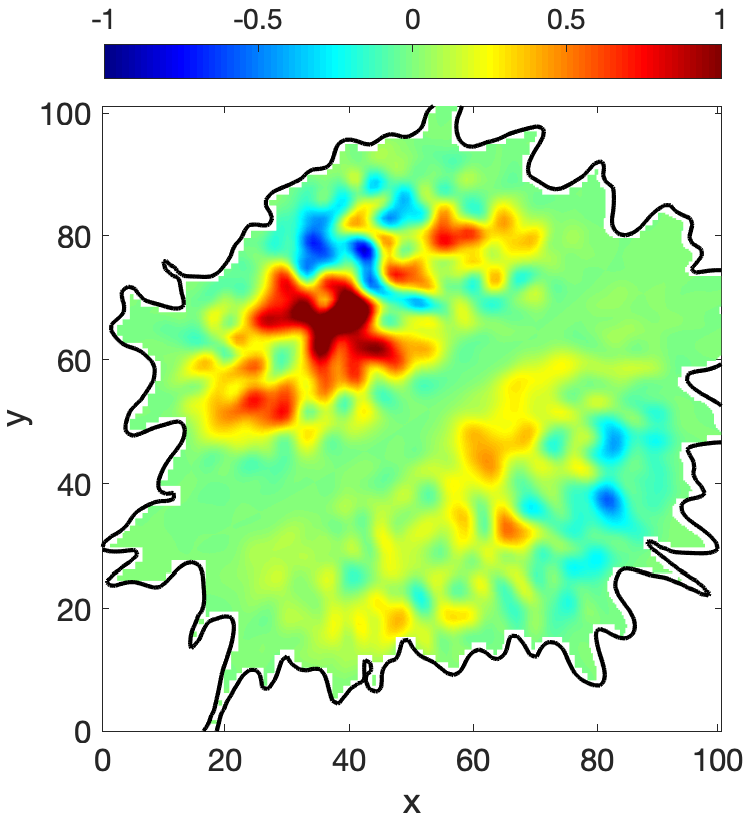}\\
\end{tabular}
\caption{This figure displays the 13$^{th}$ POD (top row, 1st panel) and DMD mode (top row, 2nd panel) with a frequency of 6 mHz, which has an azimuthal symmetry of the  fundamental slow body kink mode \cite{albidah2020RS}. The 3D visualisation of this mode is shown in Figure \ref{fig:C_3D} in the Appendix}
\label{fig:fund_kink_Circular}
\end{figure*}

 Let us start with the sunspot with a circular cross-section shape shown in the upper left panel of Figure \ref{fig:snapshot_circular&E}. The analysis was applied on the ROI represented by the blue box on the upper right panel of the same figure where the umbra/penumbra boundary is shown by a solid black line with an intensity threshold level at 0.85. In addition to the \textit{fundamental slow body sausage mode} (shown in Figure \ref{fig:fund_sausage_Circular}), and the \textit{ fundamental slow body kink mode} (Figure \ref{fig:fund_kink_Circular}) identified previously  \citep[see e.g.][]{albidah2020RS}, the POD and DMD analysis reveals the existence of the higher-order MHD wave modes.

\begin{figure*}[!t]
\centering
 \begin{tabular}{ccc}
 \includegraphics[width=40mm]{C_blank_white.png}&
 \includegraphics[width=40mm]{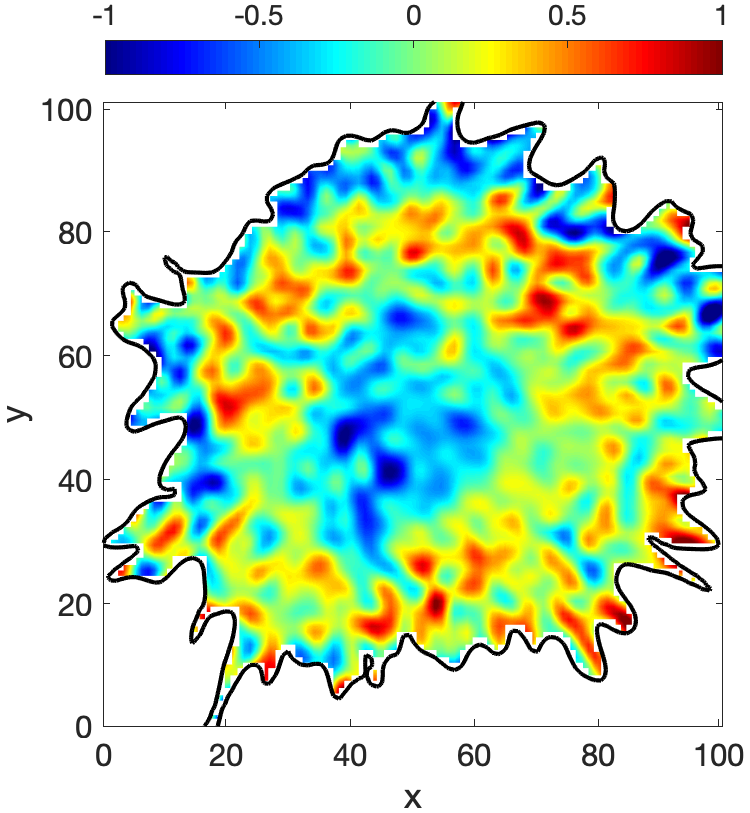}&
 \includegraphics[width=40mm]{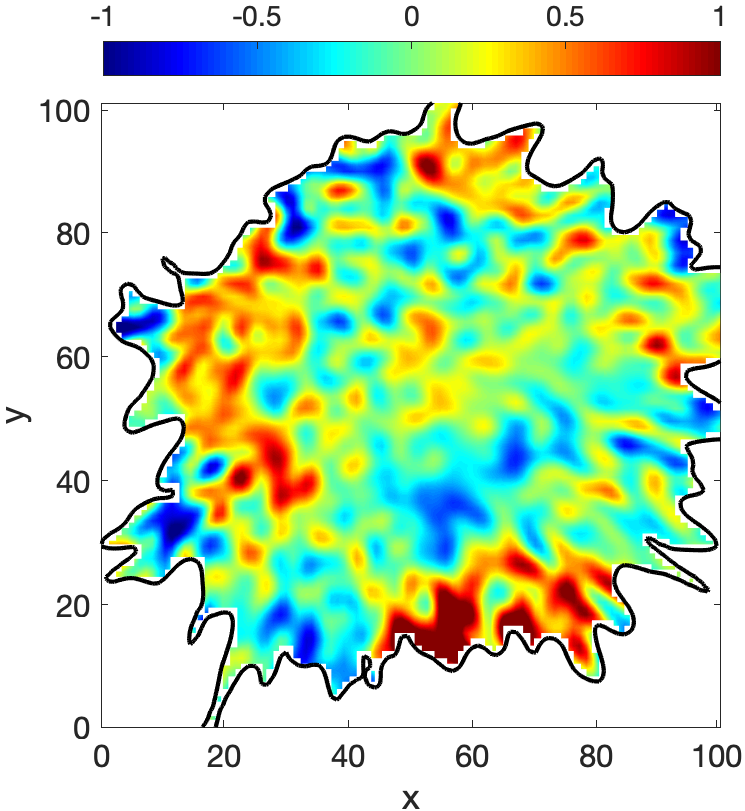}\\
 \includegraphics[width=40mm]{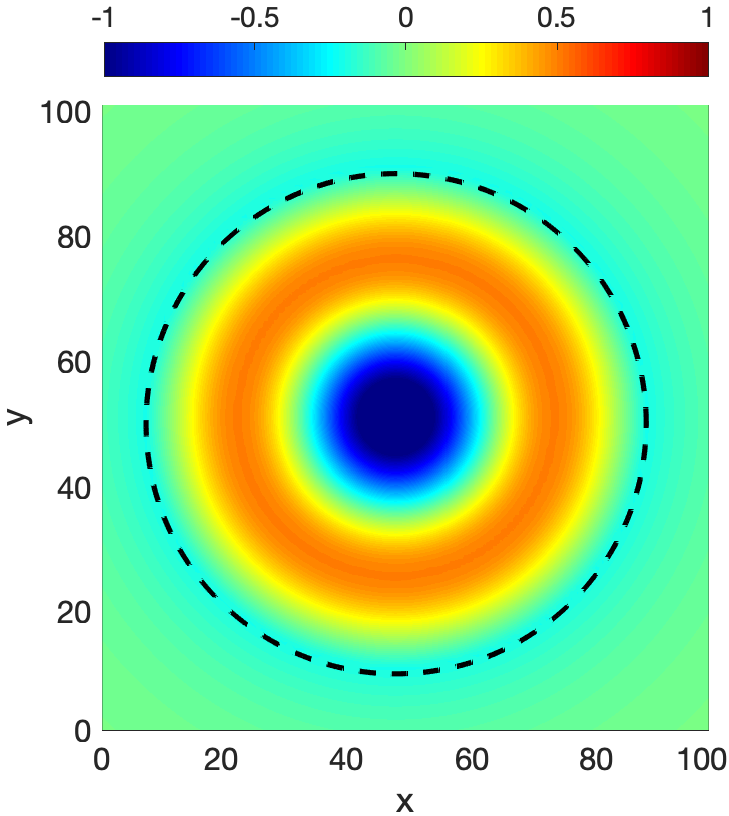}&
 \includegraphics[width=40mm]{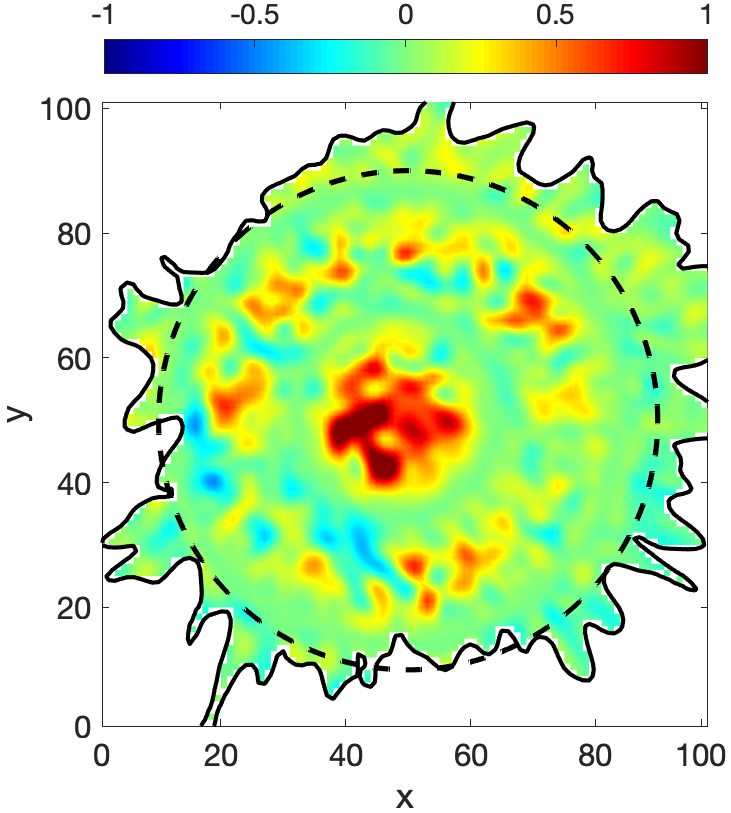}&
 \includegraphics[width=40mm]{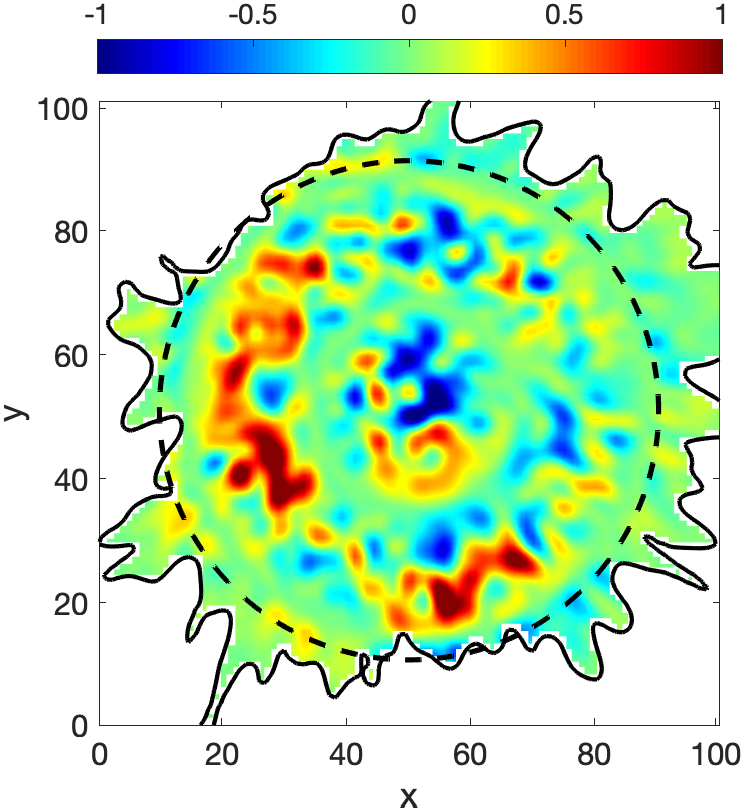}\\
 \includegraphics[width=40mm]{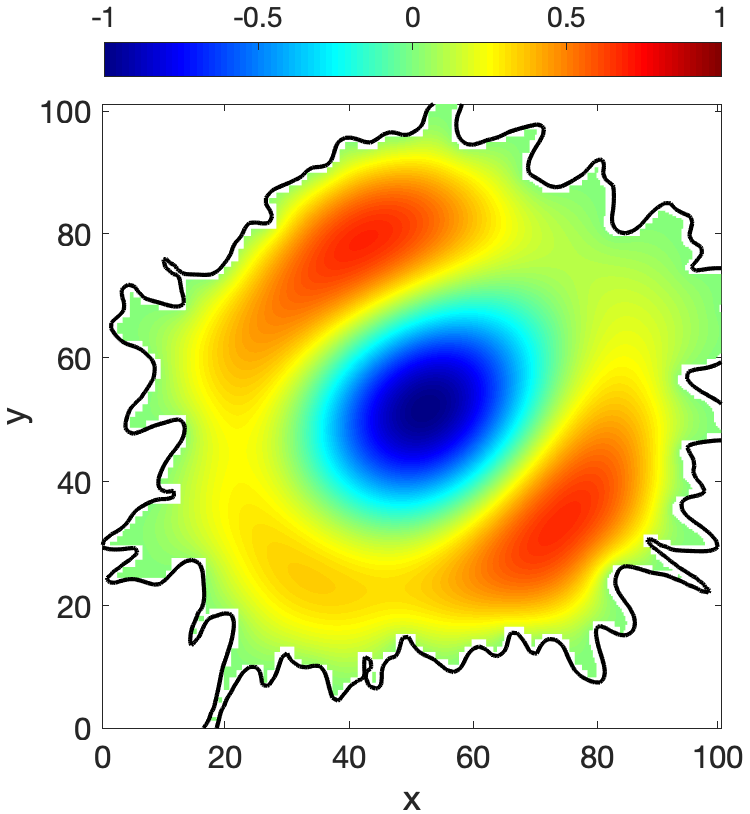}&
 \includegraphics[width=40mm]{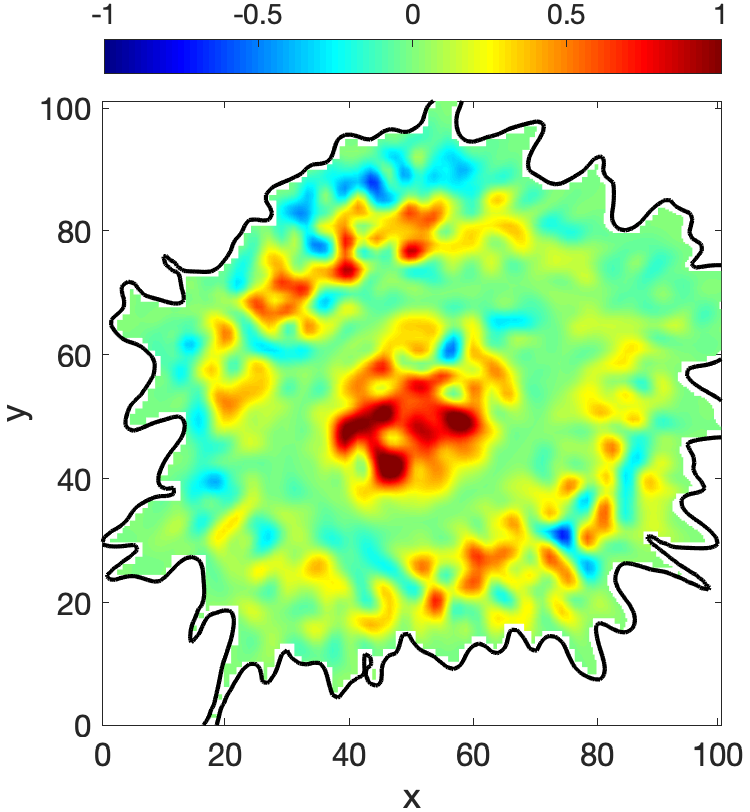}&
 \includegraphics[width=40mm]{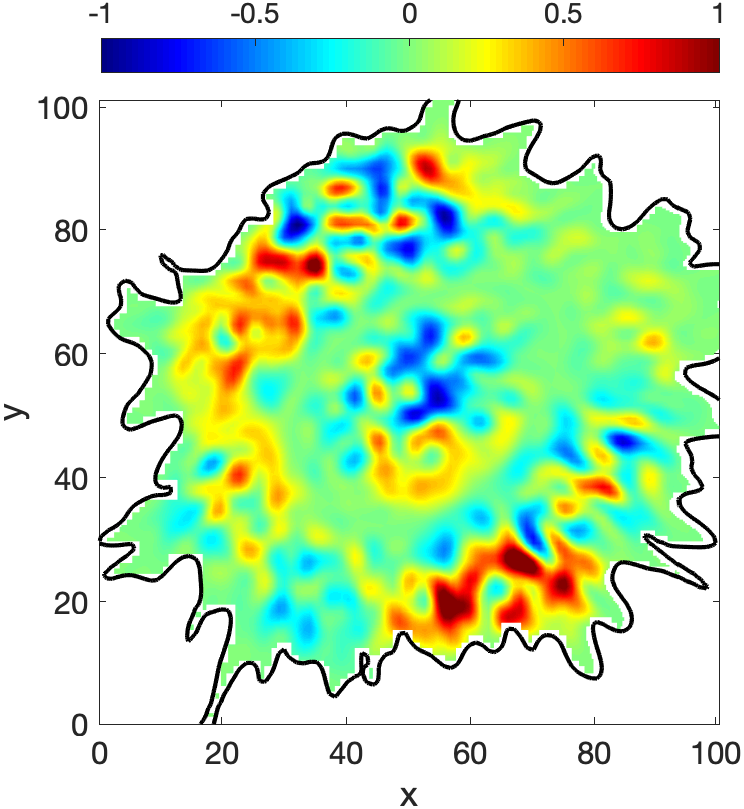}\\
\end{tabular}
\caption{This figure displays the 19$^{th}$ POD (top row, 1st panel) and DMD mode (top row, 2nd panel) with a frequency of 5.6 mHz which has an azimuthal symmetry of the slow body sausage overtone mode. The 3D visualisation of this mode is shown in Figure \ref{fig:C_3D} in the Appendix}
\label{fig:saus_overtone_Circular}
\end{figure*}

\begin{figure*}[!t]
  \centering
  \begin{tabular}{ccc}
     \includegraphics[width=55mm]{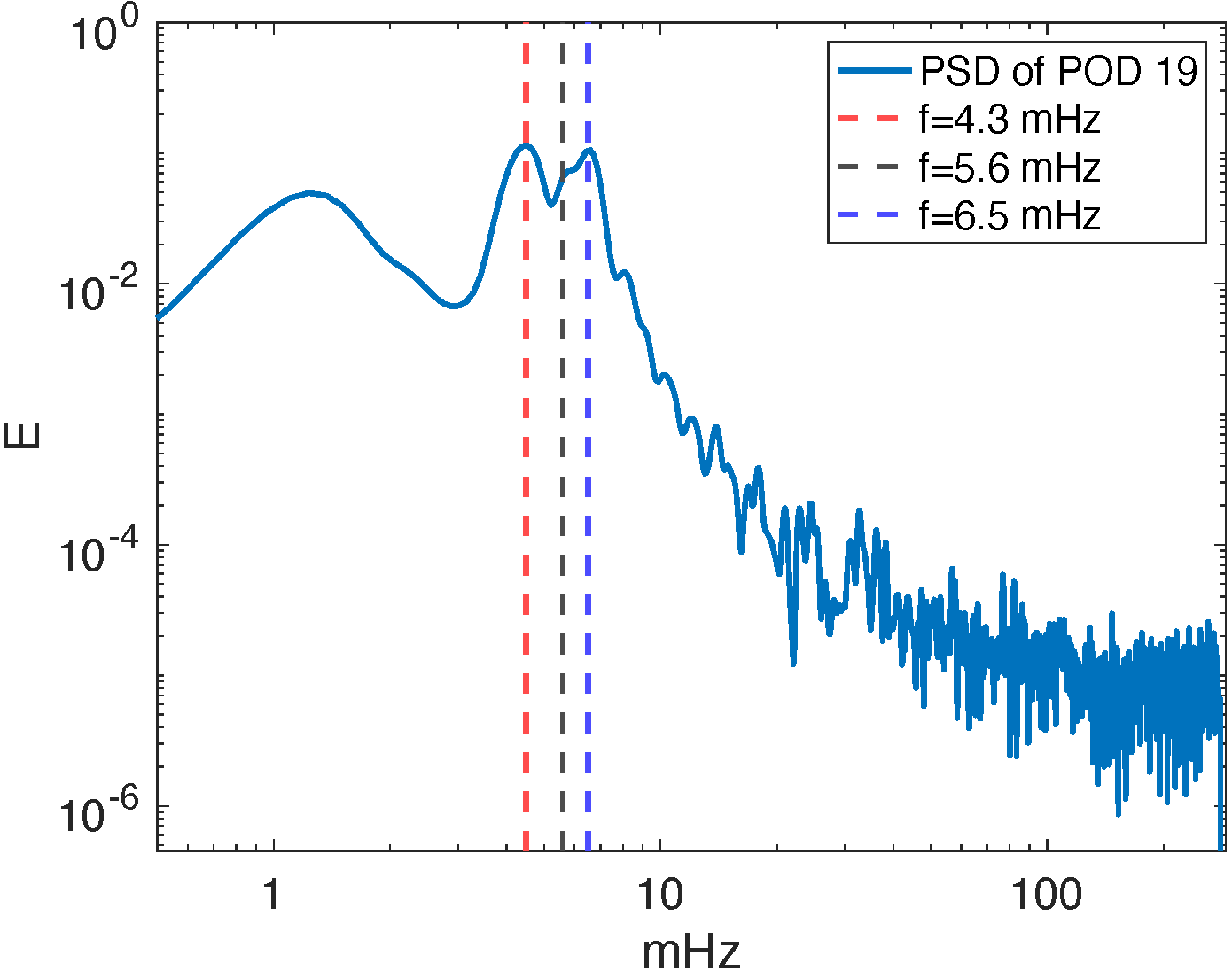}&
     \includegraphics[width=55mm]{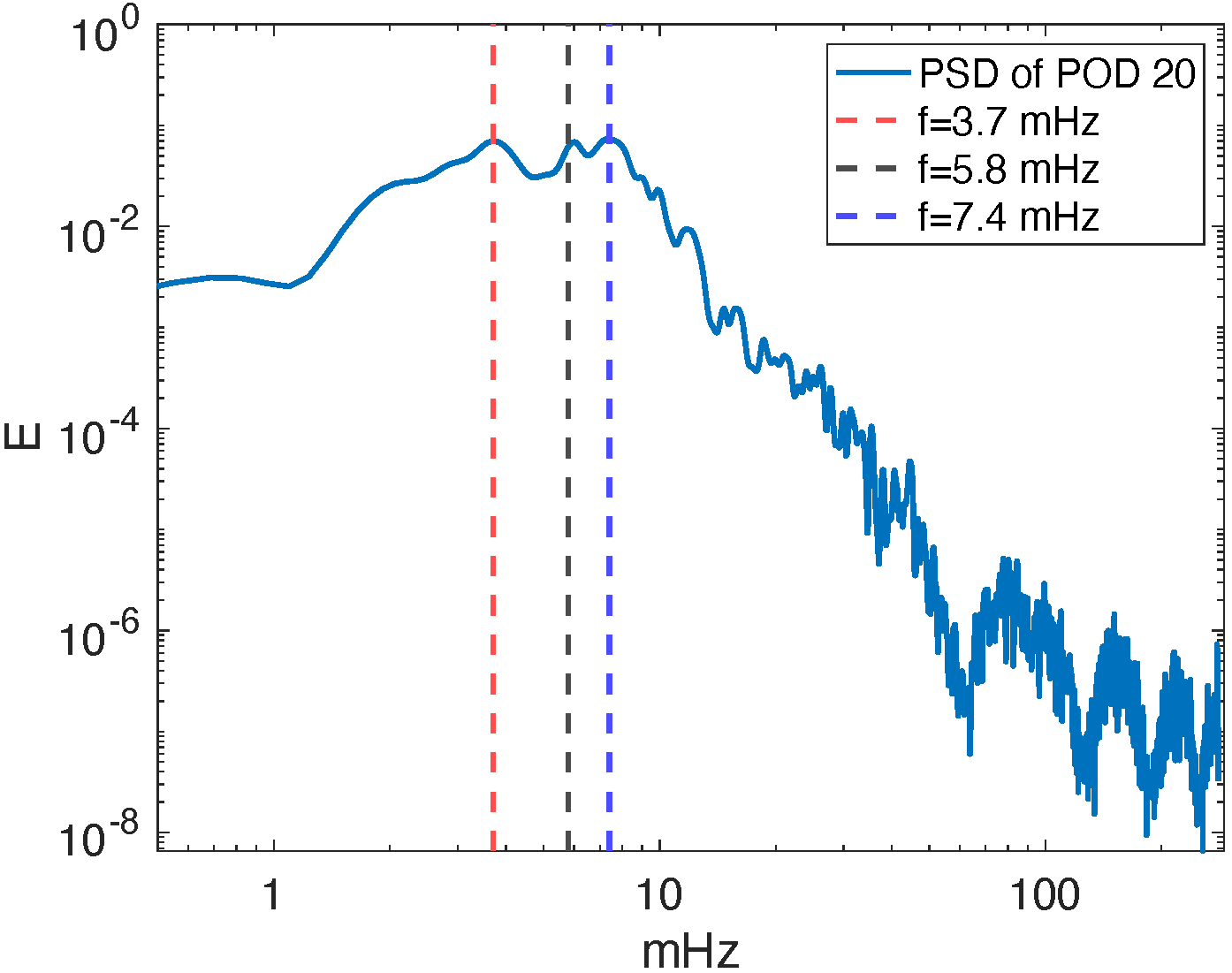}&
     \includegraphics[width=55mm]{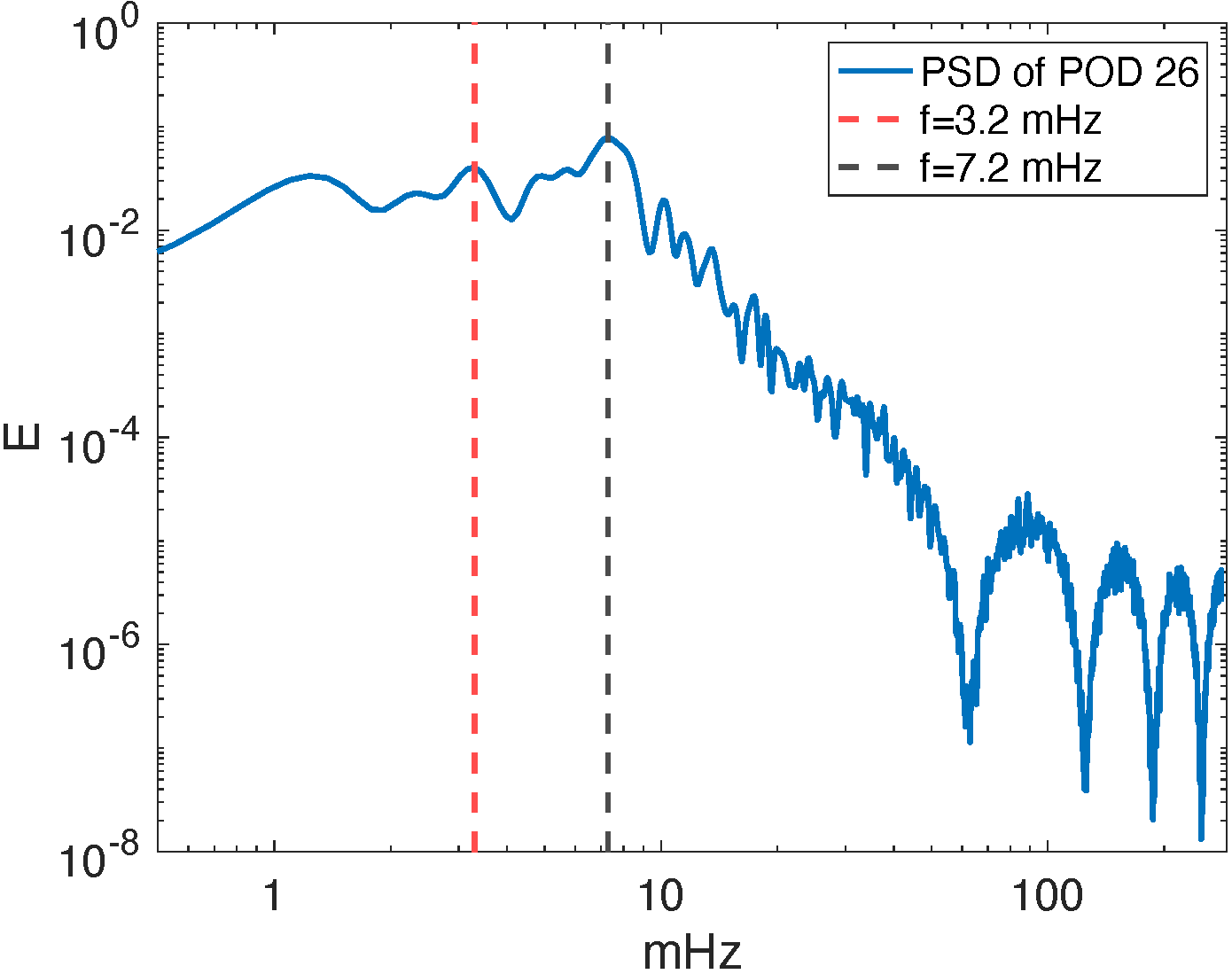}\\
    \end{tabular}
 \caption{This figure shows the power spectrum density (PSD) of the time coefficients of POD 19 (left panel), POD 20 (middle panel) and POD 26 (right panel) modes. The vertical coloured-dash lines represent the values in the frequency domain that correspond to the peaks of the PSD, and the values of the peaks' location are shown in the legend of each figure. The PSD of POD 1 and POD 13 (the fundamental sausage and kink modes) can be found in our earlier study \citep[][]{albidah2020RS}.} \label{PSD_circular}
\end{figure*}

The POD mode that can be interpreted as a MHD wave mode is the 19$^{th}$ mode which has the azimuthal symmetry of a \textit{slow body sausage overtone mode}, i.e. a mode with more than one radial node, and the DMD mode that corresponds to the spatial structure has a frequency of 5.6 mHz, as shown in Figure \ref{fig:saus_overtone_Circular}. The power spectrum density (PSD) of the time coefficient of POD 19 has a mix of peaks around 4.3 mHz, 5.4 mHz and 6.5 mHz on its frequency domain, as shown in the left panel of Figure \ref{PSD_circular}. 

\begin{figure*}[!t]
 \centering
 \begin{tabular}{ccc}
 \includegraphics[width=40mm]{C_blank_white.png}&
 \includegraphics[width=40mm]{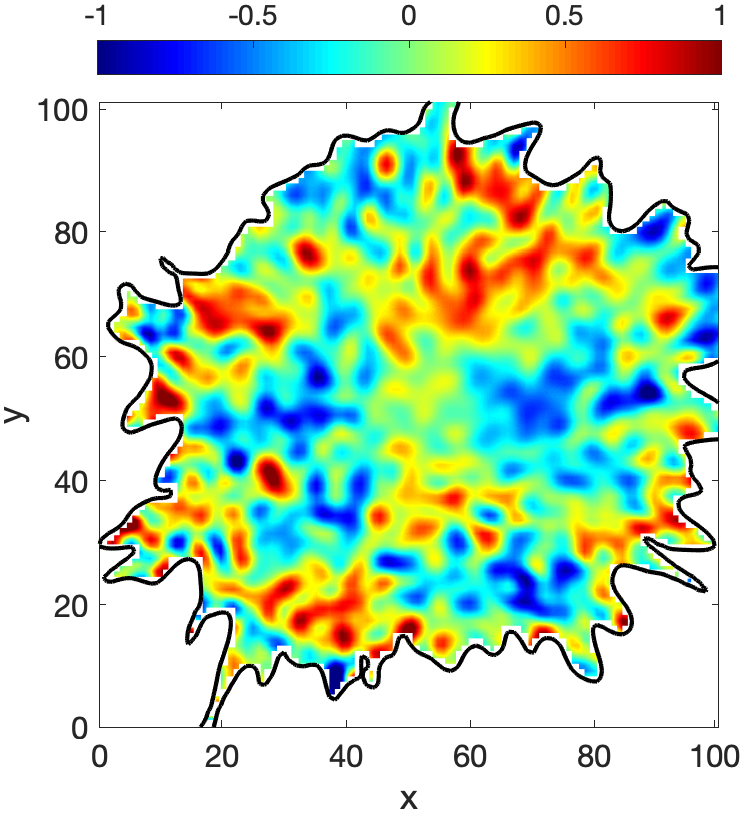}&
 \includegraphics[width=40mm]{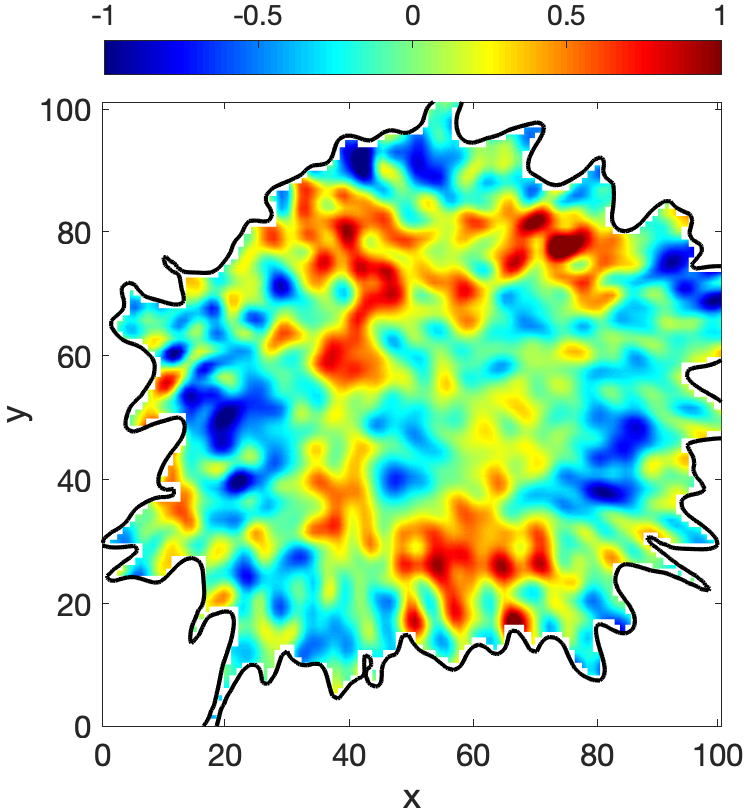}\\
 \includegraphics[width=40mm]{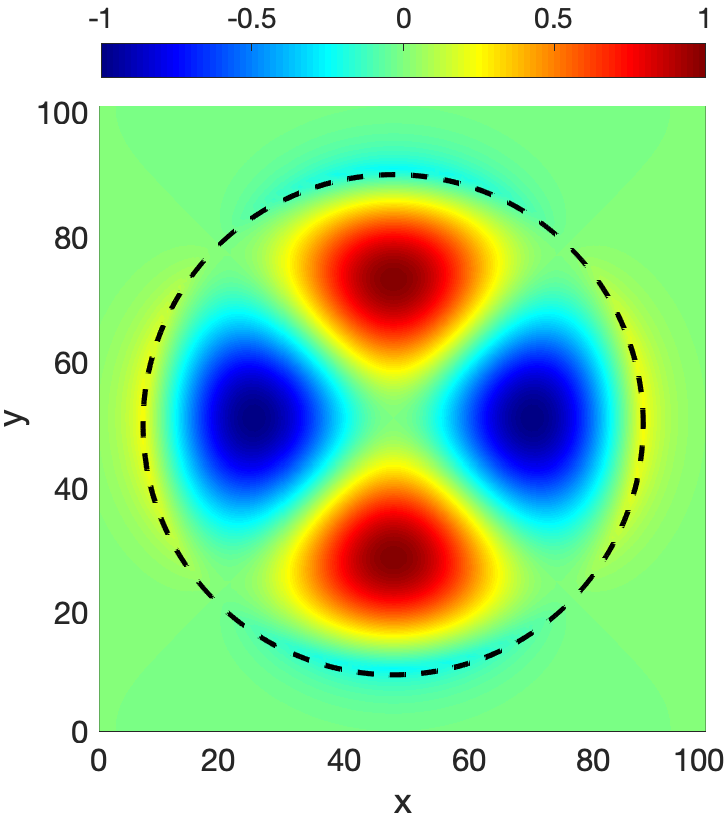}&
 \includegraphics[width=40mm]{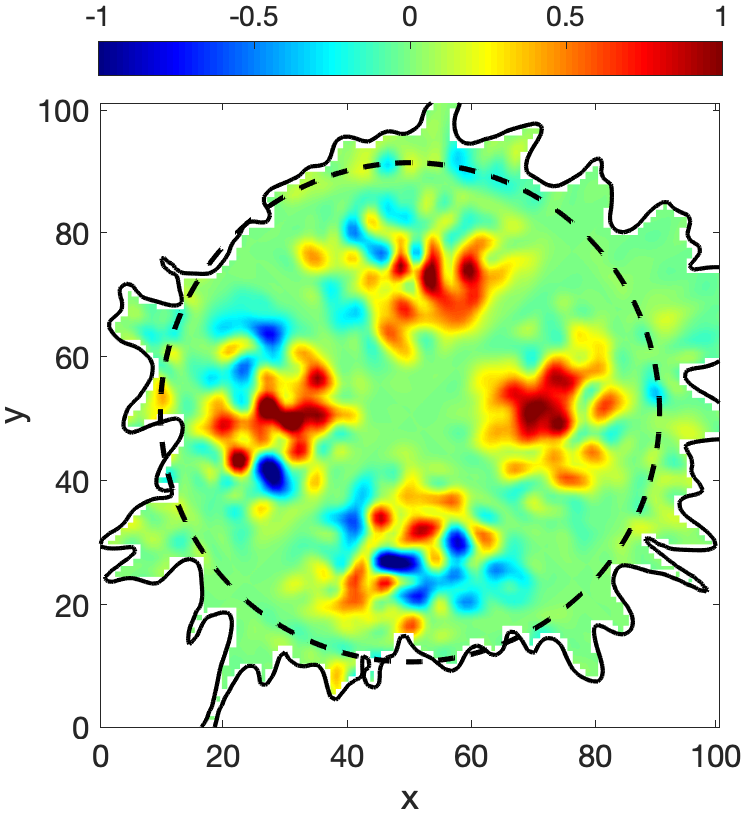}&
 \includegraphics[width=40mm]{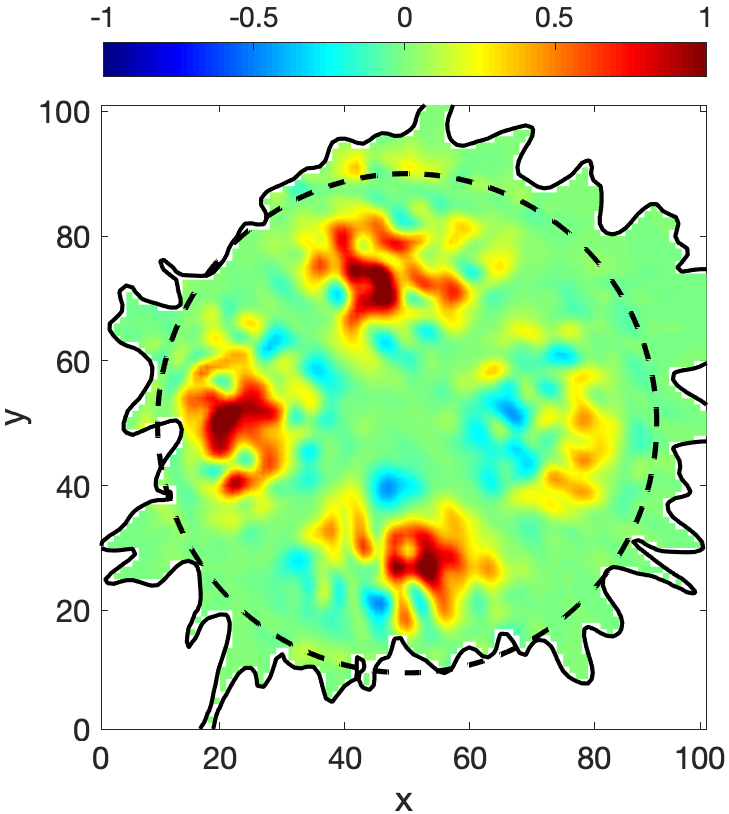}\\
 \includegraphics[width=40mm]{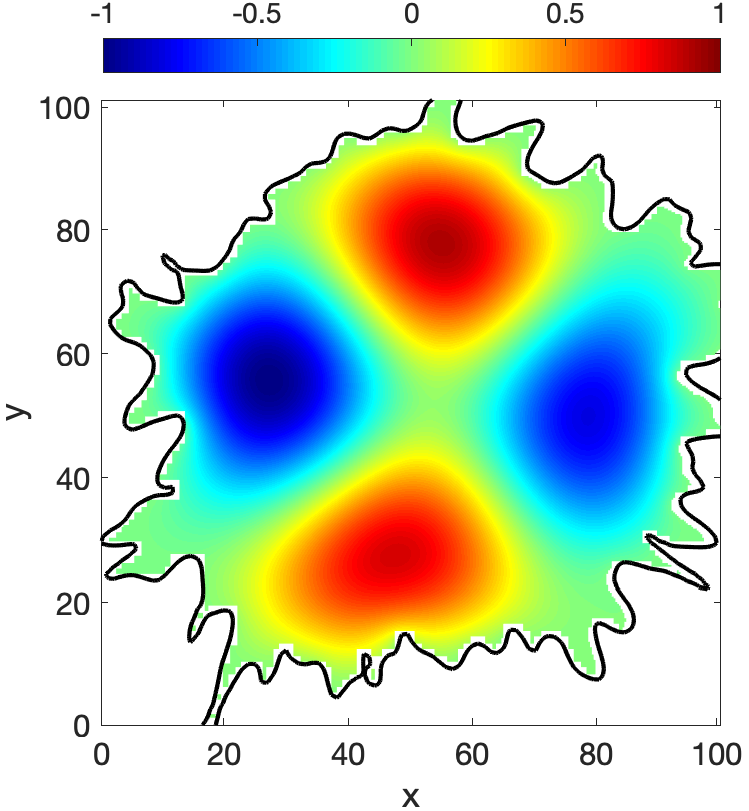}&
 \includegraphics[width=40mm]{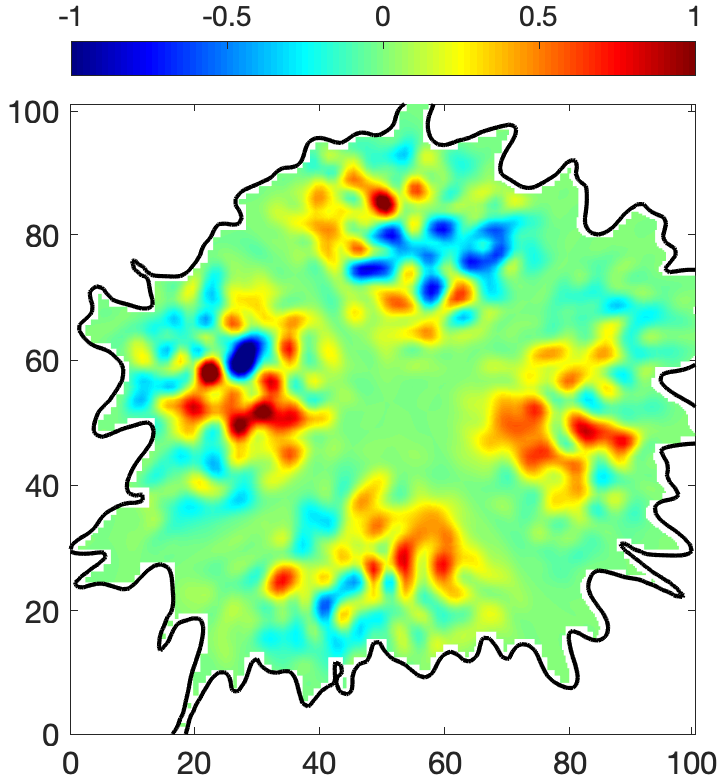}&
 \includegraphics[width=40mm]{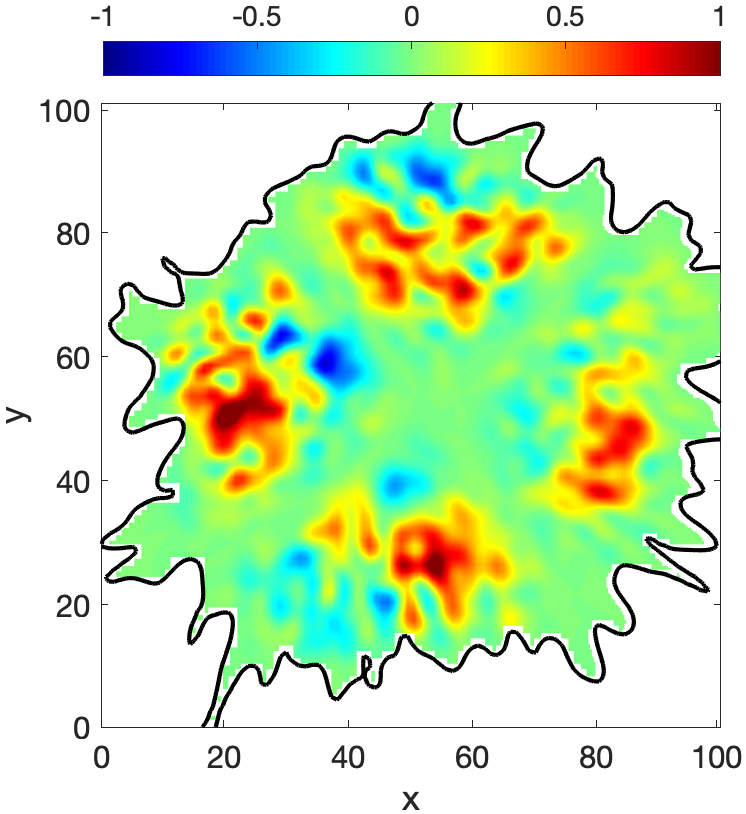}\\
\end{tabular}
\caption{This figure displays the 20$^{th}$ POD (top row, 1st panel) and DMD mode (top row, 2nd panel) with a  frequency of 7.6 mHz, which has an azimuthal symmetry of the slow body fluting mode ($n=2$). The 3D visualisation of this mode is shown in Figure \ref{fig:C_3D} in the Appendix.}
\label{fig:fluting(n=2)_circular}
\end{figure*}

\begin{figure*}[!t]
\centering
\begin{tabular}{ccc}
\includegraphics[width=40mm]{C_blank_white.png}&
\includegraphics[width=40mm]{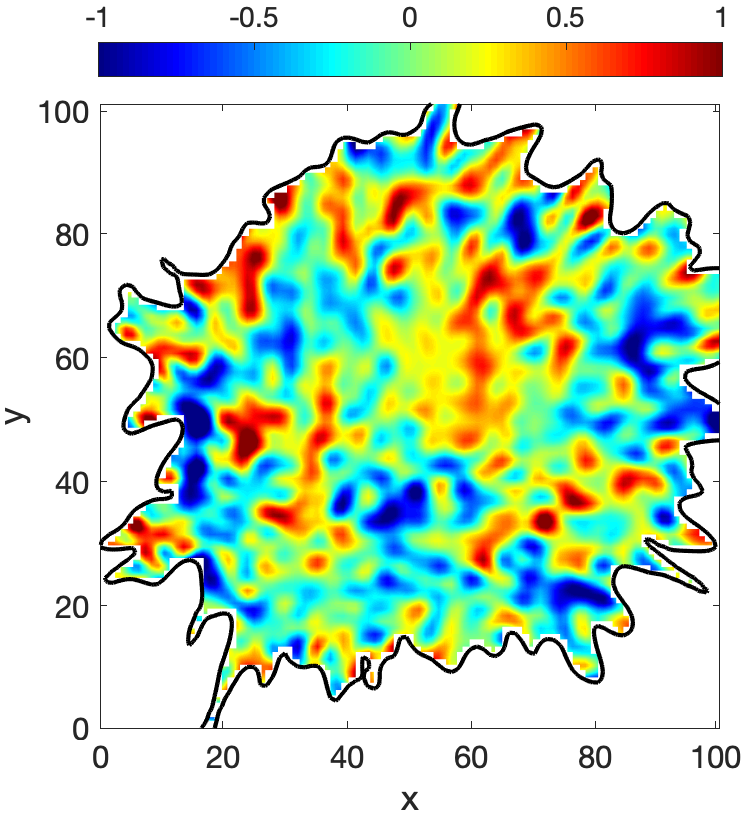}&
\includegraphics[width=40mm]{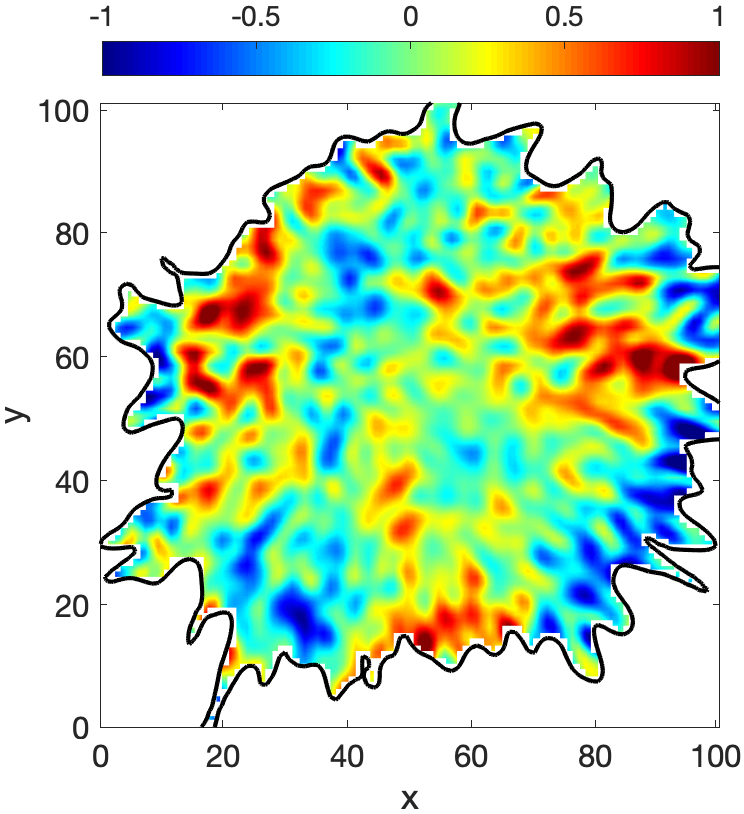}\\
\includegraphics[width=40mm]{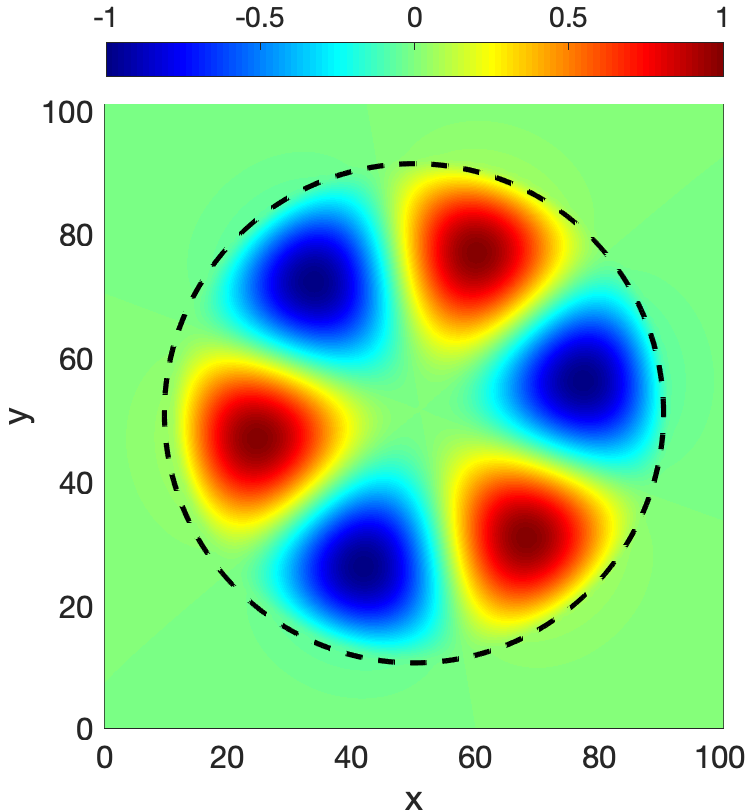}&
\includegraphics[width=40mm]{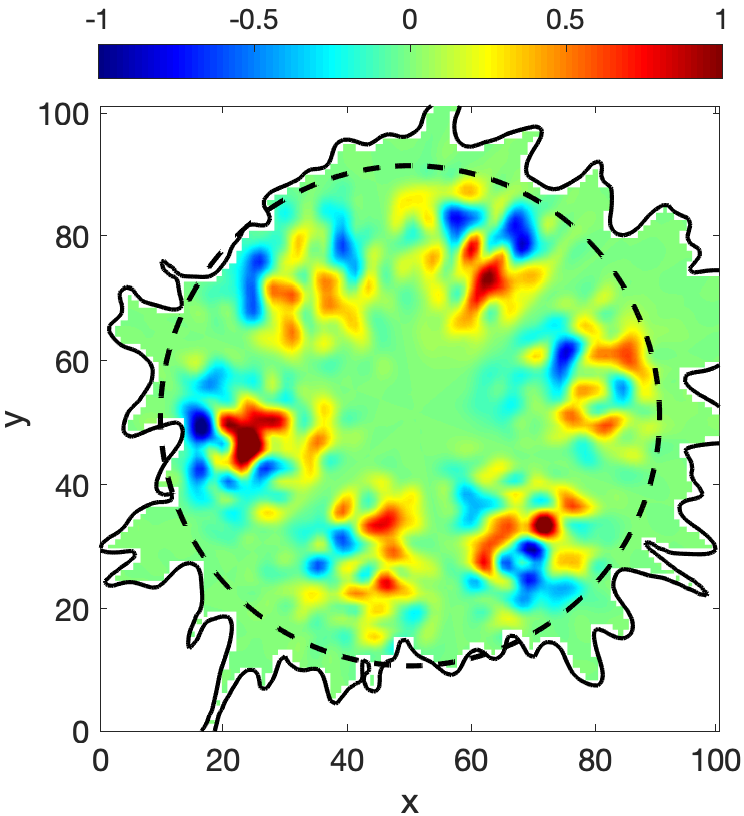}&
\includegraphics[width=40mm]{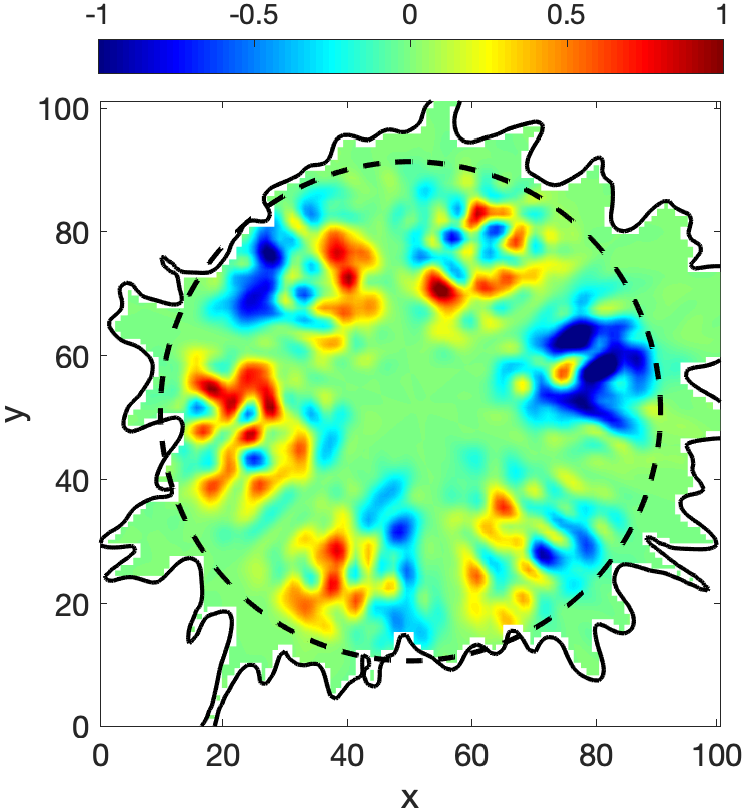}\\
\includegraphics[width=40mm]{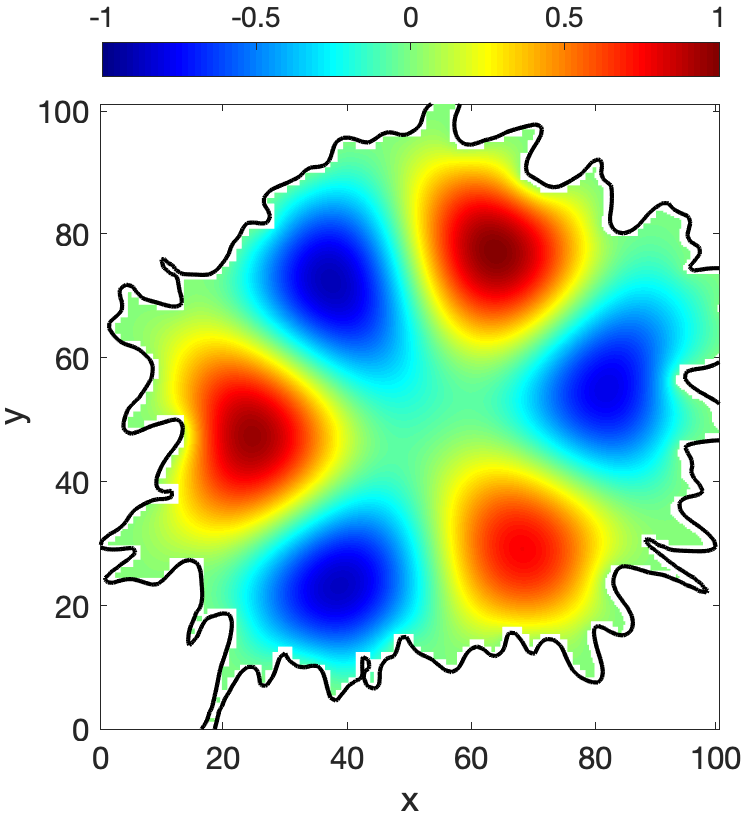}&
\includegraphics[width=40mm]{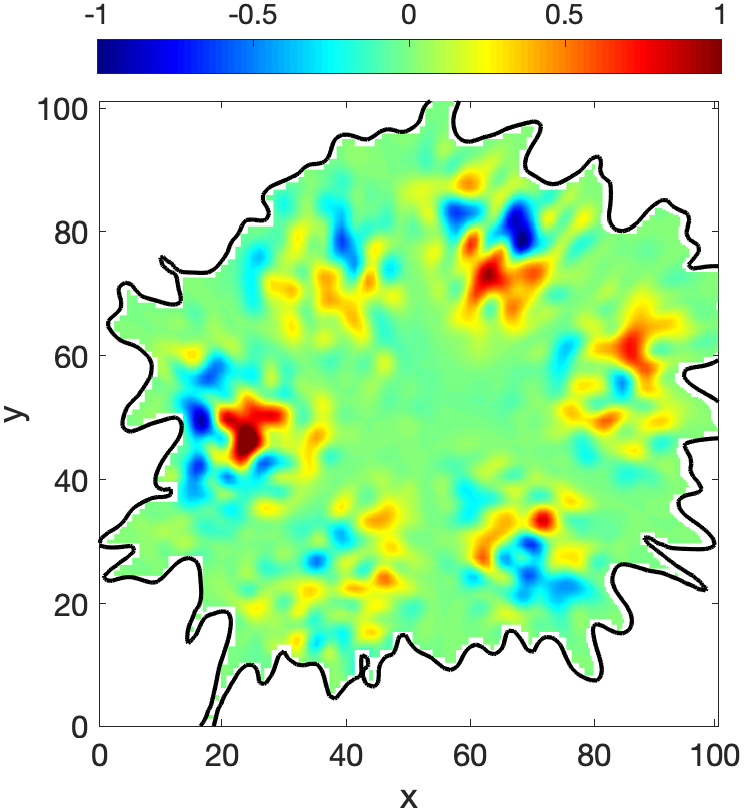}&
\includegraphics[width=40mm]{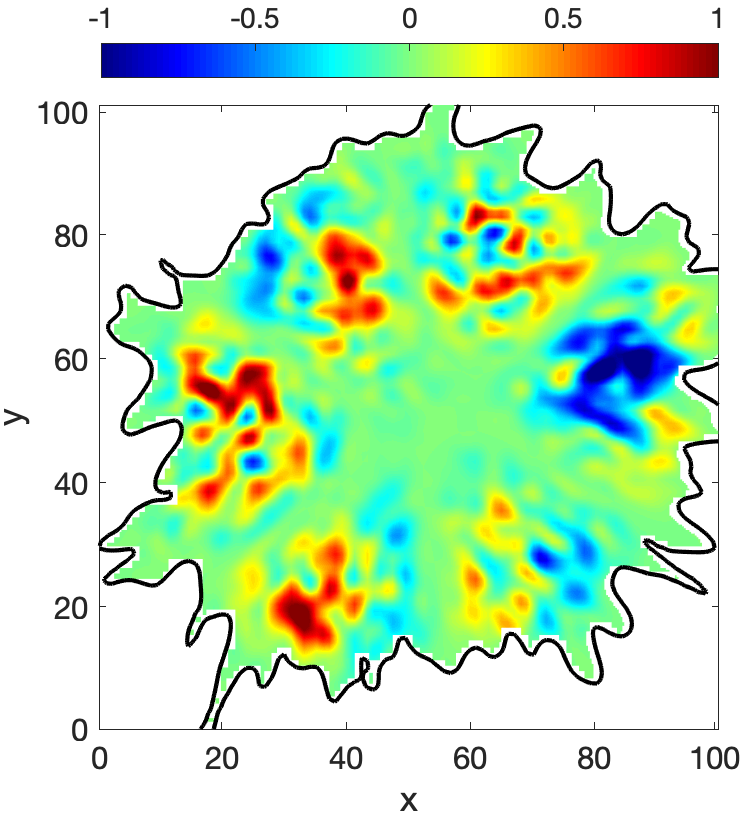}\\
\end{tabular}
\caption{This figure displays the 26$^{th}$ POD (top row, 1st panel) and DMD mode (top row, 2nd panel) with a frequency of 7.4 mHz which has an azimuthal symmetry of the slow body fluting mode ($n=3$). The 3D visualisation of this mode is shown in Figure \ref{fig:C_3D} in the Appendix.}
\label{fig:fluting(n=3)_circular}
\end{figure*}

The next POD mode that can be interpreted as a MHD wave mode is the 20$^{th}$ component of the POD ranking and it is visible in the DMD decomposition, too. This mode has a frequency of 7.6 mHz, and has an azimuthal symmetry corresponding to a \textit{slow body fluting mode (n=2)}, as shown in Figure \ref{fig:fluting(n=2)_circular}. The PSD of the 20$^{th}$ POD mode shows a mix of peaks around 3.7 mHz, 5.8 mHz and 7.4 mHz, as shown in the middle panel of Figure \ref{PSD_circular}. Finally, the last mode that was detected by the POD and DMD analysis that shows an azimuthal symmetry of a \textit{slow body fluting mode ($n=3$)} is the 26$^{th}$ POD component and the DMD mode that corresponds to 7.5 mHz, as shown in Figure \ref{fig:fluting(n=3)_circular}. The PSD of the 26$^{th}$ POD mode shows a mix of peaks at around 3.2 mHz and 7.2 mHz, as shown in the right panel of Figure \ref{PSD_circular}. 

It is expected that some of the PSD of POD modes may have a mix of peaks in their frequency domain, and this is one disadvantage of the POD technique, therefore, making it difficult to decide which frequency is relevant for the mode identification. However, this ambiguity is resolved by the DMD technique by taking the peaks and finding the DMD mode that correspond to the peaks, allowing us to make a decision on which DMD mode has a spatial structure similar to the mode recovered by means of POD, and hence we consider the distinct frequency of that DMD mode.



\begin{table}
\begin{center}
\begin{tabular}{l C C C C C}
\hline
MHD wave mode  &$m_{i}$ & $f$ (mHz) & $k_z$ (Mm$^{-1}$)  &  $\lambda$ (Mm) & $V_{ph}$ (Km/s)\\
\noalign{\smallskip}
\hline
\hline
Fundamental slow body sausage & $1.0353$ & $4.8$ & $3.0022$ & $2.0928$  & $10.04$ \\
Fundamental slow body kink  & $1.6765$ &  $6$ & $3.73754$  & $1.6811$ & $10.08$\\
Slow body overtone sausage  & $2.3754$ &  $5.6$ & $3.395$  & $1.8507$ & $10.36$ \\
Slow body fluting ($n=2$) & $2.214$ &  $7.6$ & $4.7294$ & $1.3285$ & $10.09$  \\
Slow body fluting ($n=3$) & $2.7471$ & $7.4$ & $4.55619$  & $1.3790$ & $10.20$\\
\hline
\end{tabular}\\
 \caption{This table displays the summary of the properties of the MHD modes detected by the POD and DMD techniques in the sunspot with circular cross-section. The first column contains the name of the modes, while the second column shows the value of the magneto-acoustic parameter, $m_i$, (see Equation \ref{Eq:1}). The third column contains the frequency determined from the DMD analysis. The fourth column contains the wavenumber along the vertical direction, and these are calculated by means of Equation \ref{Eq:1}, with $\omega=2\pi f$, $c_i=0.01$ (Mm/s) and $v_{A_i}=4c_i$. The fifth column shows the wavelength ($\lambda=2\pi/k_z$), while the last column gives the phase speed ($V_{ph}=f \lambda$) of the identified modes. \label{Table1}}
\end{center}
\end{table}

\begin{figure*}[!t]
\centering
\begin{tabular}{ccc}
\includegraphics[width=40mm]{C_blank_white.png}&
\includegraphics[width=40mm]{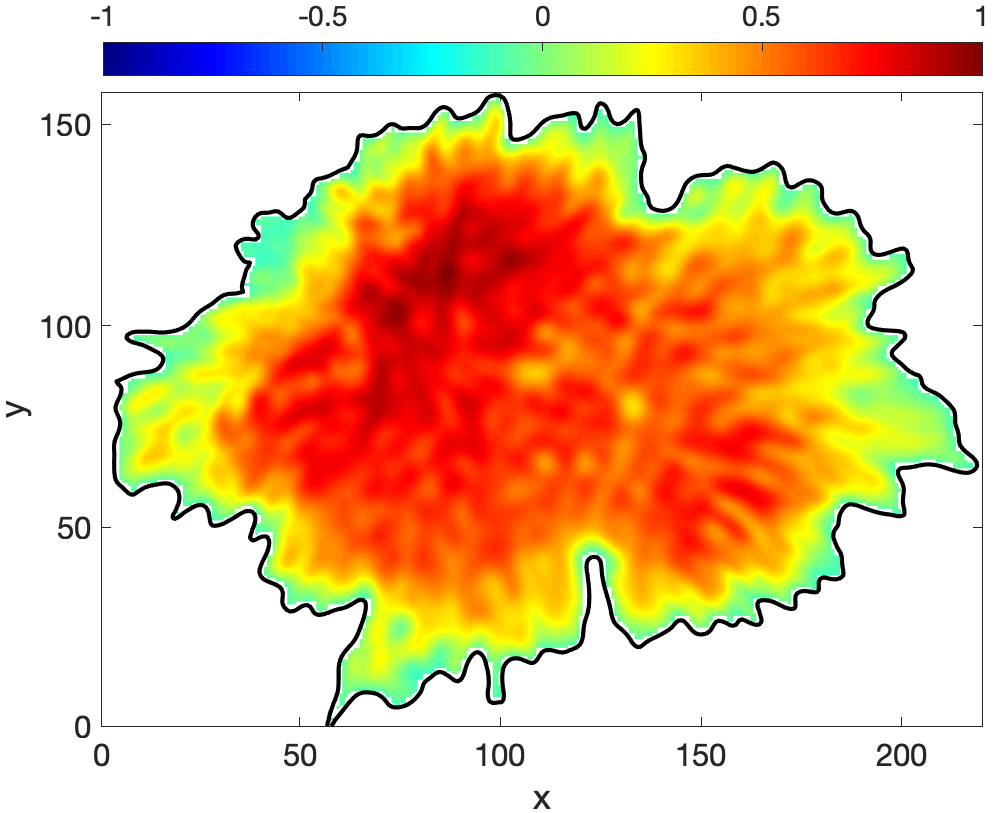}&
\includegraphics[width=40mm]{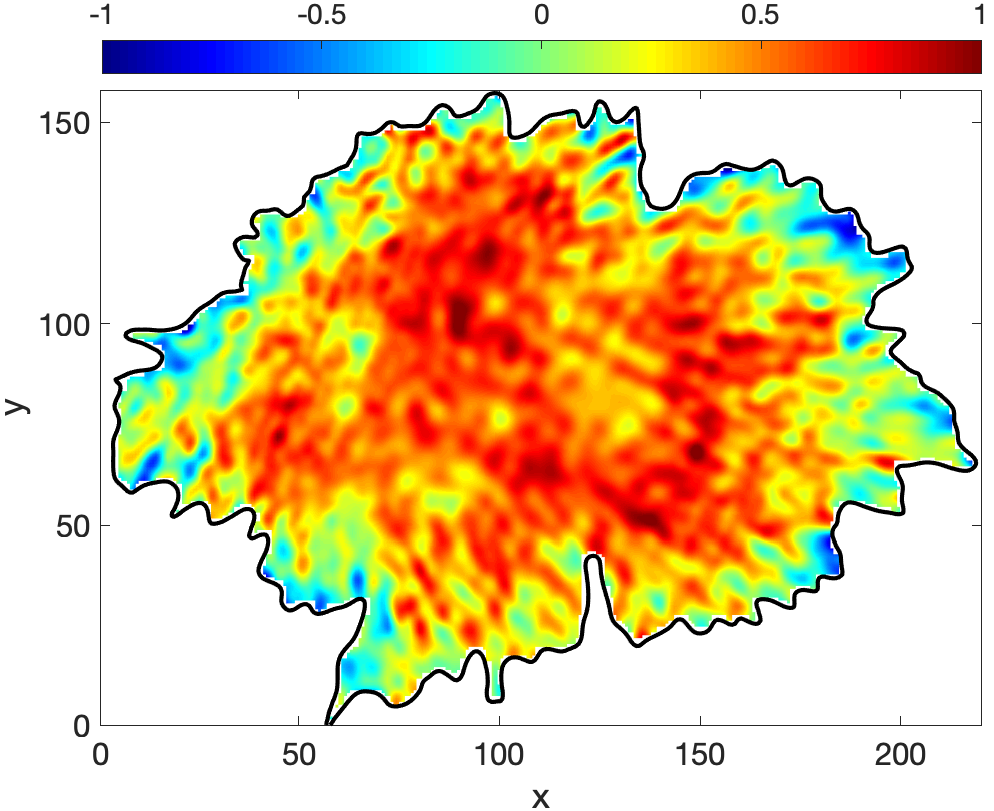}\\
\includegraphics[width=40mm]{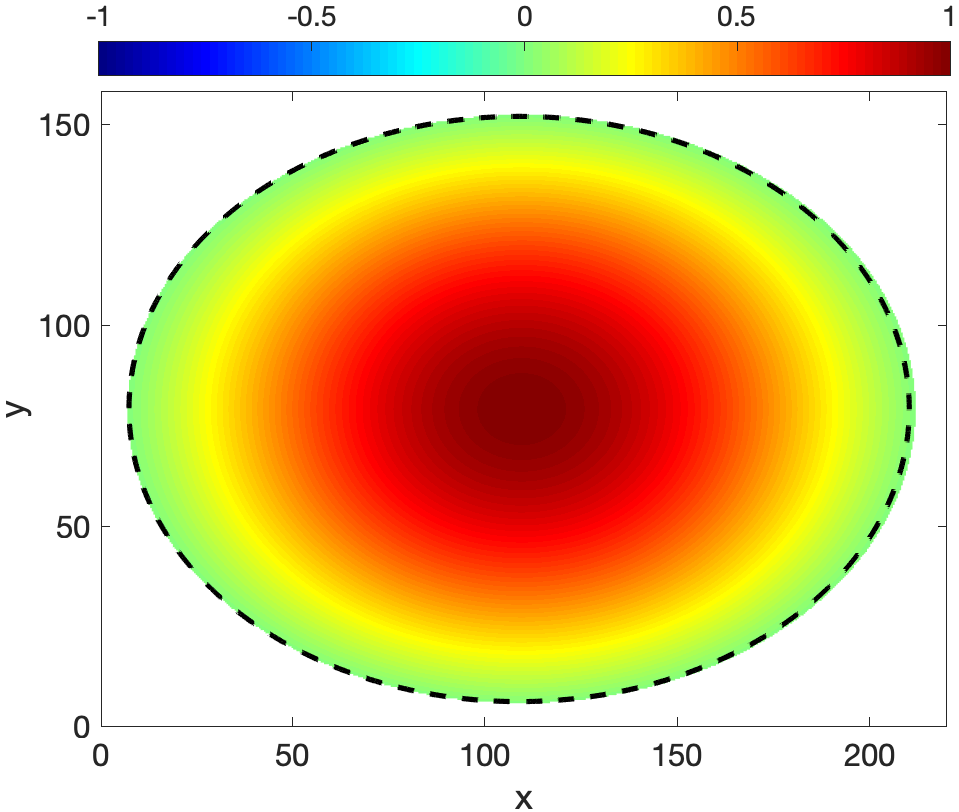}&
\includegraphics[width=40mm]{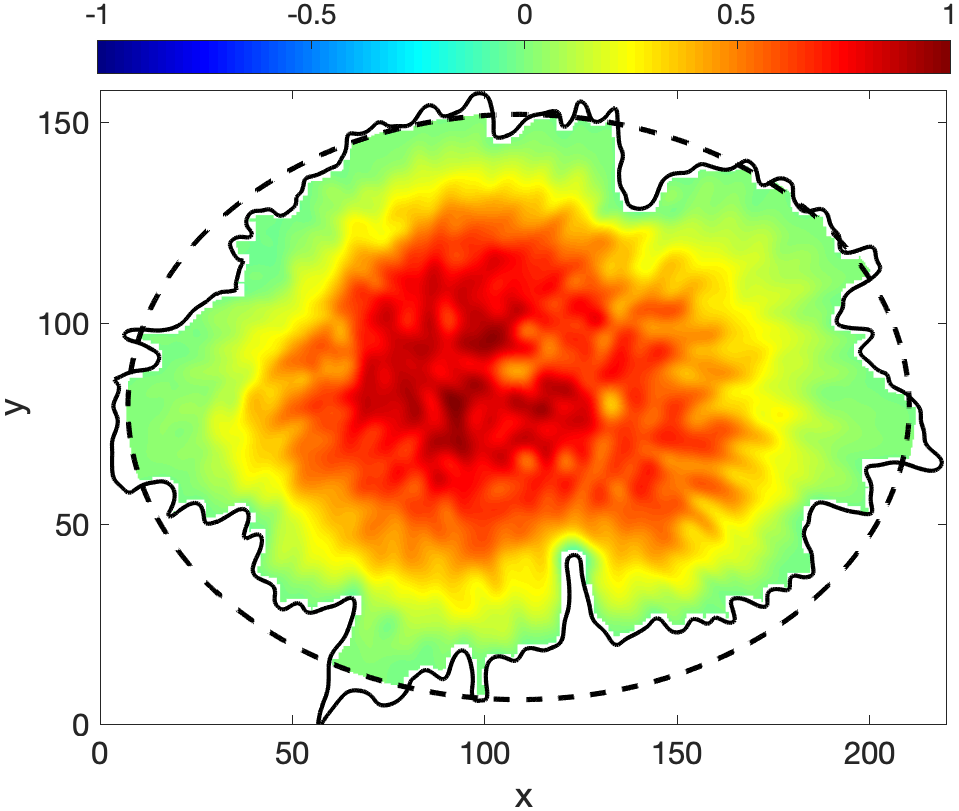}&
\includegraphics[width=40mm]{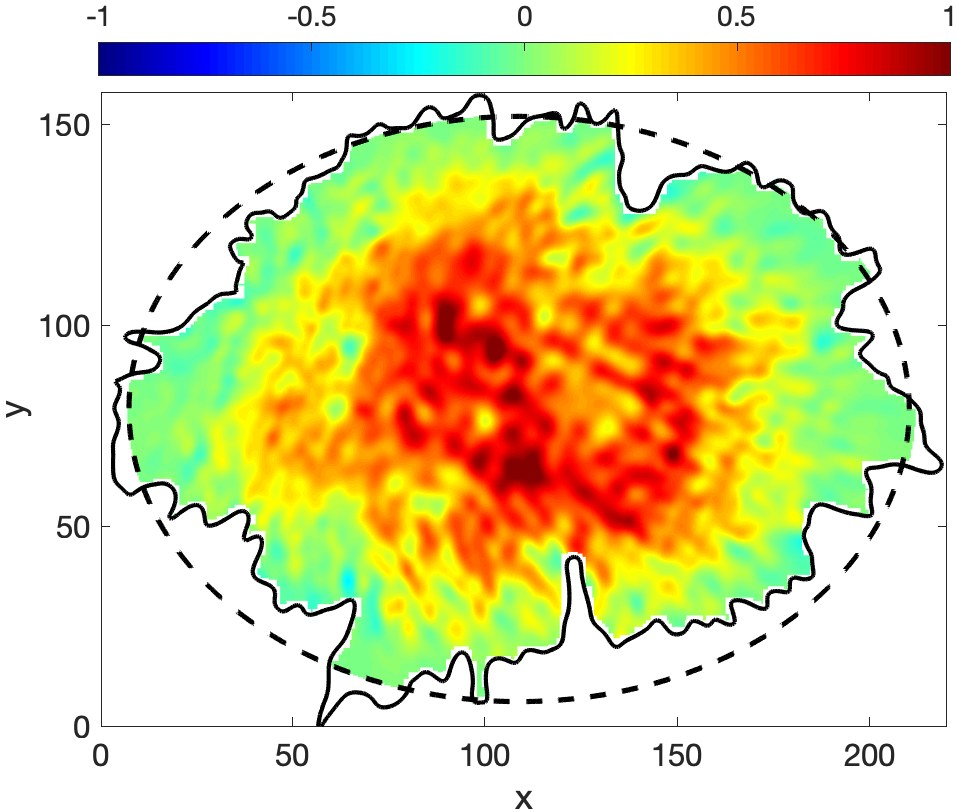}\\
\includegraphics[width=40mm]{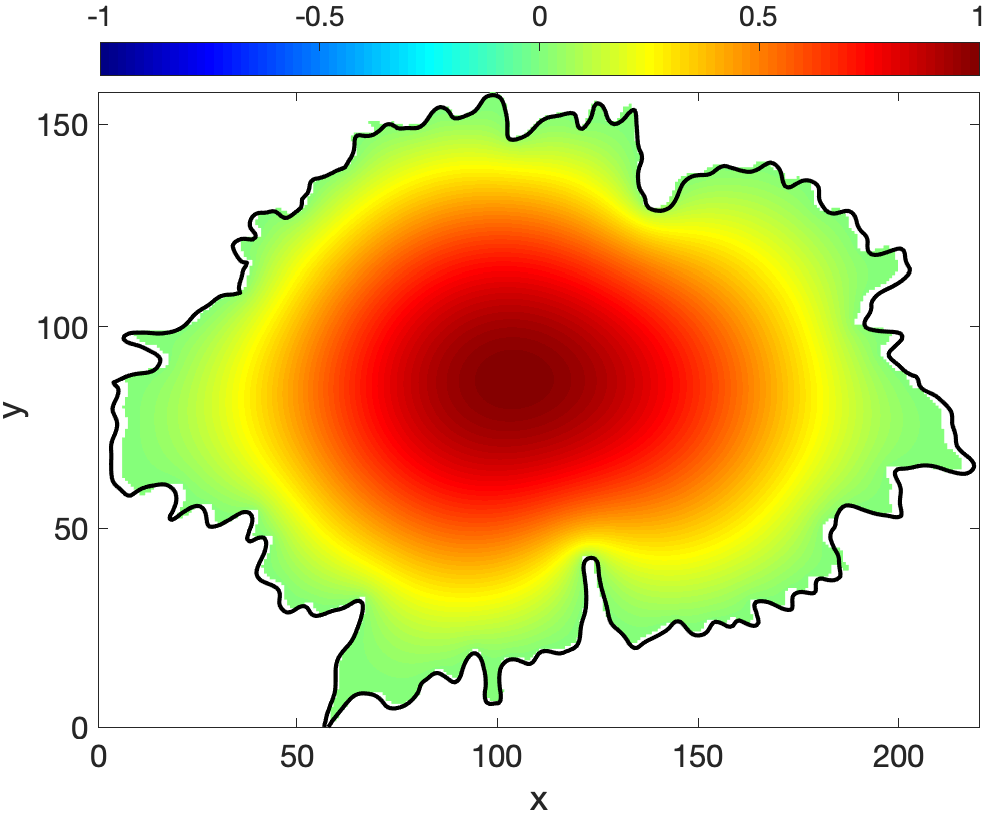}&
\includegraphics[width=40mm]{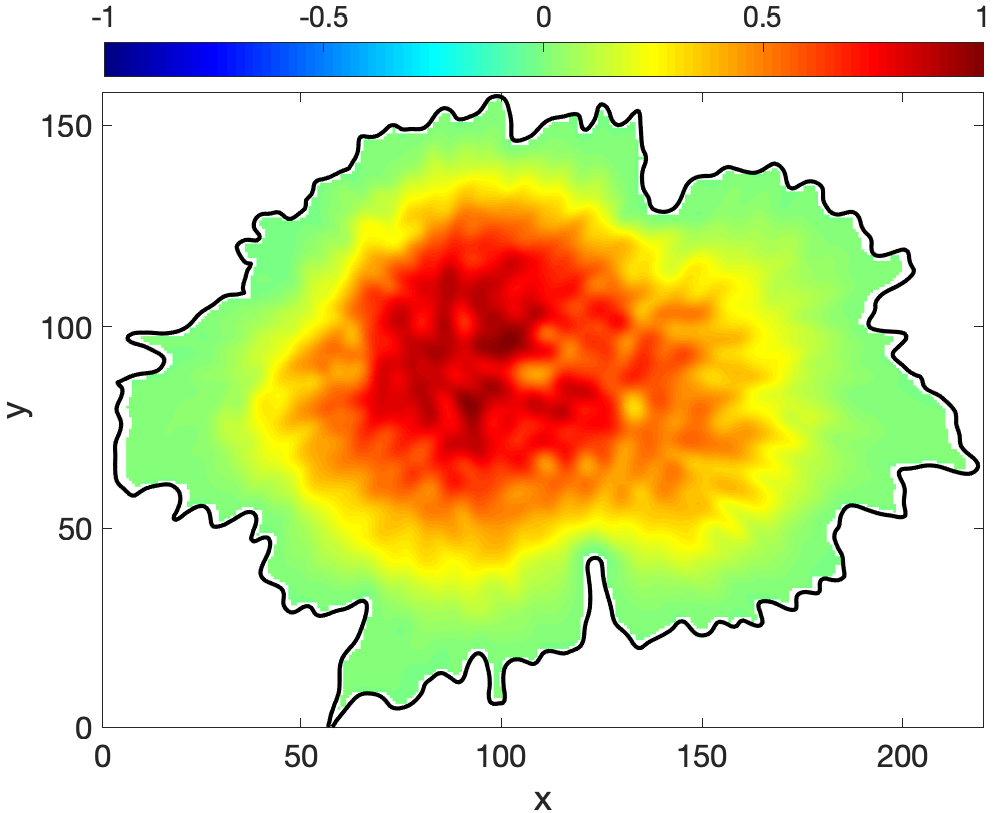}&
\includegraphics[width=40mm]{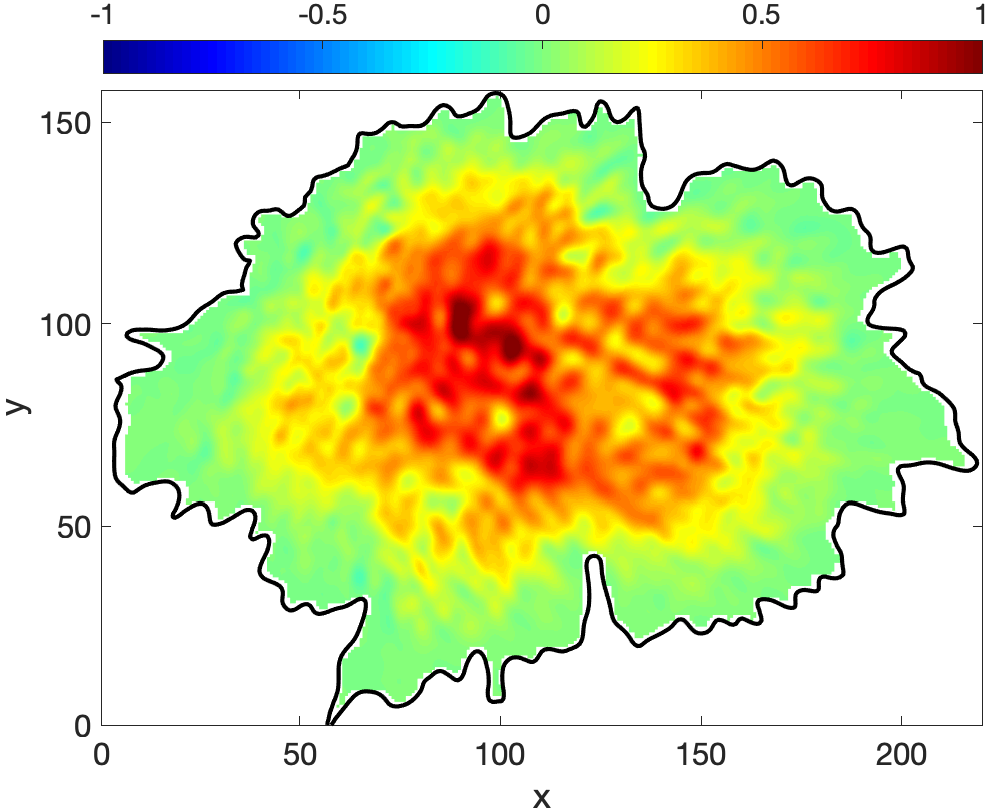}\\
\end{tabular}
\caption{The first row displays the spatial structure of the first POD mode (left panel) and the DMD mode that corresponds to 3.4 mHz (right panel). The first column displays the theoretical modes of the fundamental slow body sausage mode in an elliptical magnetic flux tube (middle) and the theoretical modes of the fundamental sausage body mode in the irregular shape that corresponds to the actual sunspot shape (bottom). The rest of the panels are showing the cross-correlation between theoretically constructed and observationally detected modes and the positive/negative numbers on the colourbar denote correlation/anti-correlation. The dashed ellipse shows the boundary of the flux tube and the solid black line shows the umbra/penumbra boundary. The 3D visualisation of this mode is displayed in Figure \ref{fig:E_3D} in the Appendix. The same configuration were used for figures \ref{fig: fun_kink_E} to \ref{fig:m=3_E}.}
\label{fig: fun_saus_E}
\end{figure*}

\begin{figure*}[!t]
\centering
\begin{tabular}{ccc}
\includegraphics[width=40mm]{C_blank_white.png}&
\includegraphics[width=40mm]{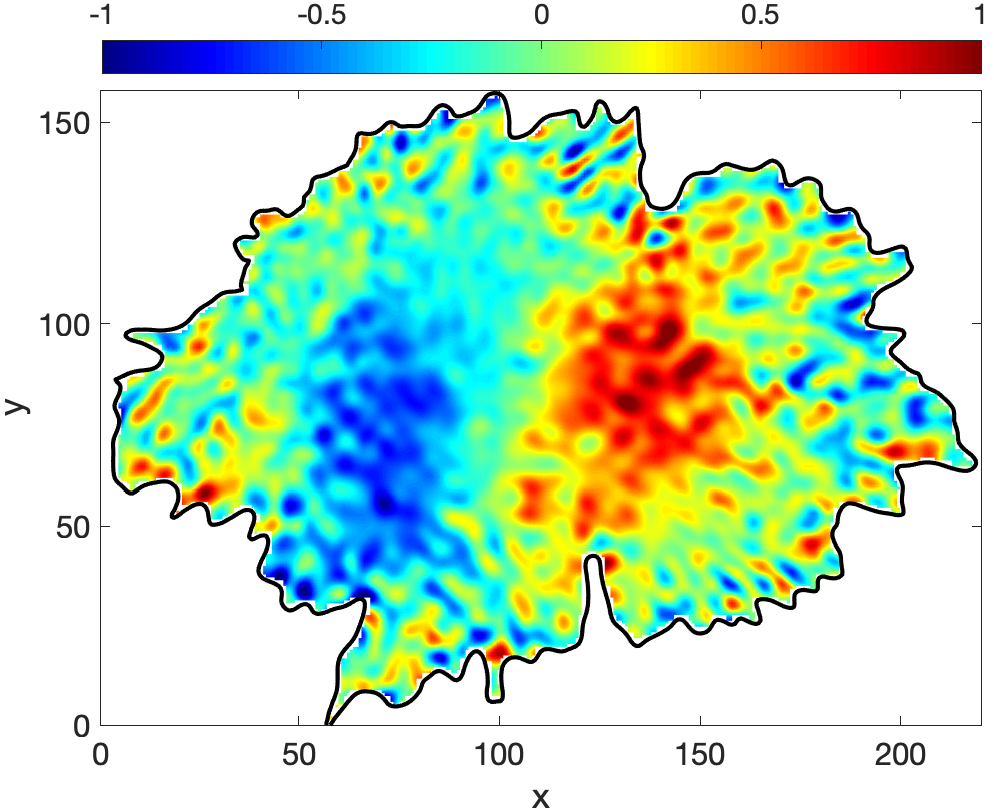}&
\includegraphics[width=40mm]{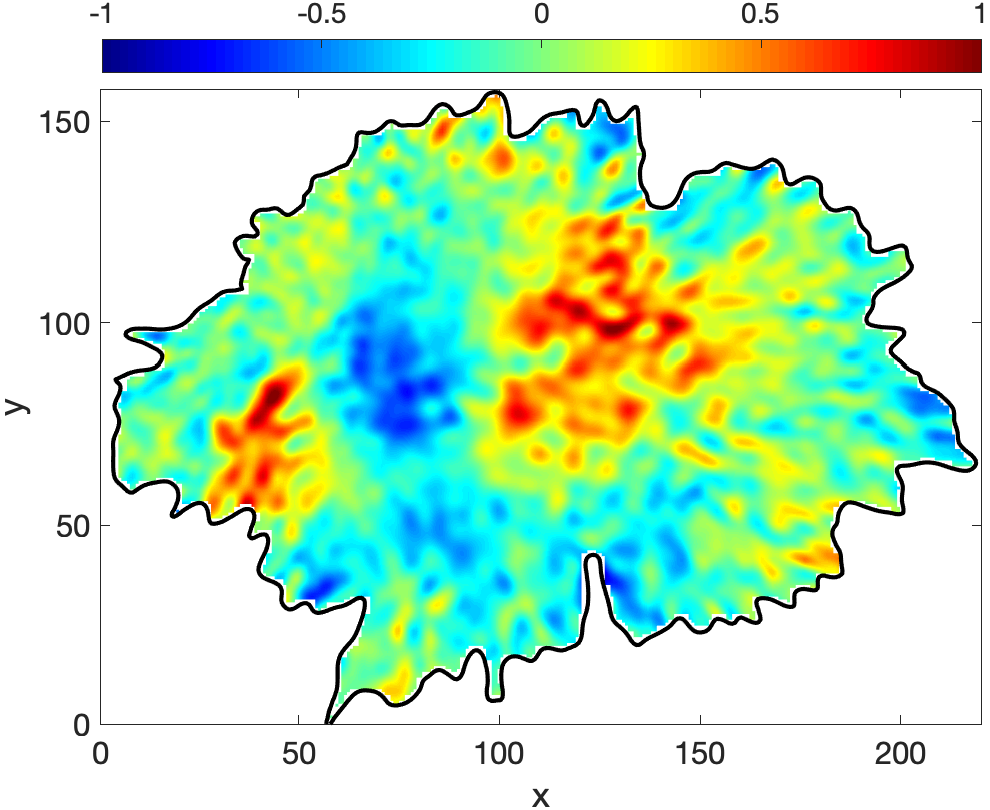}\\
\includegraphics[width=40mm]{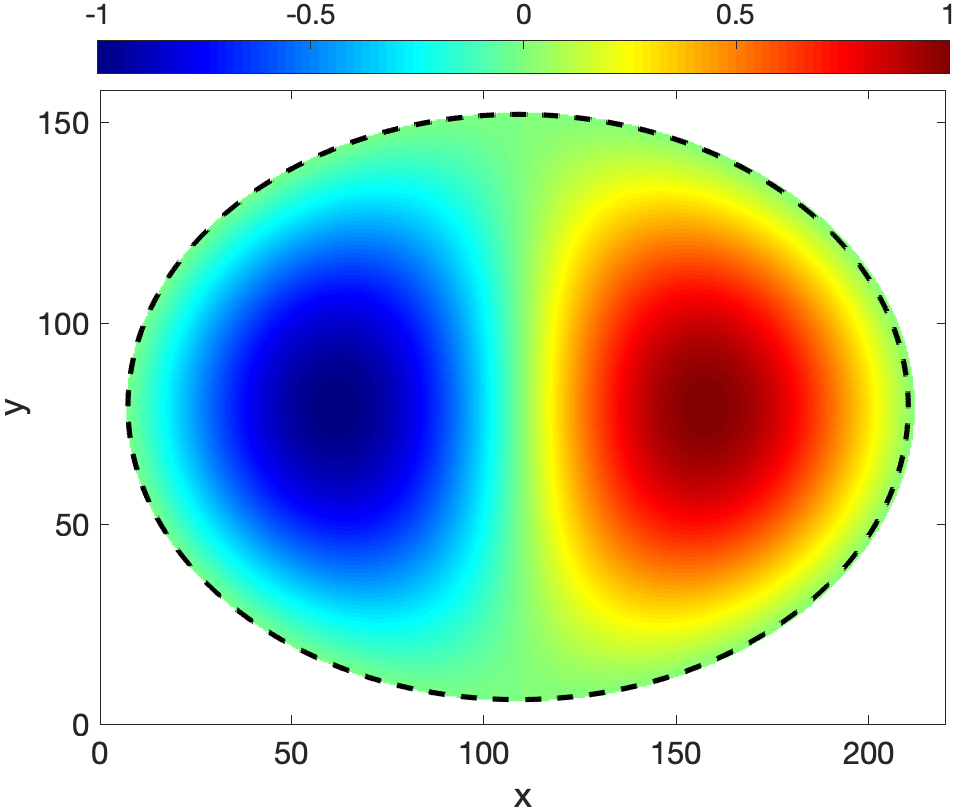}&
\includegraphics[width=40mm]{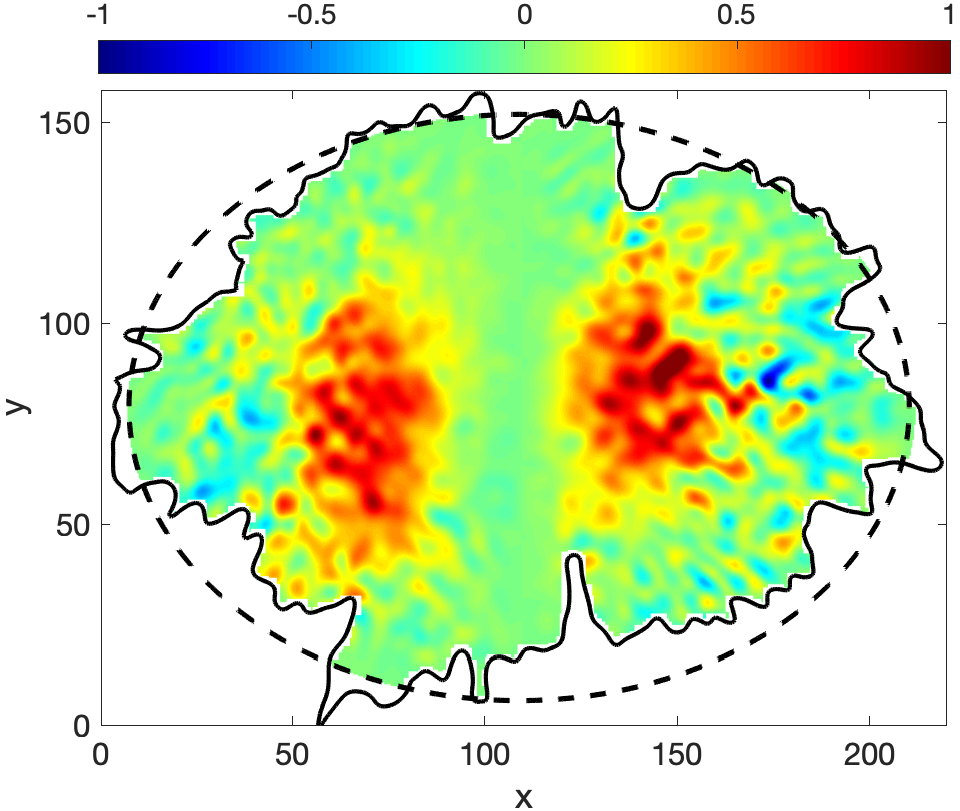}&
\includegraphics[width=40mm]{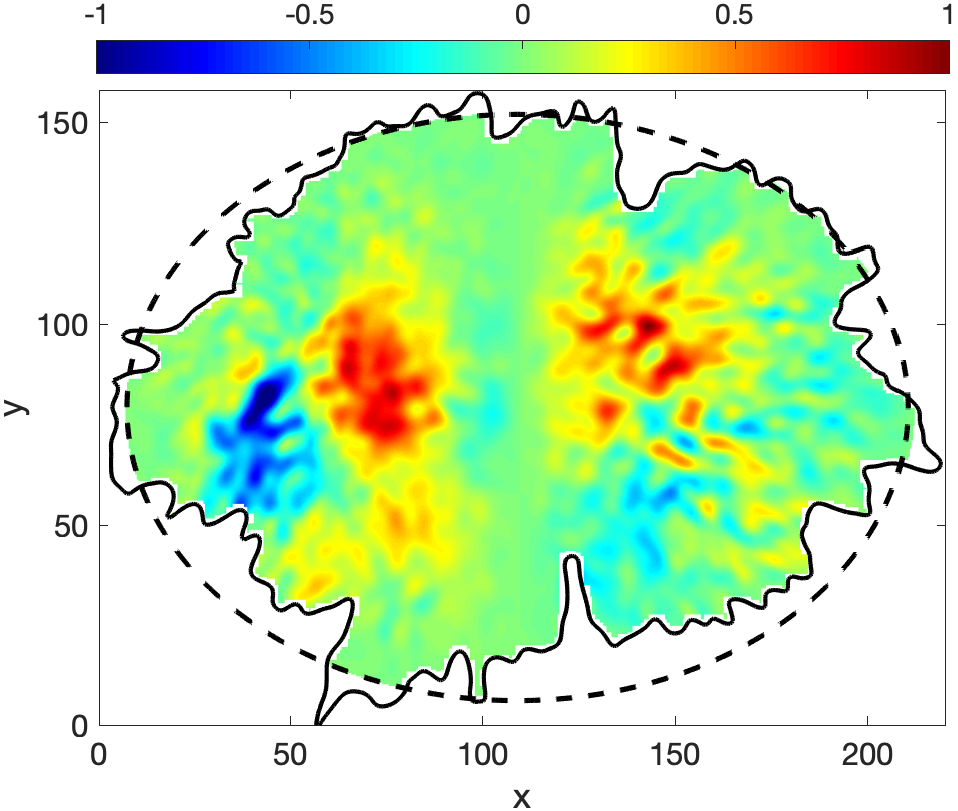}\\
\includegraphics[width=40mm]{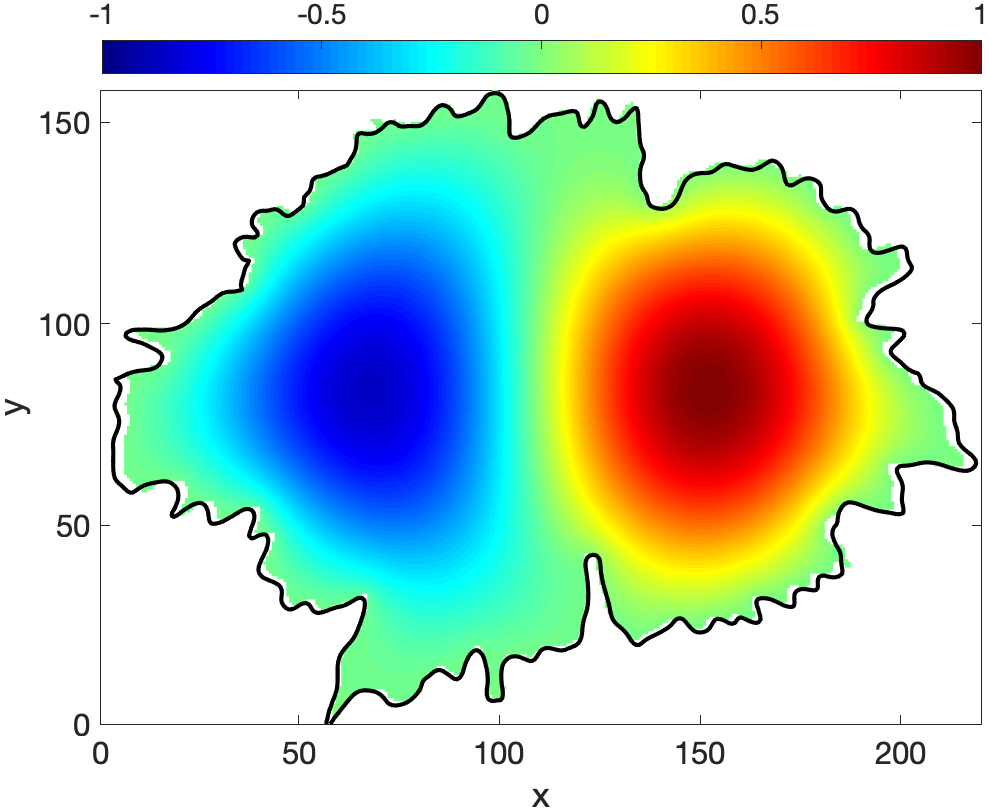}&
\includegraphics[width=40mm]{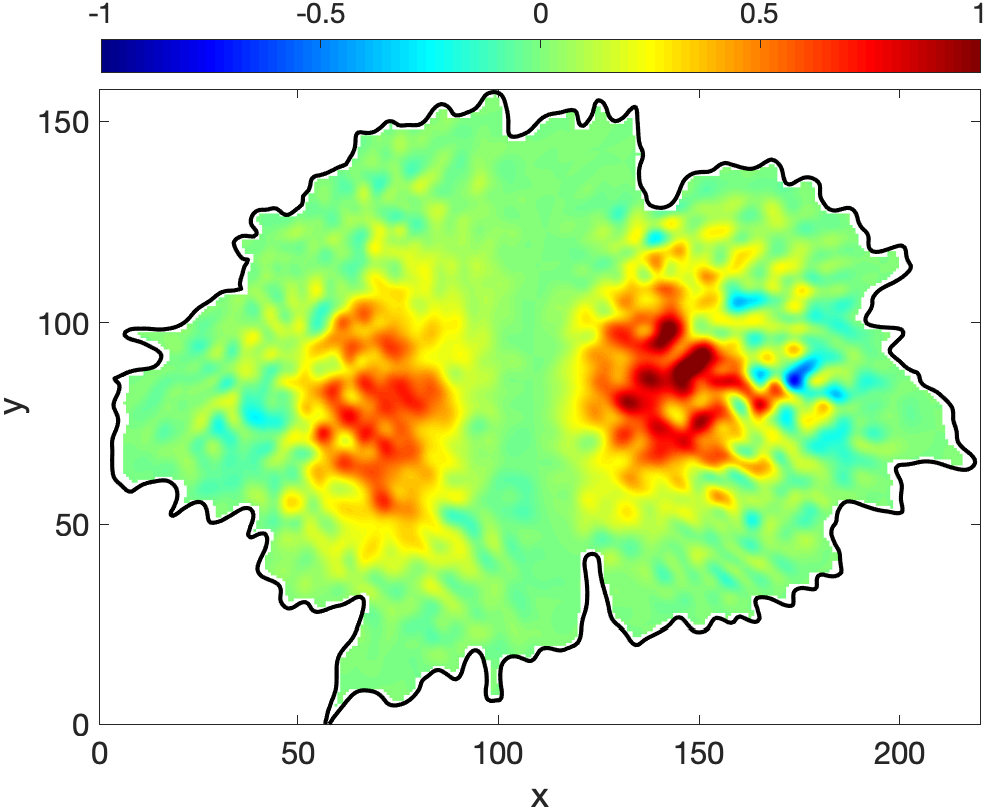}&
\includegraphics[width=40mm]{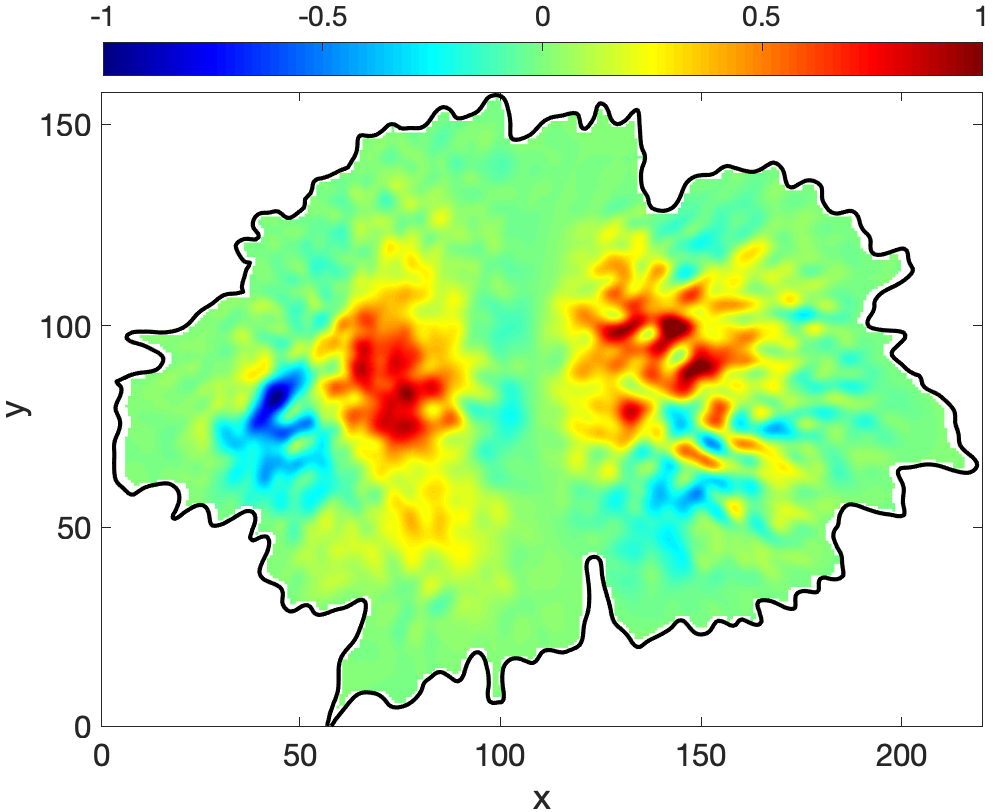}\\
\end{tabular}
\caption{This figure displays the 14$^{th}$ POD (top row, left panel) and DMD modes (top row, right panel) with a frequency of 5.8 mHz which has the azimuthal symmetry corresponding to the fundamental slow body kink mode. The 3D visualisation of this mode is shown in Figure \ref{fig:E_3D} in the Appendix.}
\label{fig: fun_kink_E}
\end{figure*}

\begin{figure*}[!t]
\centering
\begin{tabular}{cc}
\includegraphics[width=55mm]{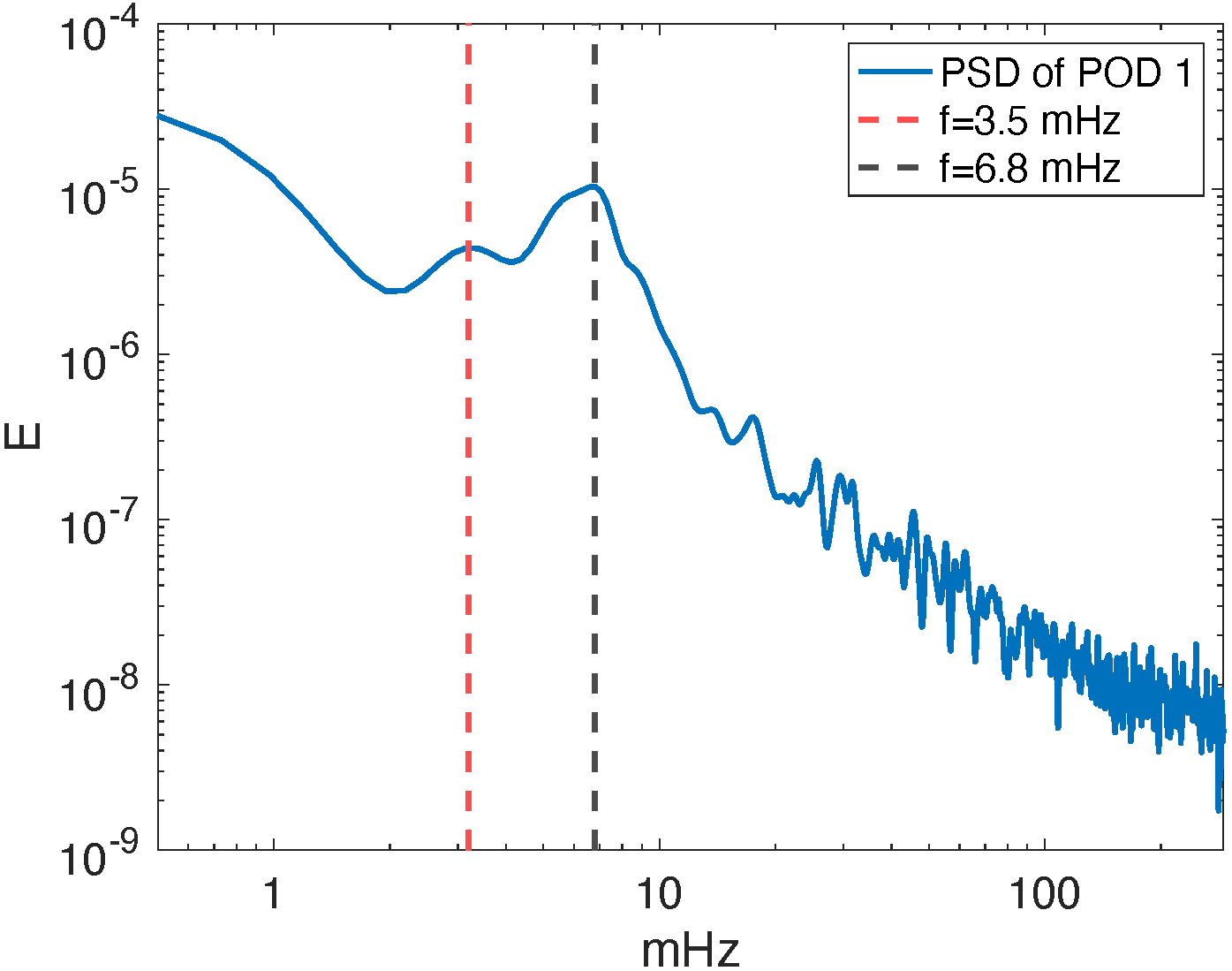}&
\includegraphics[width=55mm]{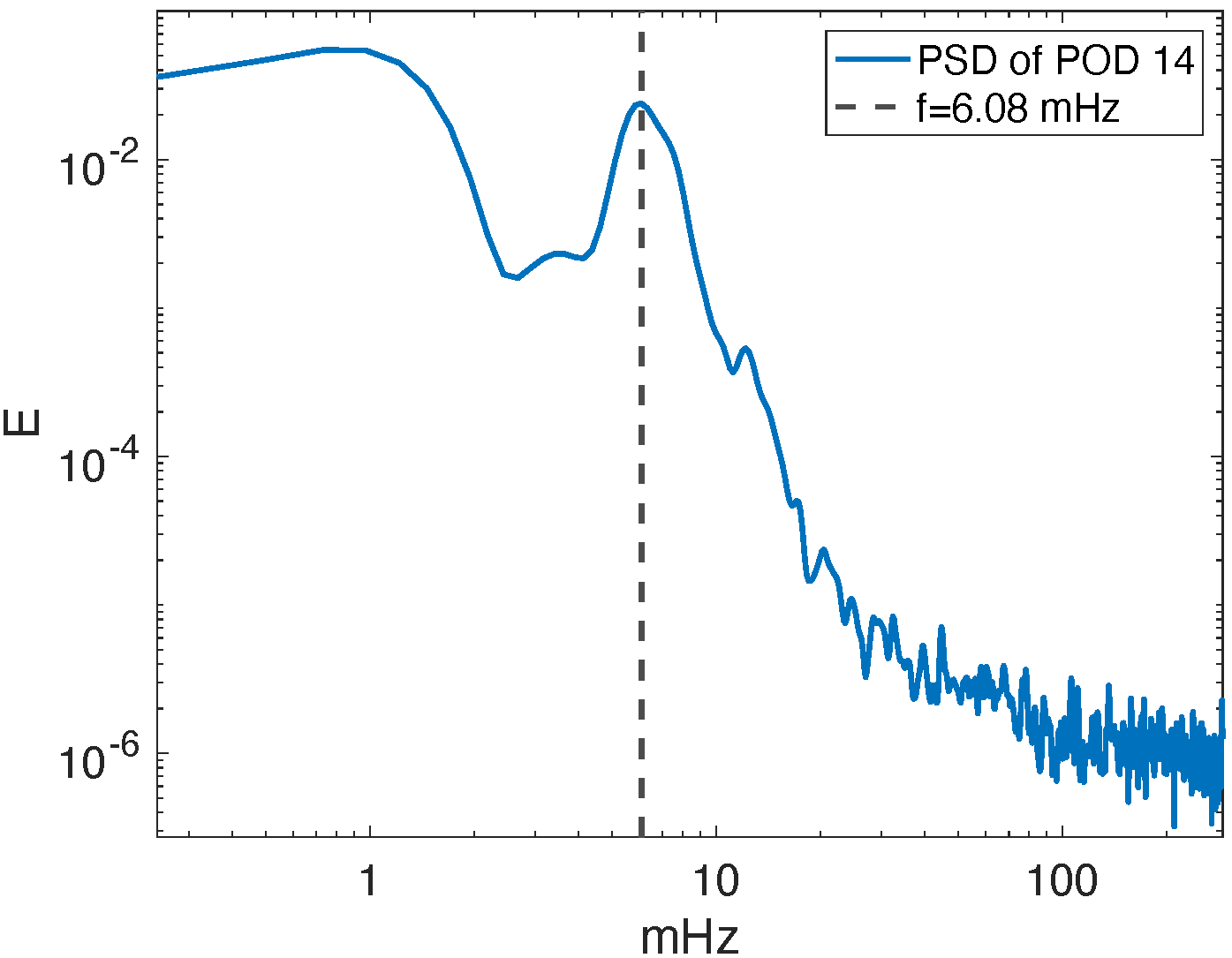}\\
\includegraphics[width=55mm]{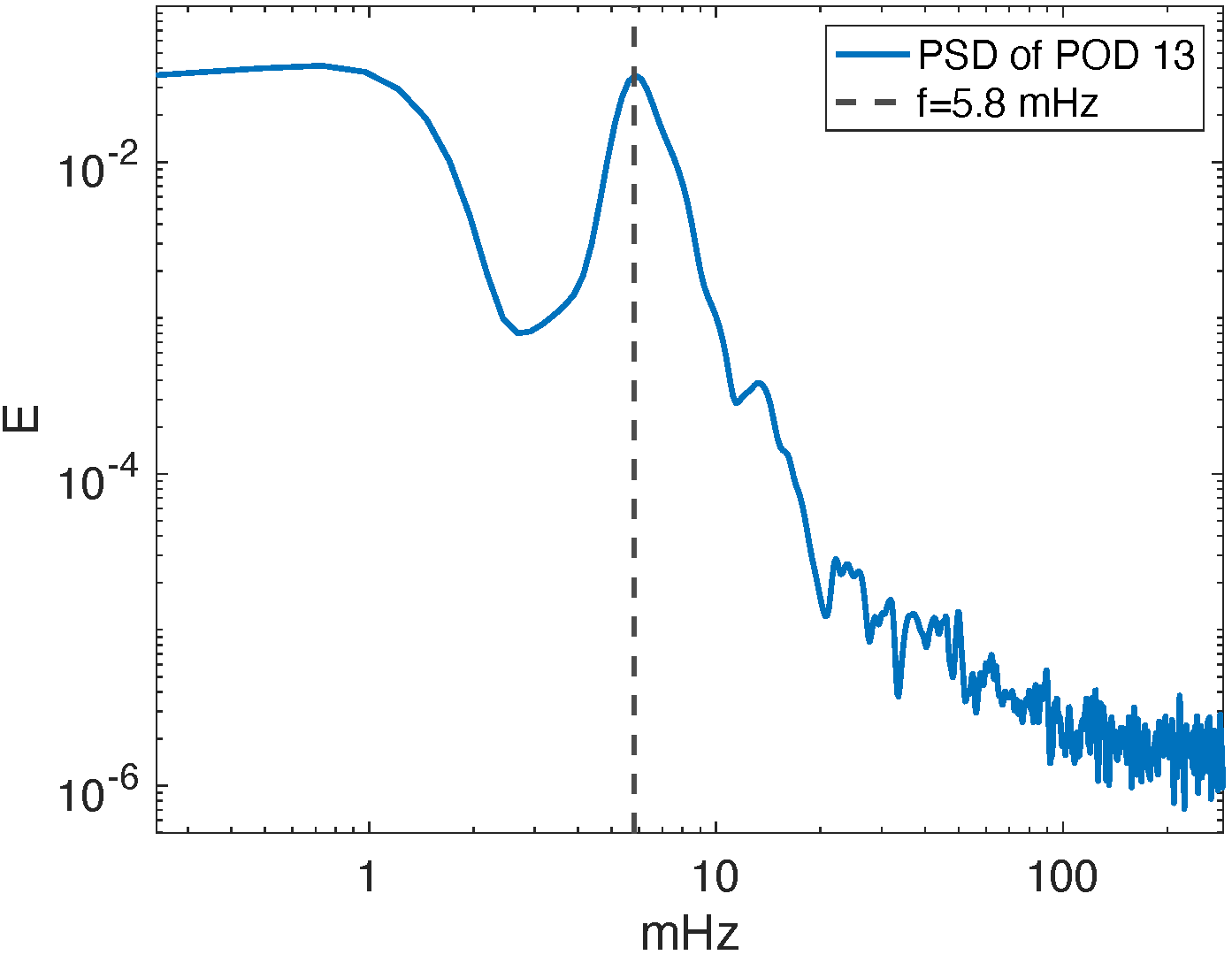}&
\includegraphics[width=55mm]{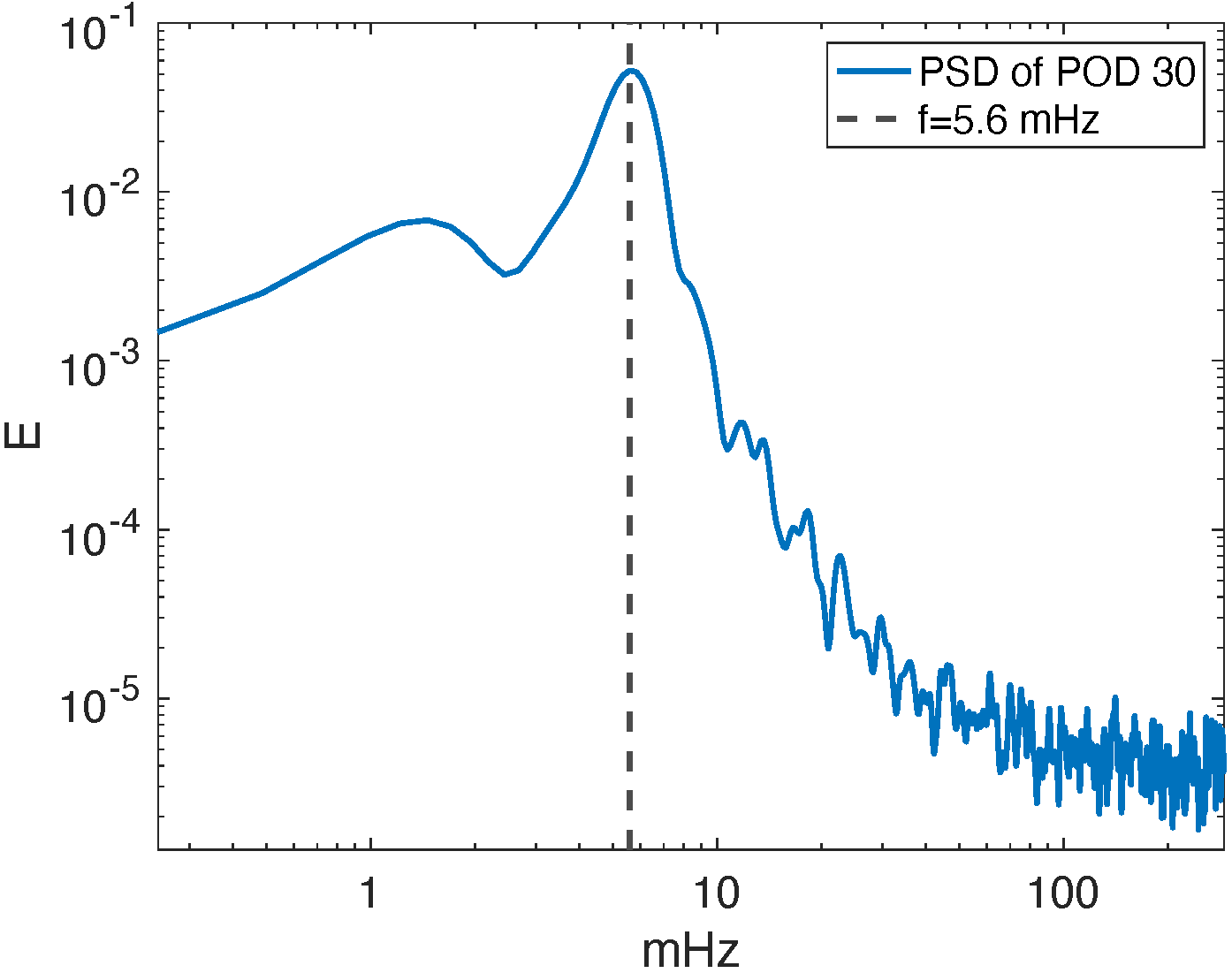} \\
\includegraphics[width=55mm]{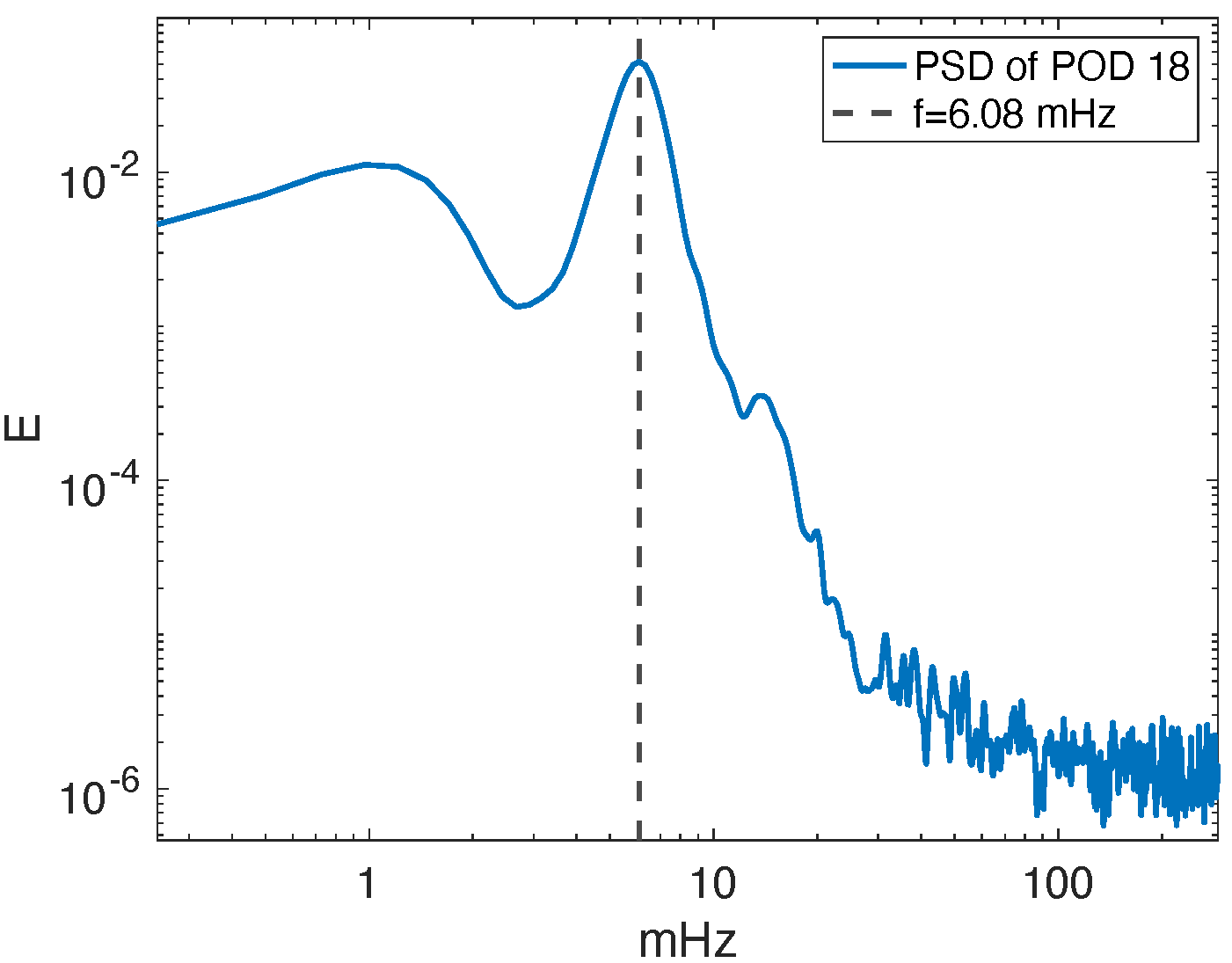}&
\end{tabular} 
\caption{This figure shows the power spectrum density (PSD) of the time coefficients of the POD 1 (upper left panel), POD 14 (upper right panel), POD 13 (middle left panel), POD 30 (middle right panel) and POD 18 (bottom left panel) modes. The  coloured-dash vertical lines represent the value in the frequency domain that corresponds to the peaks of the PSD, where the frequencies are shown in the legend.} \label{fig: PSD_E}
 \end{figure*}

The sunspot with an elliptical cross-sectional shape is shown in the lower left panel of Figure \ref{fig:snapshot_circular&E}. The POD and DMD analysis was applied to the ROI shown by the blue box in the lower right panel of the same figure where the umbra/penumbra boundary is shown by a solid black line with an intensity threshold level at 0.4. The first POD mode that can be interpreted as a MHD wave is the 1$^{st}$ POD mode that shows the symmetry of the \textit{fundamental slow body sausage mode} and the associated DMD mode corresponds 3.4 mHz, as shown in Figure \ref{fig: fun_saus_E}. The PSD of the time coefficient of POD 1 shows peaks around 3.5 mHz and 6.8 mHz, as shown on the top left panel of Figure \ref{fig: PSD_E}. The next mode that can be identified in our data is the \textit{fundamental slow body kink mode}, and the POD mode that shows a high correlation with this mode of oscillation is POD 14, as shown in Figure \ref{fig: fun_kink_E}. The PSD of time coefficient of POD 14 shows a clear peak at 5.88 mHz (the top right panel of Figure \ref{fig: PSD_E}). The DMD mode that shows an azimuthal symmetry with the fundamental kink is the DMD mode that corresponds to 5.8 mHz, as shown in Figure \ref{fig: fun_kink_E}.

\begin{figure*}[!t]
  \centering
  \begin{tabular}{ccc}
    \includegraphics[width=40mm]{C_blank_white.png}&
    \includegraphics[width=40mm]{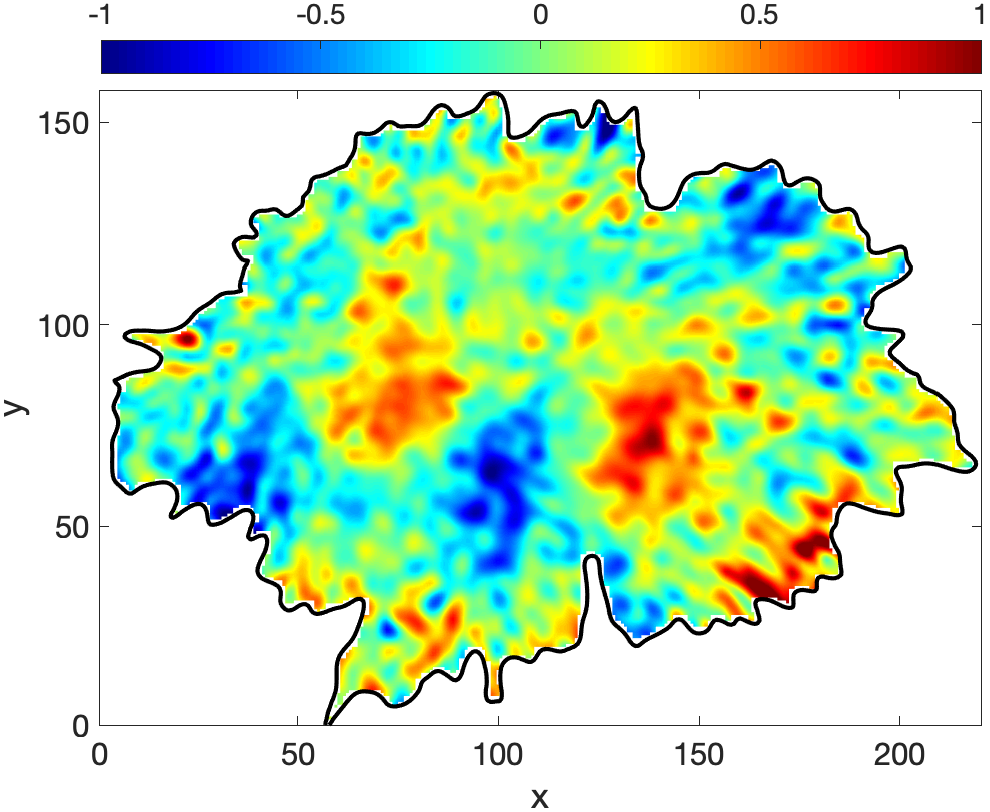}&
    \includegraphics[width=40mm]{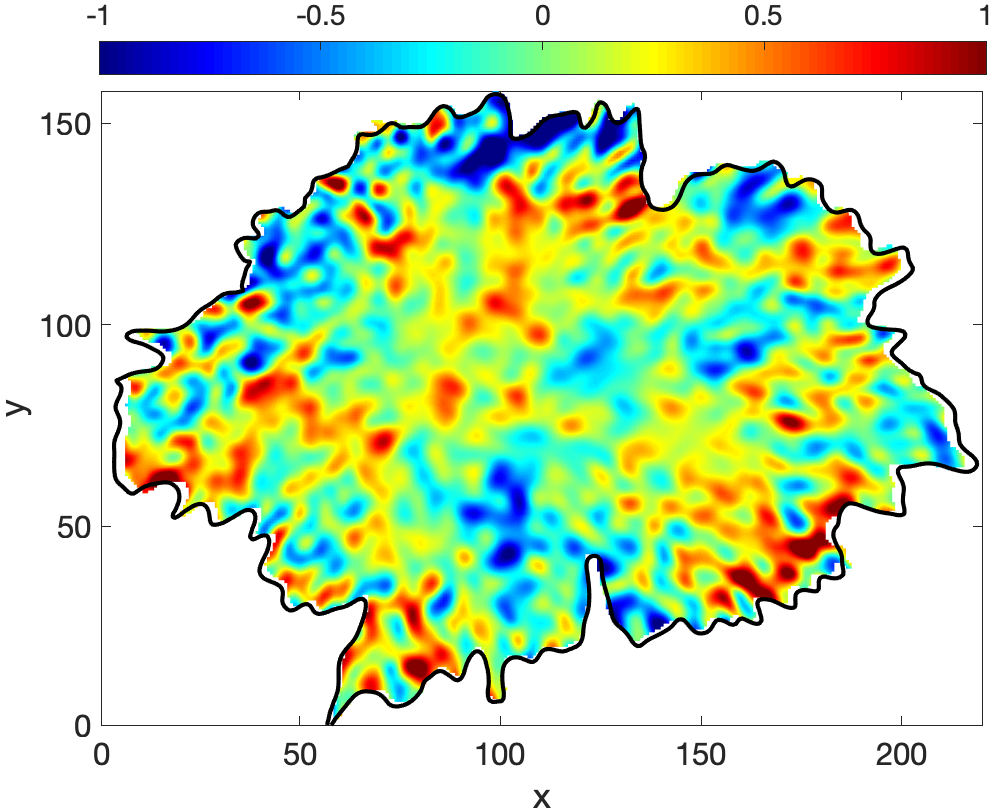}\\
    \includegraphics[width=40mm]{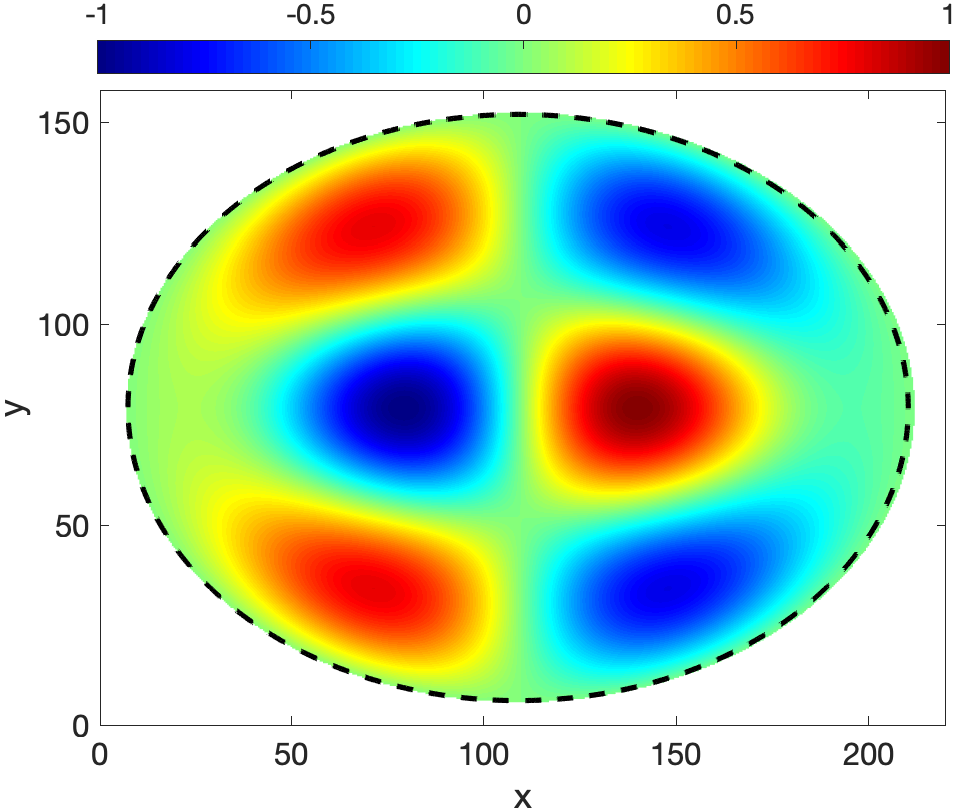}&
    \includegraphics[width=40mm]{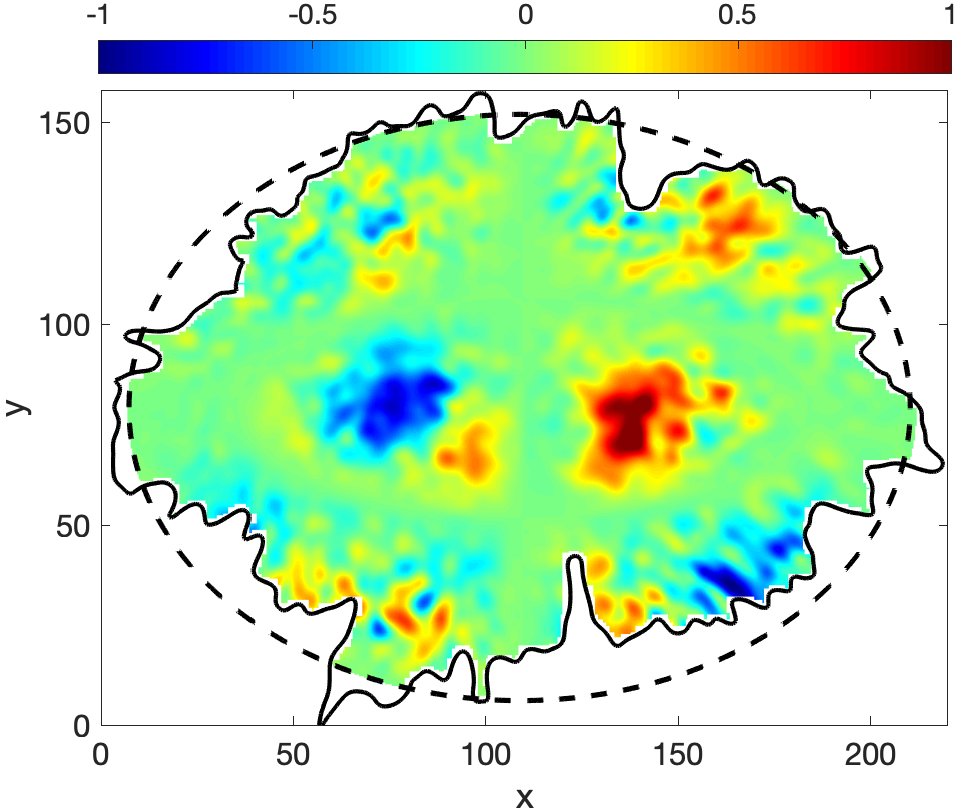}&
    \includegraphics[width=40mm]{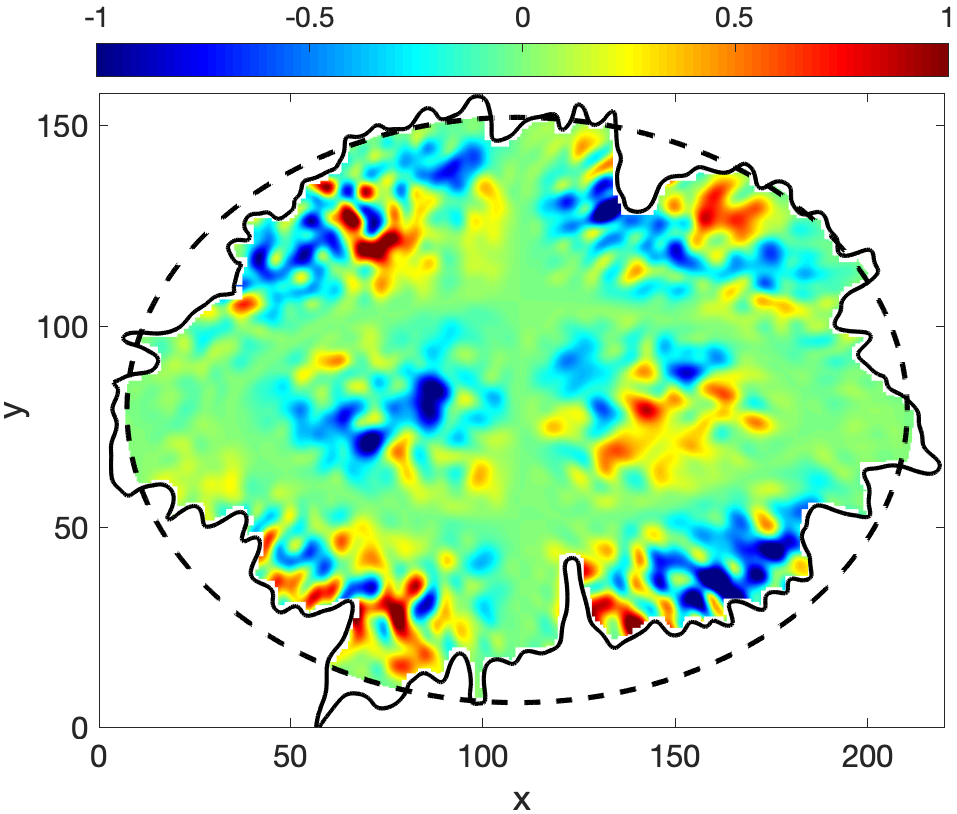}\\
    \includegraphics[width=40mm]{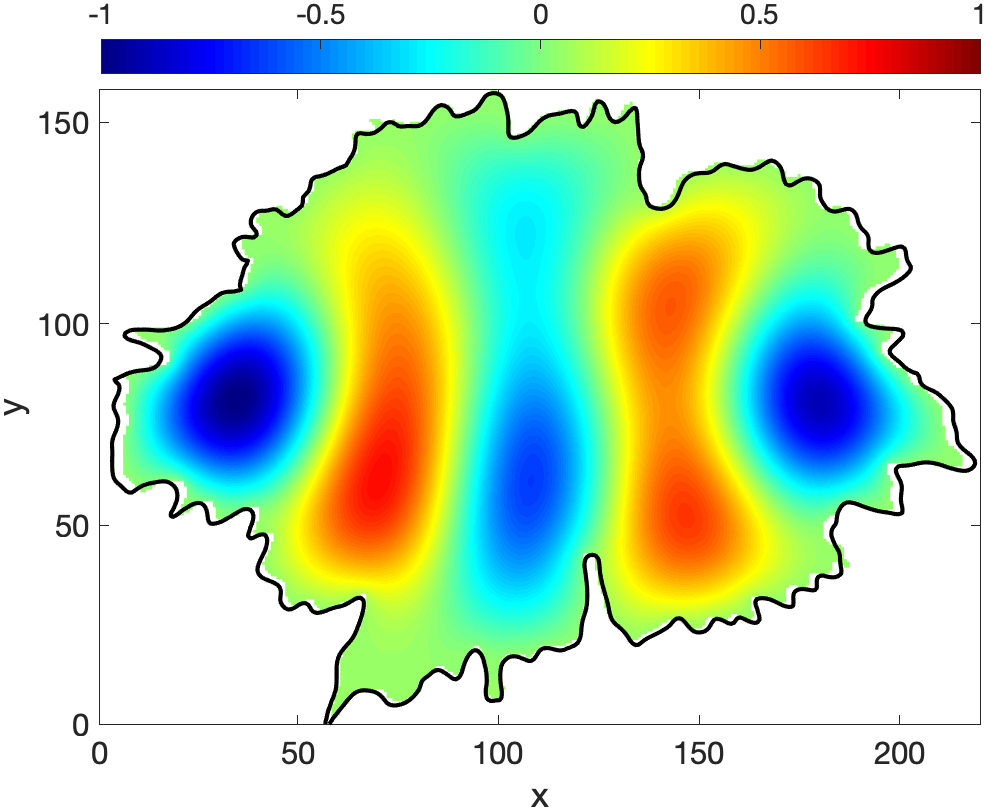}&
    \includegraphics[width=40mm]{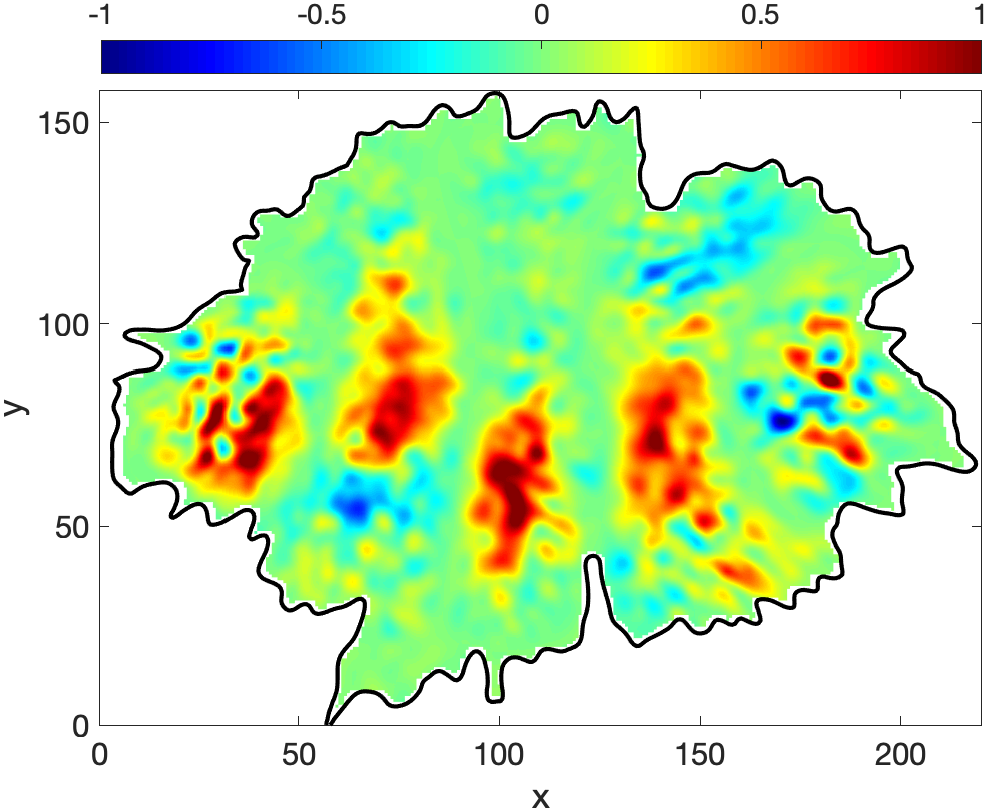}&
    \includegraphics[width=40mm]{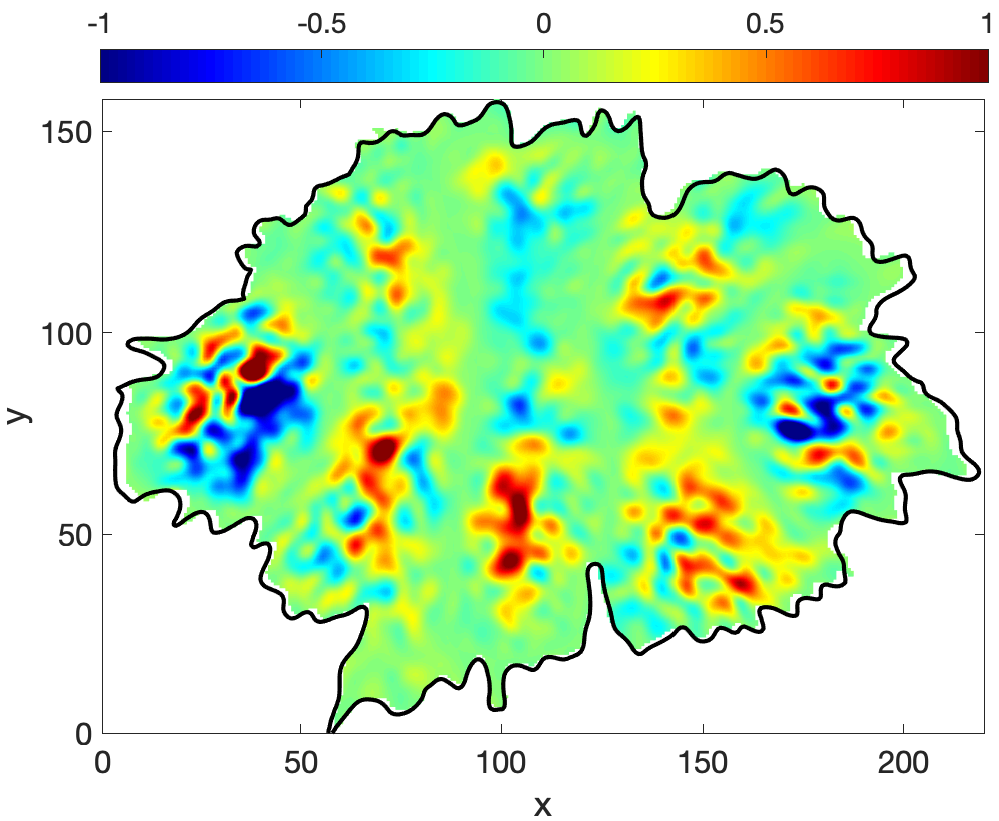}\\
  \end{tabular}
\caption{This figure displays the 30$^{th}$ POD (top row, left panel) and DMD mode (top row, right panel) with a frequency of 5.3 mHz, which has an azimuthal symmetry of the slow body kink overtone mode. The 3D visualisation of this mode is shown in Figure \ref{fig:E_3D} in the Appendix.}
\label{fig:kink_overtone_E}
\end{figure*}

\begin{figure*}[!t]
  \centering
  \begin{tabular}{ccc}
    \includegraphics[width=40mm]{C_blank_white.png}&
    \includegraphics[width=40mm]{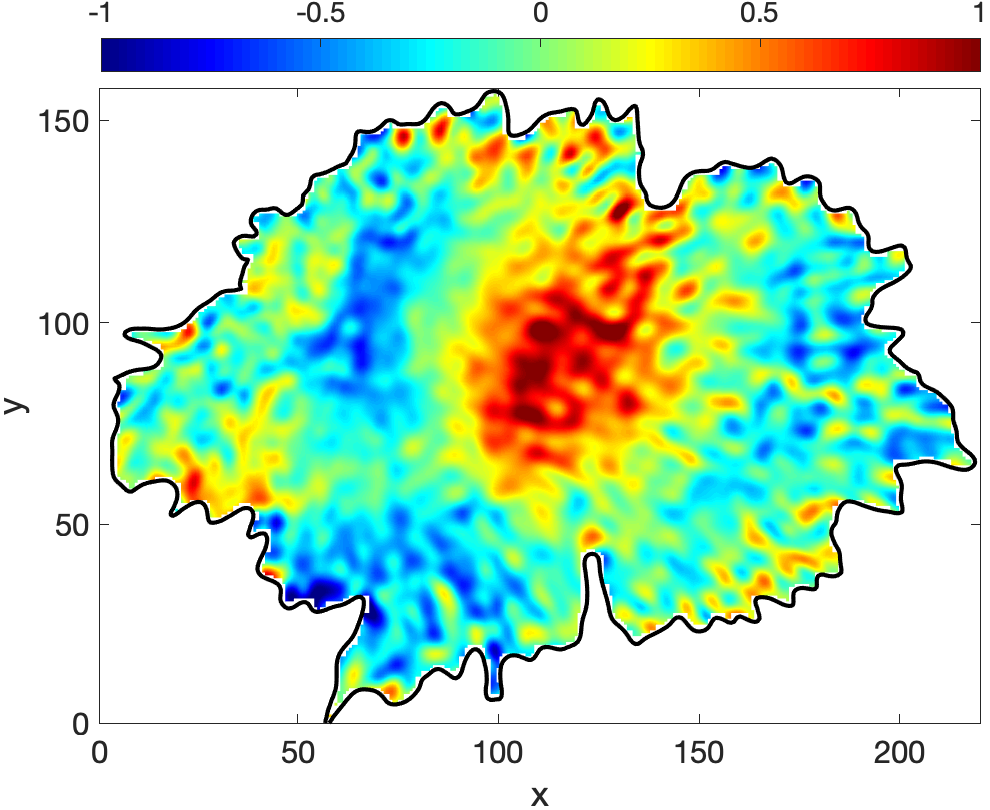}&
    \includegraphics[width=40mm]{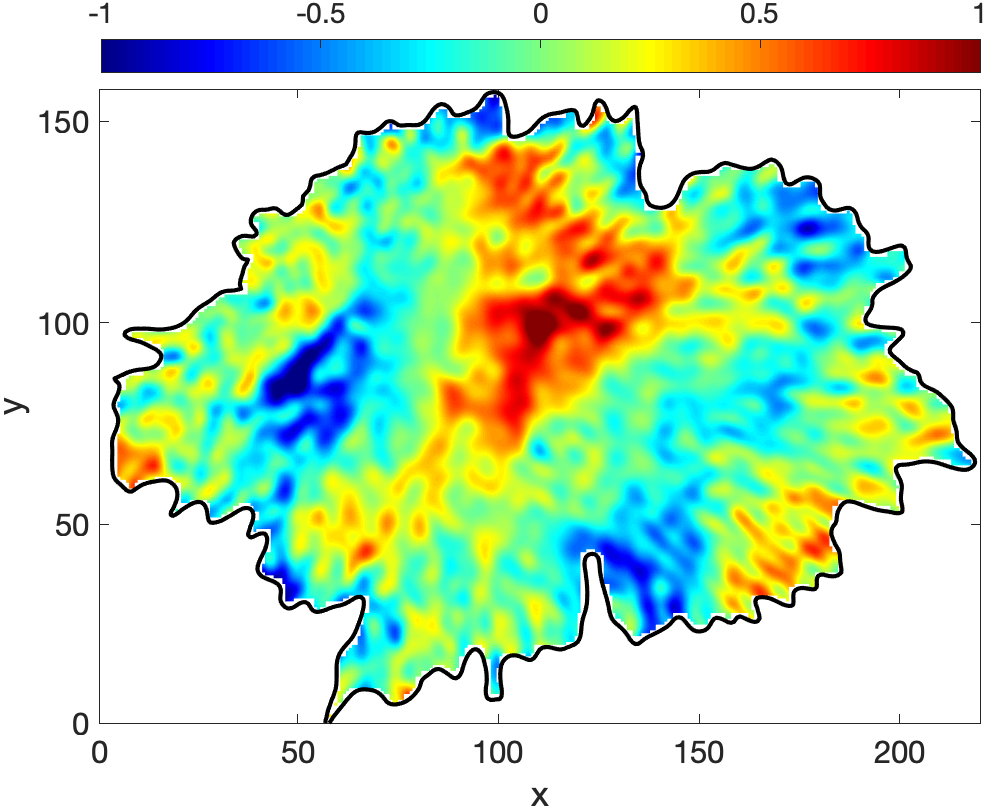}\\
    \includegraphics[width=40mm]{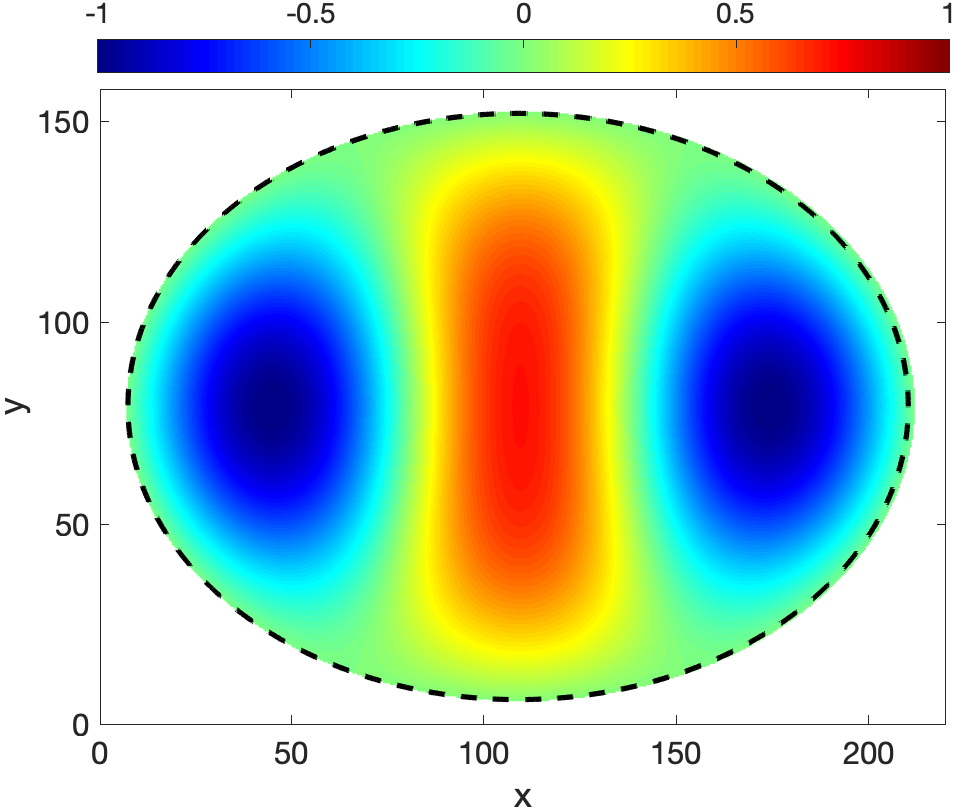}&
    \includegraphics[width=40mm]{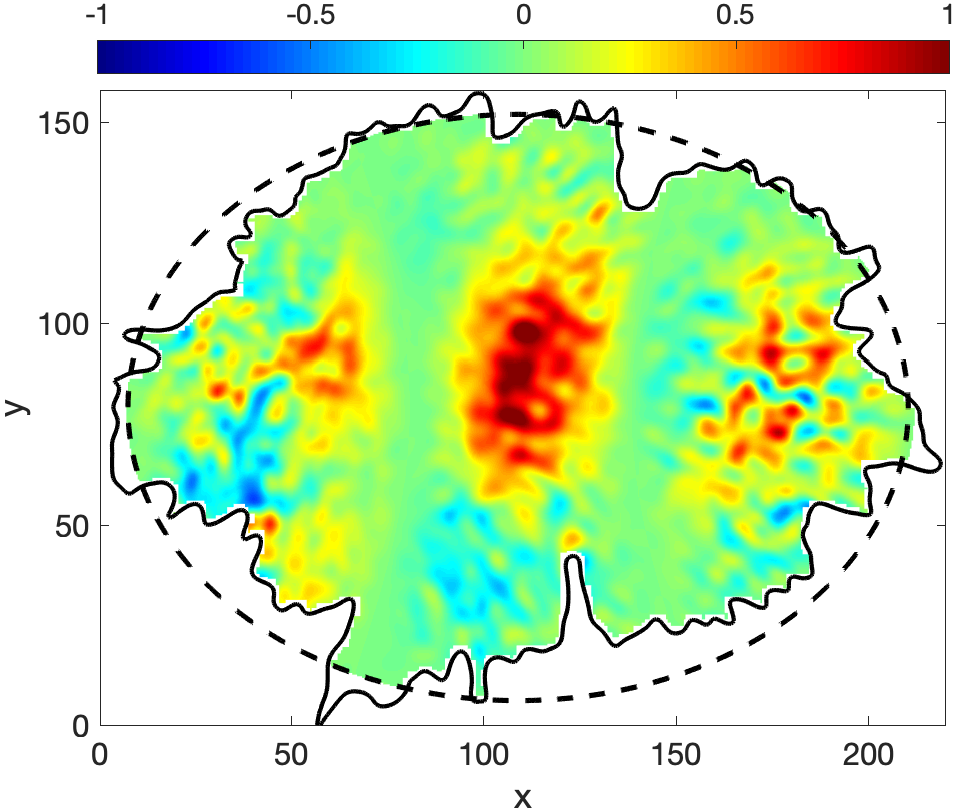}&
    \includegraphics[width=40mm]{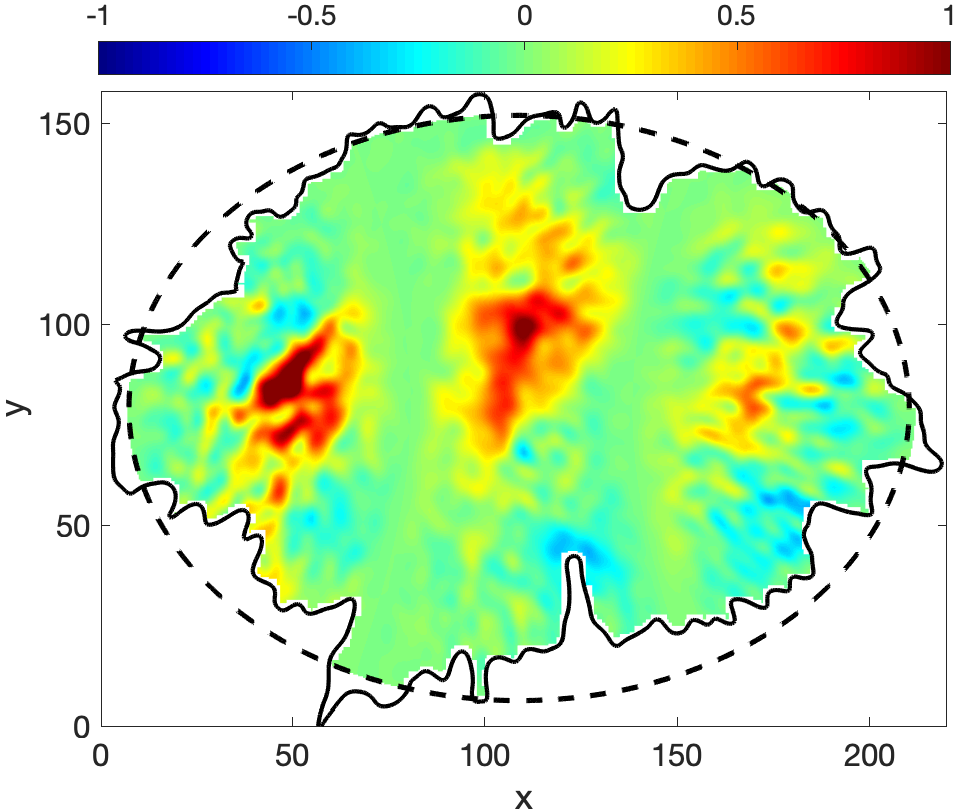}\\
    \includegraphics[width=40mm]{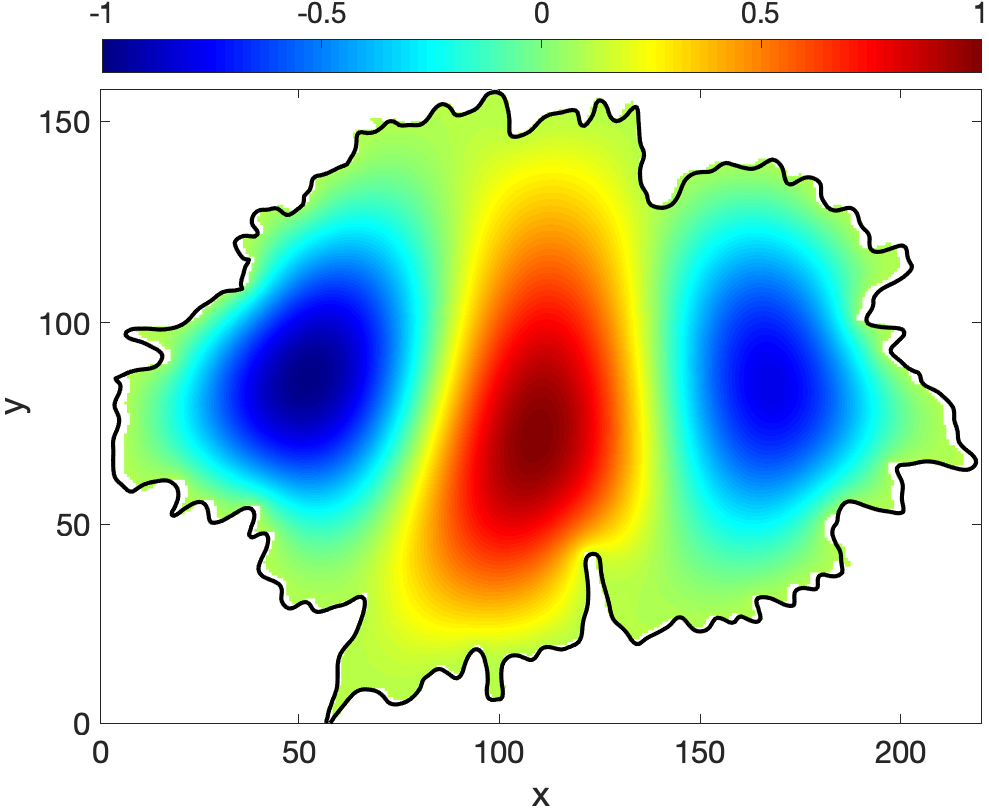}&
    \includegraphics[width=40mm]{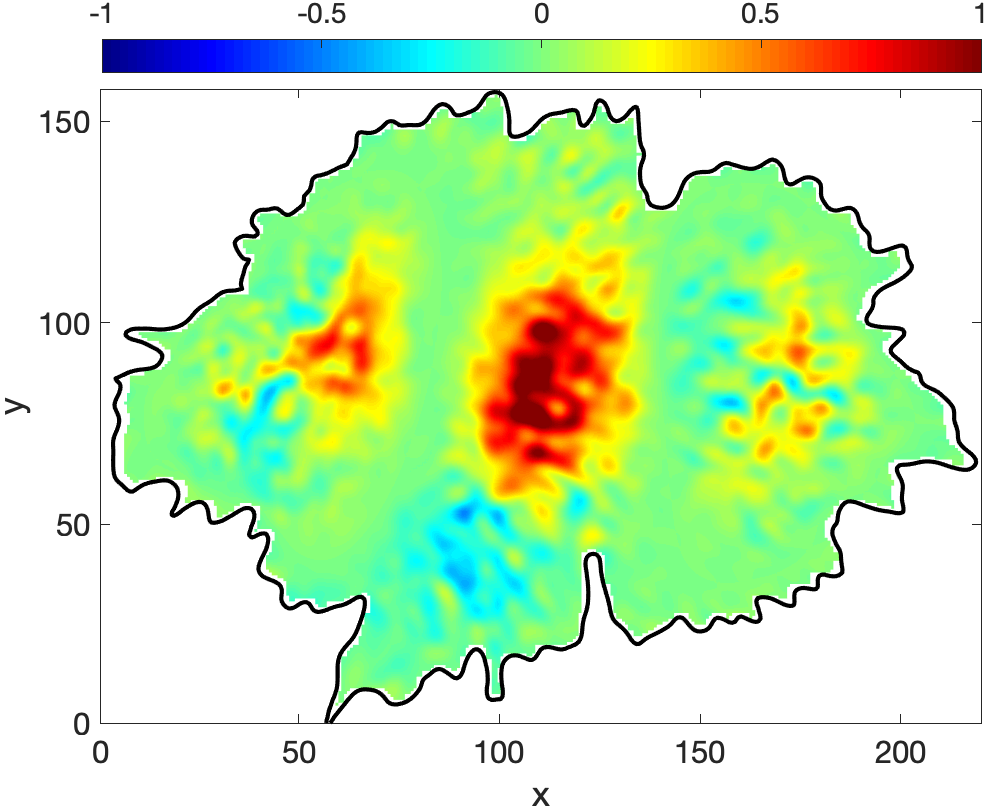}&
    \includegraphics[width=40mm]{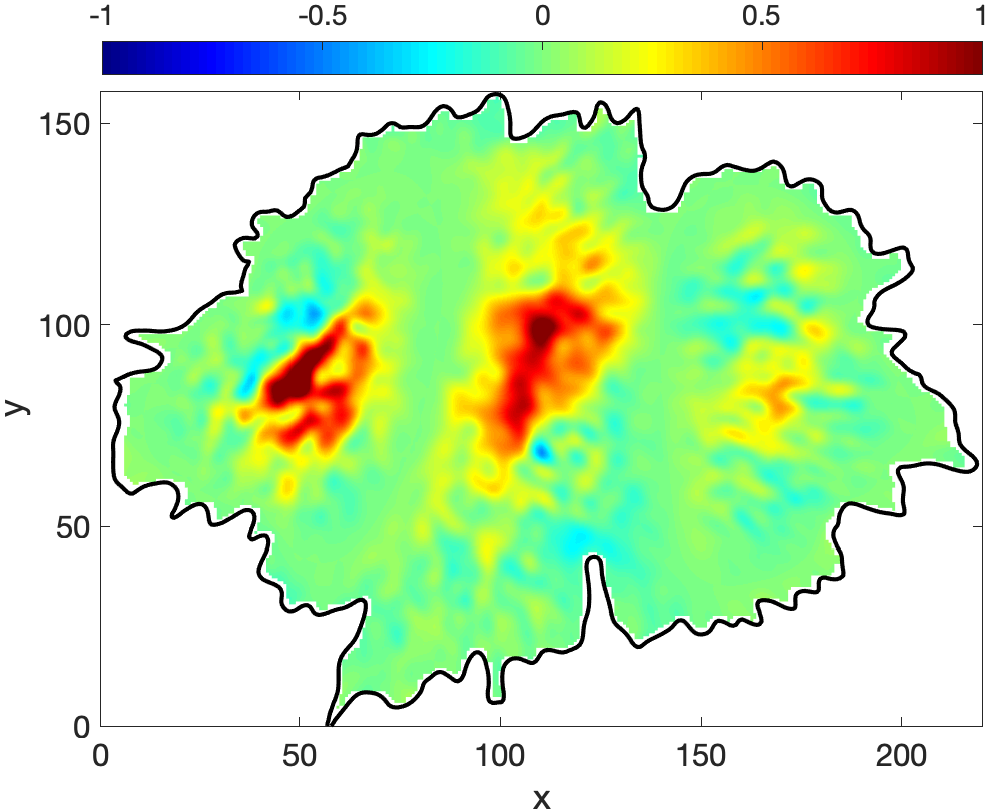}\\
\end{tabular}
\caption{This figure displays the 13$^{th}$ POD (top row, 1st panel) and DMD mode (top row, 2nd panel) with a  frequency of 5.6 mHz, which has an azimuthal symmetry of the slow body fluting mode ($n=2$). The 3D visualisation of this mode is shown in Figure \ref{fig:E_3D} in the Appendix.} \label{fig: n=2}
\end{figure*}

\begin{figure*}[!t]
  \centering
  \begin{tabular}{ccc}
    \includegraphics[width=40mm]{C_blank_white.png}&
    \includegraphics[width=40mm]{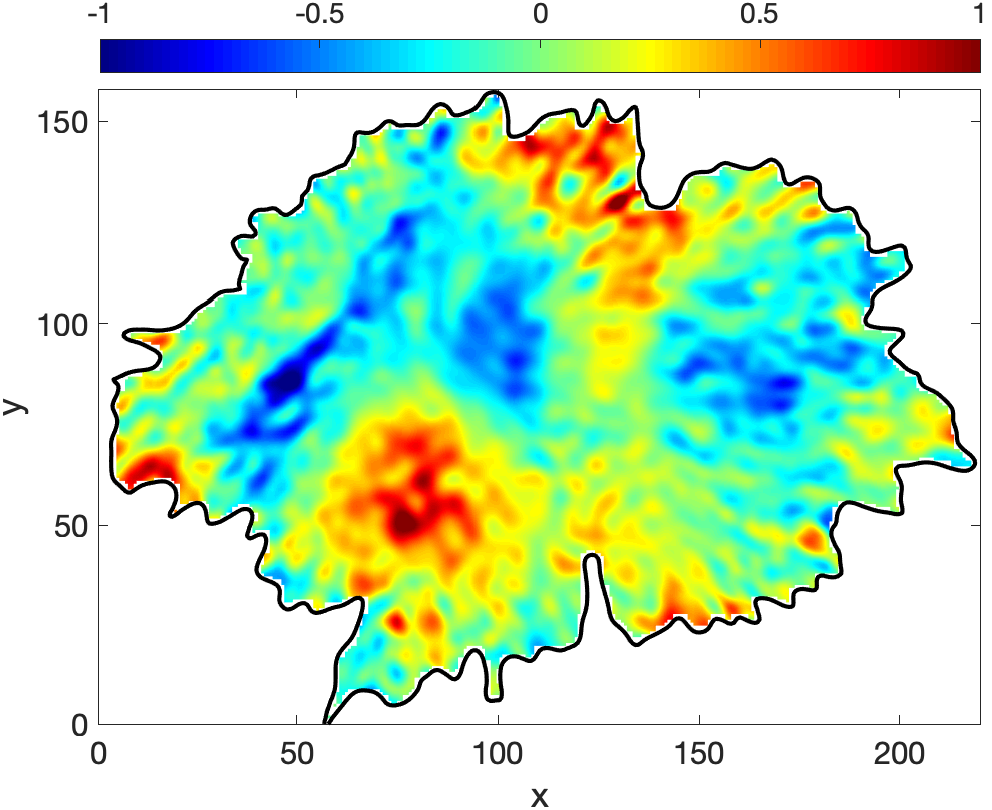}&
    \includegraphics[width=40mm]{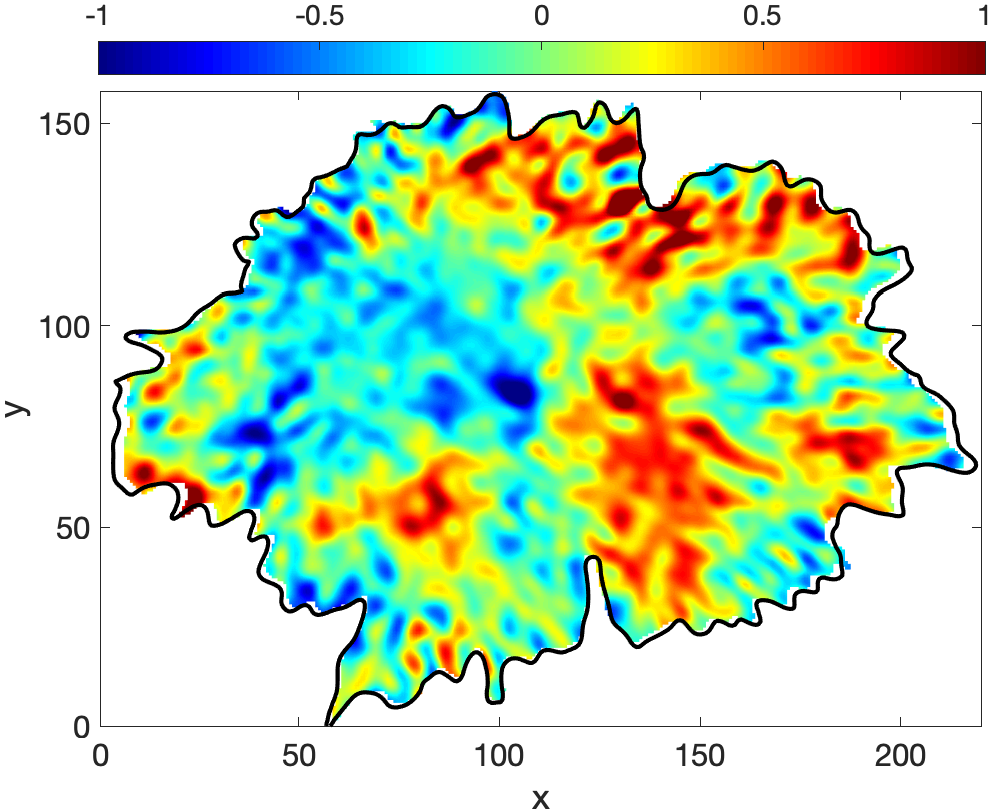}\\
    \includegraphics[width=40mm]{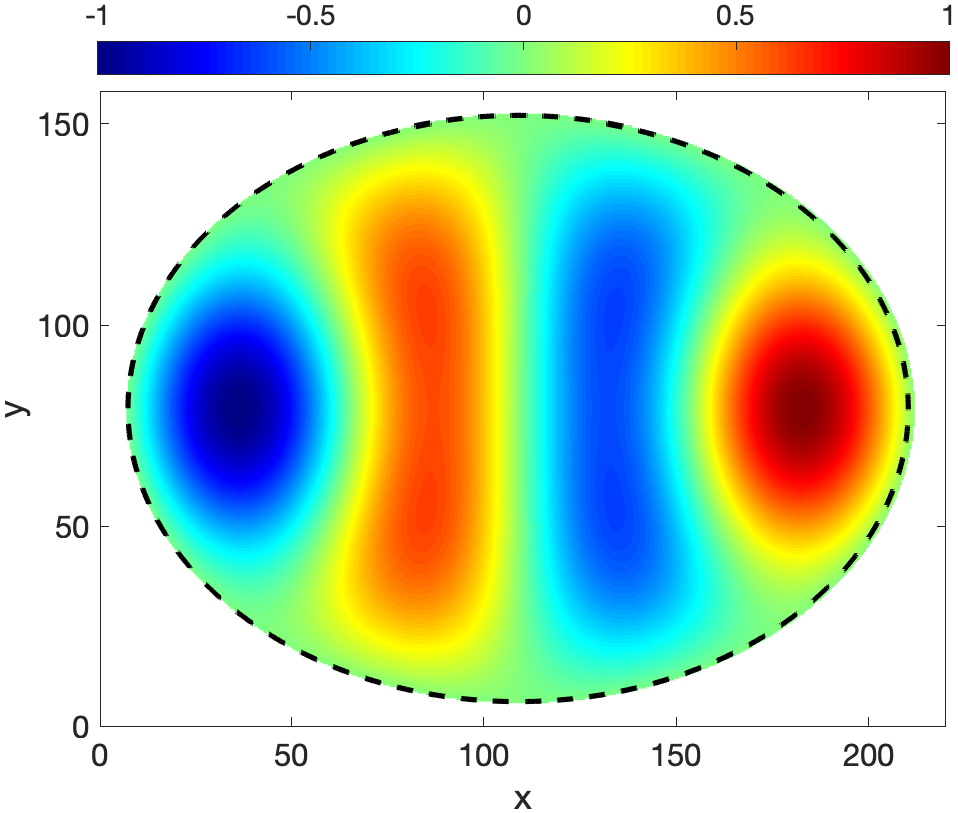}&
    \includegraphics[width=40mm]{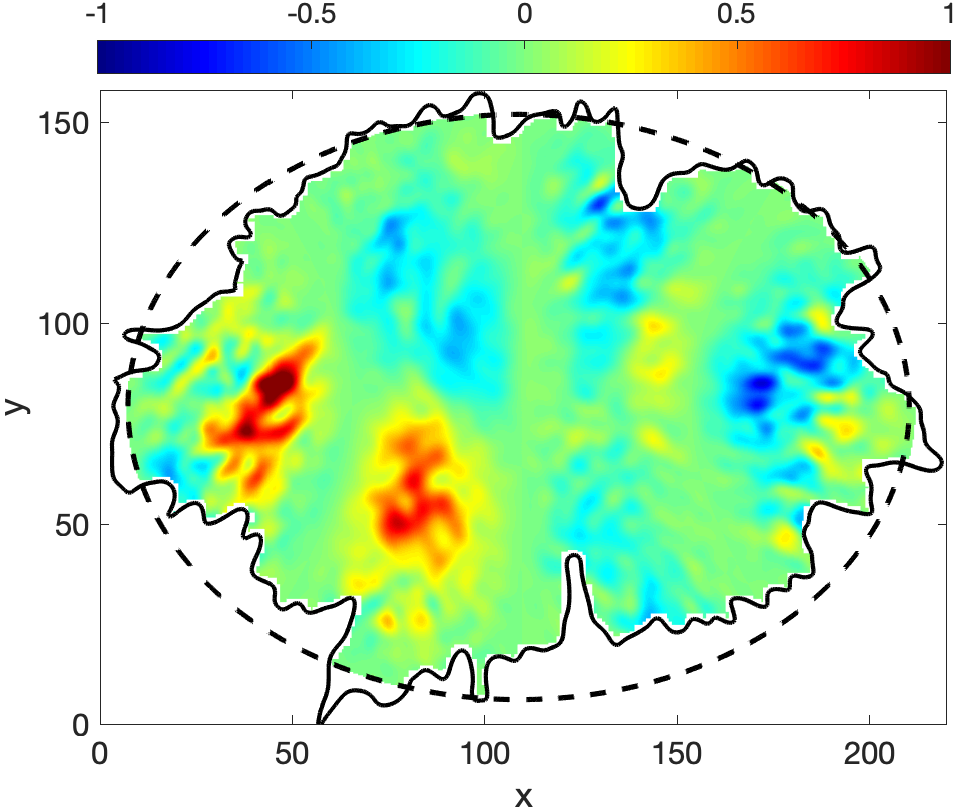}&
    \includegraphics[width=40mm]{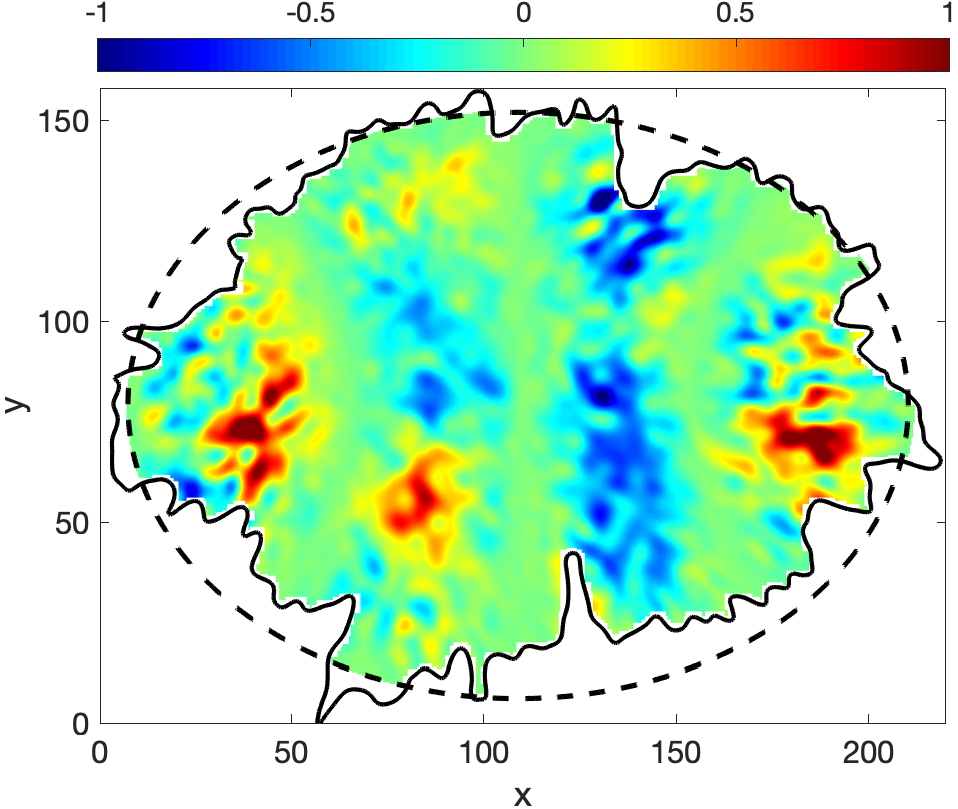}\\
    \includegraphics[width=40mm]{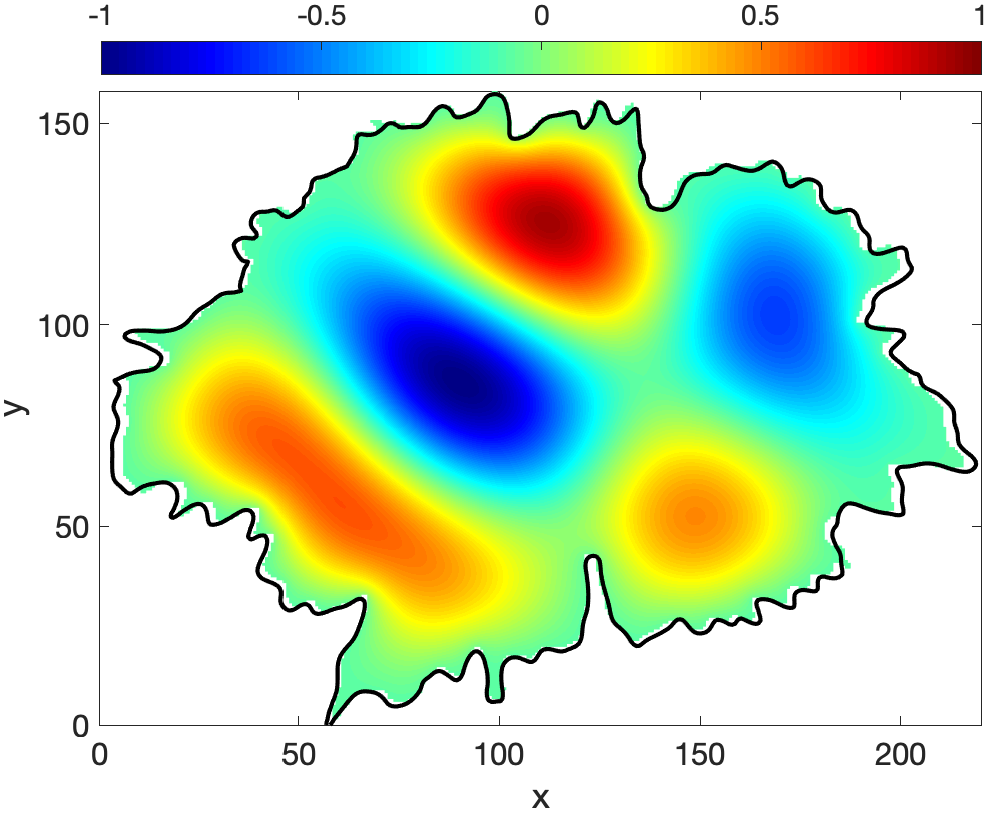}&
    \includegraphics[width=40mm]{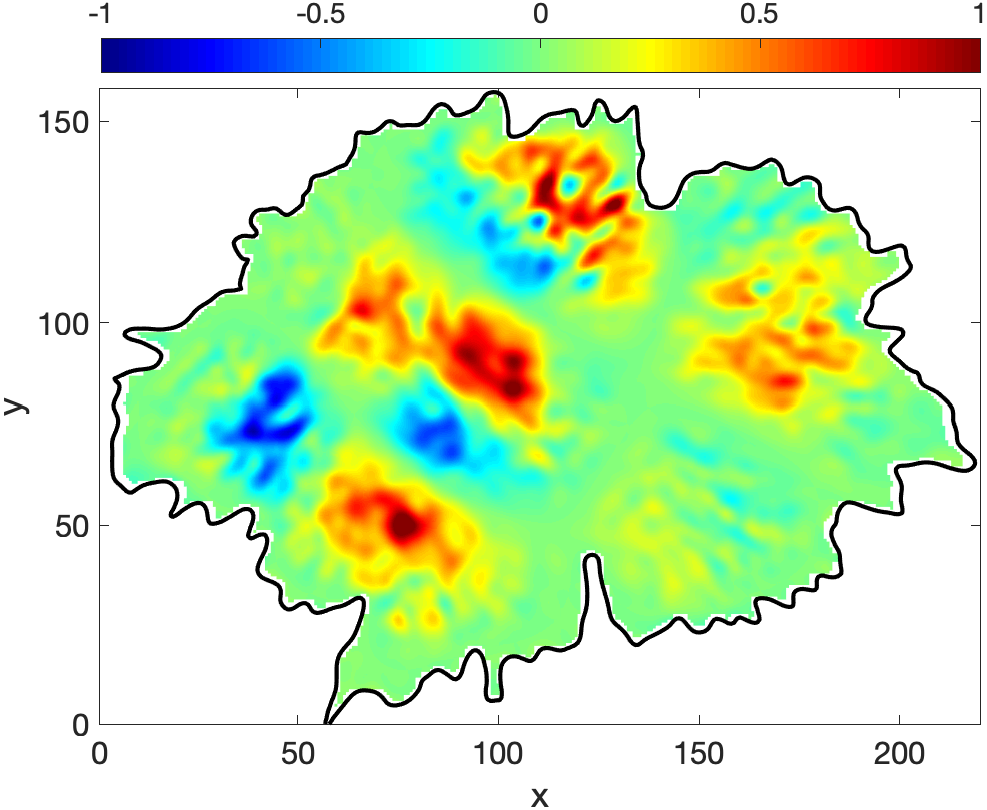}&
    \includegraphics[width=40mm]{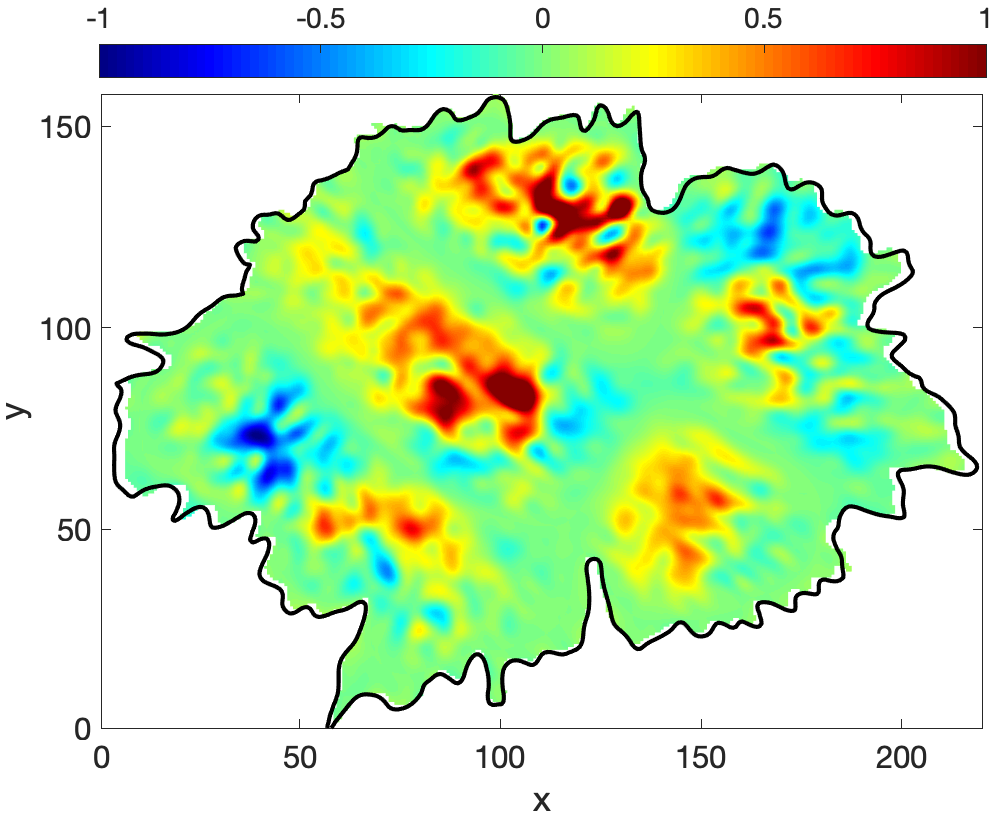}
\end{tabular}
\caption{This figure displays the 18$^{th}$ POD (top row, 1st panel) and DMD mode (top row, 2nd panel) with a  frequency of 6.2 mHz, which has an azimuthal symmetry of the slow body fluting mode ($n=3$). The 3D visualisation of this mode is shown in Figure \ref{fig:E_3D} in the Appendix.} \label{fig:m=3_E}
 \end{figure*}

It is important to note that in the case of fundamental modes (both, sausage and kink body modes), the change in the shape of the cross-section (from cylindrical, to elliptical and, finally, to a irregular shape) does not introduce significant changes in the morphology of waves, meaning that these modes can be confidently studied in regular shapes.   
The above statement proves to be incorrect for higher order modes. The 30$^{th}$ POD mode and the DMD mode that corresponds to 5.3 mHz show a high correlation with the \textit{slow body kink overtone mode}, as shown in Figure \ref{fig:kink_overtone_E}. The PSD of the time coefficients of the 30$^{th}$ POD shows a peak around 5.6 mHz, as displayed in the right panel, middle row of Figure \ref{fig: PSD_E}. The difference in the morphology of this wave between the pattern prediction of an elliptical and realistic (irregular) waveguide is remarkable, making the identification of the mode from observation misleading.  
Next, the 13$^{th}$ POD mode and the DMD mode that corresponds to 5.6 mHz are the modes that can be interpreted to have a higher correlation with the \textit{slow body fluting mode ($n=2$)}, as shown in Figure \ref{fig: n=2}. The PSD of POD 13 shows a peak at 5.8 mHz, as visible in the left panel, middle row of Figure \ref{fig: PSD_E}. Finally, the last mode is the \textit{slow body fluting mode} ($n=3$) identified as the 18$^{th}$ POD mode and the associated DMD mode with a frequency of 6.2 mHz, as shown in Figure \ref{fig:m=3_E}. The PSD of POD 18 mode has a peak around 6.08 mHz, as shown in the bottom left panel of Figure \ref{fig: PSD_E}. 

\begin{figure*}
\centering
\includegraphics[width=1\textwidth]{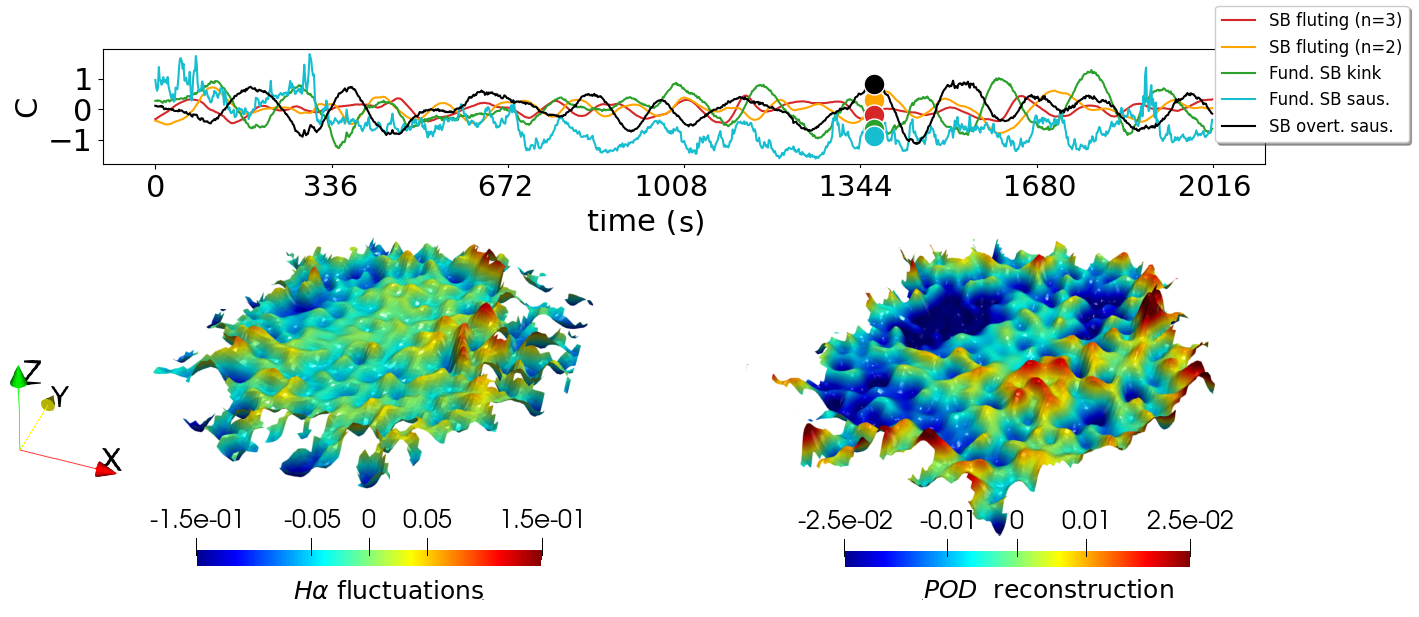}
\caption{Intensity fluctuations in the circular sunspot. The top panel shows the time coefficient, $C$, for the POD modes identified as MHD waves: slow body (SB) sausage overtone, SB fundamental sausage, SB fundamental kink, SB fluting (n=2), SB fluting (n=3). The colors of the lines and circles depict the detected MHD wave modes and the position of the circle indicates the time used for the plots in the bottom panels. The left bottom panel presents a 3D surface plot of the umbra where the $z$-direction describes the oscillations in the H$\alpha$ observations and it is colored by the observed intensity fluctuations at time $t=1450.7$ s. The right bottom panel is the 3D surface of the POD reconstruction of the intensity fluctuations using only the POD modes identified as MHD waves. The 3D surface is colored by the intensity fluctuations at time $t=1450.7$ s. The video of this 3D visualisation can be found at \href{https://sites.google.com/sheffield.ac.uk/pdg/visualisations\#h.dlw9y3ggtg2a}{PDG visualisations} web-page.}
\label{fig:POD_C_rec}
\end{figure*}
\begin{figure*}
\centering
\includegraphics[width=1\textwidth]{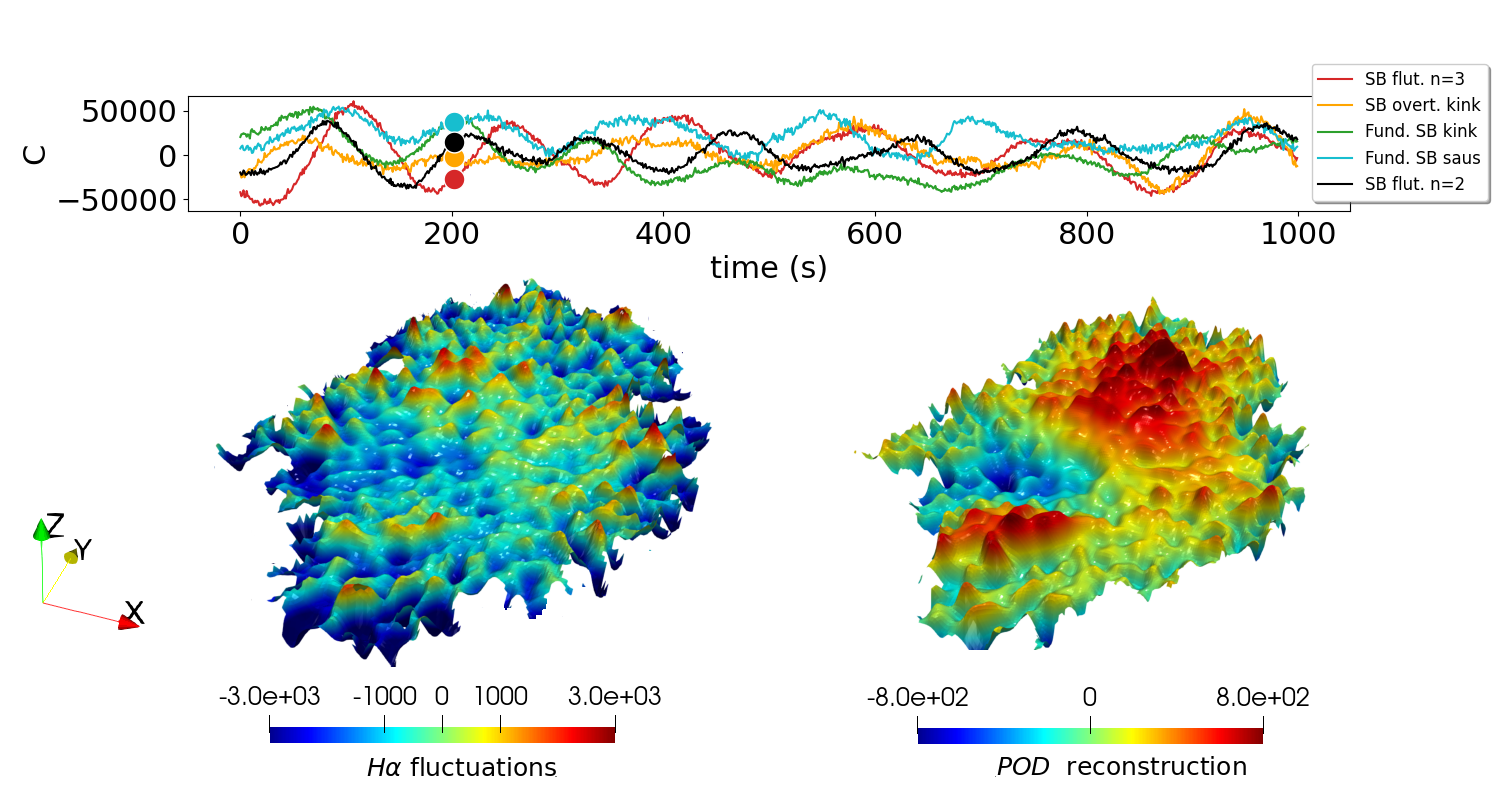}
\caption{Intensity fluctuations in the elliptical sunspot. The top panel shows the time coefficient, $C$, for the POD modes identified as MHD waves: slow body (SB) fluting (n=2), SB fundamental sausage, SB fundamental kink, SB overtone kink, SB fluting (n=3). The colors of the lines and circles depict the detected MHD wave modes and the position of the circle indicates the time used for the plots in the bottom panels. The left bottom panel presents a 3D surface plot of the umbra where the $z$-direction describes the oscillations in the H$\alpha$ observations and it is colored by the observed intensity fluctuations at time $t=202$ s. The right bottom panel is the 3D surface of the POD reconstruction of the intensity fluctuations using only the POD modes identified as MHD waves. The 3D surface is colored by the intensity fluctuations at time $t=202$s. For the POD reconstruction, we only used the  POD modes identified as MHD waves. The video of this 3D visualisation can be found at \href{https://sites.google.com/sheffield.ac.uk/pdg/visualisations\#h.dlw9y3ggtg2a}{PDG visualisations} web-page.}
\label{fig:POD_E_rec}
\end{figure*}
 
Apart from the sensitivity of the modes on the cross-sectional shape of the magnetic waveguide, it is also important to note that there is a higher correlation between the observed modes and the predictions of the model corresponding to the irregular shape. We should also note that there is no complete agreement between the observed modes in either in the circular or elliptical cross-section sunspot and the theoretical models and this disagreement can be attributed to the assumptions made in theoretical models regarding the constant values of the temperature, density, pressure, and magnetic field inside the magnetic flux tube. In reality the magnetic flux tubes are inhomogeneous in the transversal direction (see, e.g. observations of intensity provided by high-resolution observations by, to name but a few, \cite{gopalakrishnan2013study} and \cite{fritts2017high}). Furthermore, in the theory of guided MHD waves (see, e.g. \citet[][]{edwin1983wave}) modes are monochromatic and the lower order MHD wave modes have a lower frequency than higher-order modes, however, this holds true only when the wavenumber, $k_z$, is constant. In the family of identified modes in the present study, there are some higher-order modes having a lower frequency than the lower-order modes, however, they have different wavenumber, as shown in Table \ref{Table1} and \ref{Table2}.

One advantage of using the POD and DMD techniques for the identification of MHD modes in sunspot is the opportunity to detect a high number of MHD wave modes in one single sunspot, as these techniques provide a number of modes equal the number of snapshots of the data set. The challenge is, however, to select and identify those modes that are physical. Other techniques have their own limitation in identifying MHD waves in a sunspot. For example, the limitation of Fourier filtering is that it applies a wide range of bandpass filter. In the case of the circular sunspot, our earlier study \citep{albidah2020RS} identified the fundamental slow body kink mode with a frequency of 5.88 mHz and the slow body sausage overtone with a frequency of 5.61 mHz. However, in the original analysis by \citep{2017ApJ...842...59J}, using the same sunspot they applied a $k-\omega$ Fourier filter ($0.45 - 0.90$ arcsec$ ^{-1}$ and $5 - 6.3$ mHz) that resulted in the identification of only the slow body kink mode. In general  higher order modes cannot be detected, and this can be attributed to the fact that these modes have less energy than the fundamental modes, so it is expected that the spectrum is dominated by the fundamental modes. The POD and DMD techniques are able to address this shortcoming. 

Figures \ref{fig:POD_C_rec} (for the circular sunspot), and \ref{fig:POD_E_rec} (for the elliptical sunspot), illustrate the ability of the POD technique to capture the portion of the umbral oscillations that are due to MHD wave modes among the observed intensity fluctuations. The top panel displays the POD time coefficient for the five detected MHD wave modes within a given time interval: 1000 seconds for the elliptical sunspot and 2016 seconds for the circular one. The lines and circles are color coded by the MHD wave mode, and the position of the circles indicate the value of time used for the plots in the bottom panels. The bottom left panels show the 3D surface representation of the original sunspot oscillations, while in the right panels we see the reconstructed oscillations using the POD modes identified as MHD wave modes. The POD  technique separates the effects of oscillations that are due to wave propagation, enhancing the expected wave pattern in the umbra. An animated movies for Figures \ref{fig:POD_C_rec} and \ref{fig:POD_E_rec} is available at \href{https://sites.google.com/sheffield.ac.uk/pdg/visualisations#h.dlw9y3ggtg2a}{PDG visualisations} web-page. The analysed sunspots present a considerable discrepancy for the values of the time coefficients as the dynamics of the spatial modes changes considerably for different umbra geometry and size. Although there is a considerable difference between the POD reconstructed oscillations intensity and the original perturbations, this discrepancy is expected as the POD modes have less energy than the other "non-physical" modes detected by POD. The low contribution of wave propagation to the observed oscillations may be a consequence of different reasons (i.e. global modes, locally excited fluctuations) to the global variance of the oscillatory field in a sunspot. This reinforces the POD as a valuable tool to apply to wave detection in sunspot as other methodologies require properly filtering the data in order to disentangle and detect the resonant modes.

\begin{table}
\begin{center}
\begin{tabular}{l c c c c c}
\hline
MHD wave mode  & $\tilde{m}_{i}$ & $f$ (mHz) & $k_z$ (Mm$^{-1}$)  &  $\lambda$ (Mm) & $V_{ph}$ (km/s)\\
\hline 
\hline 
Fundamental slow body sausage & $1.1644$ & $3.5$ & $2.24887$ & $2.7939$& $9.77$\\
Fundamental slow body kink & $1.6368$ & $5.88$ & $3.7696$ & $1.6667$& $9.80$\\
Slow body overtone kink & $3.1698$ & $5.3$ & $3.412798$ & $1.8410$& $9.75$\\
Slow body fluting ($n=2$)   & $2.6328$ & $5.61$ & $3.6166$ & $1.7373$& $9.74$\\
Slow body fluting ($n=3$)  &  $2.7837$ & $6.2$ & $3.995553$ & $1.5725$& $9.74$\\
\hline
\end{tabular}
 \caption{This table shows the summary of MHD waves' properties that were detected by the POD and DMD techniques in the sunspot with elliptical cross-section. The first column contains the type of the modes, and the value of the magnetoacoustic parameter, $\tilde{m}_{i}$, is shown in the second column (see Equation \ref{Eq:2}). The third column contains the frequency of waves, as determined from the DMD analysis. The fourth column contains the wavenumber along the vertical direction of the sunspot, and it is calculated using Equation \ref{Eq:2}, with $\omega=2\pi f$, $c_i=10$ km s$^{-1}$, $v_{A_i}=4c_i$ and $\sigma^{2}=0.4174$. The fifth column shows the wavelength ($\lambda=2\pi/k_z$) of waves, while the last column contains the phase speed ($V_{ph}=f \lambda$) of waves. All units of physical quantities are shown in the table.\label{Table2}}
\end{center}
\end{table}

\newpage
\subsection{Surface wave identification}
\label{sec:surface}

\begin{figure*}[!t]
\centering
\begin{tabular}{ccc}
\includegraphics[width=40mm]{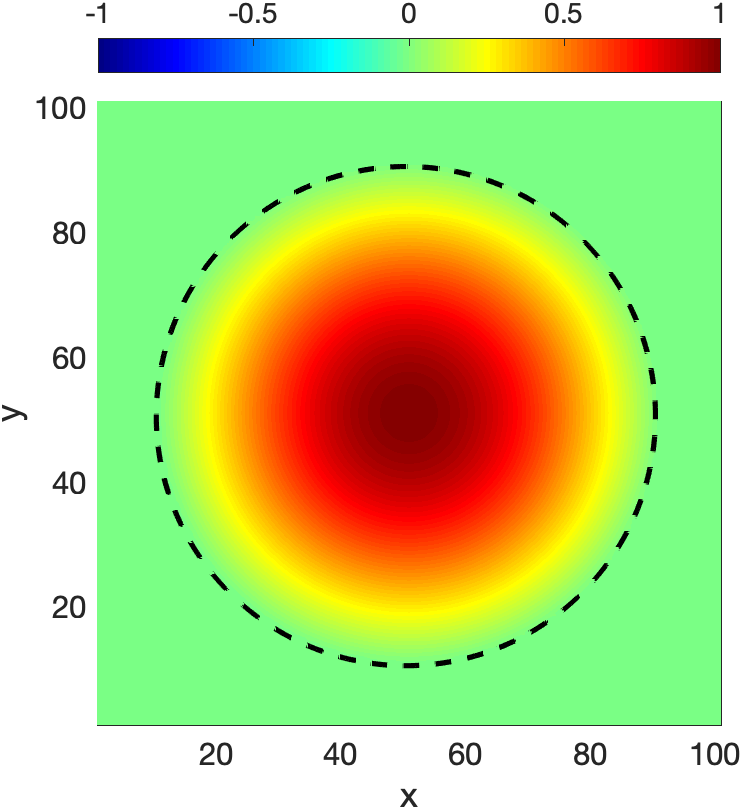}&
\includegraphics[width=40mm]{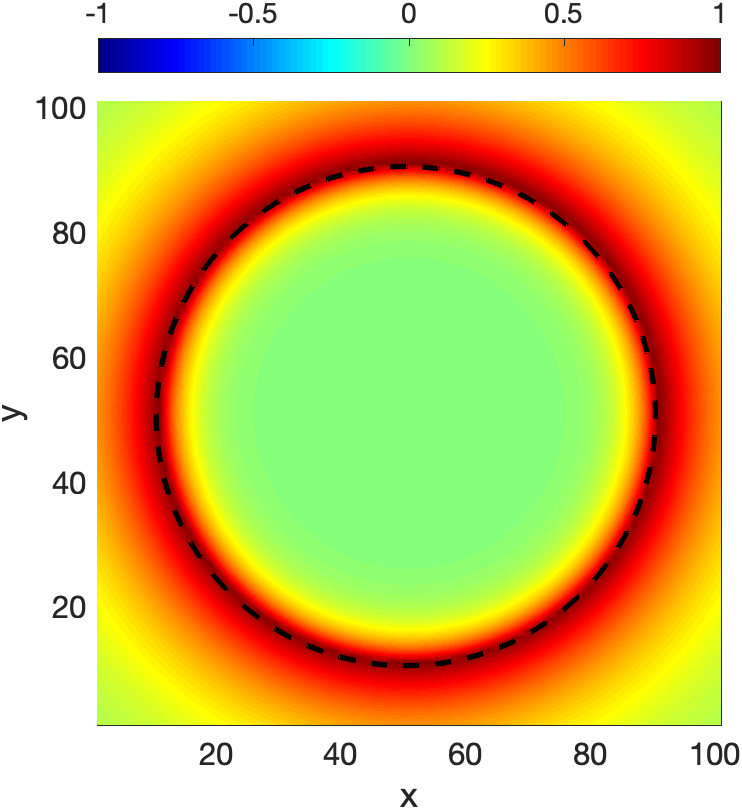}&
\includegraphics[width=40mm]{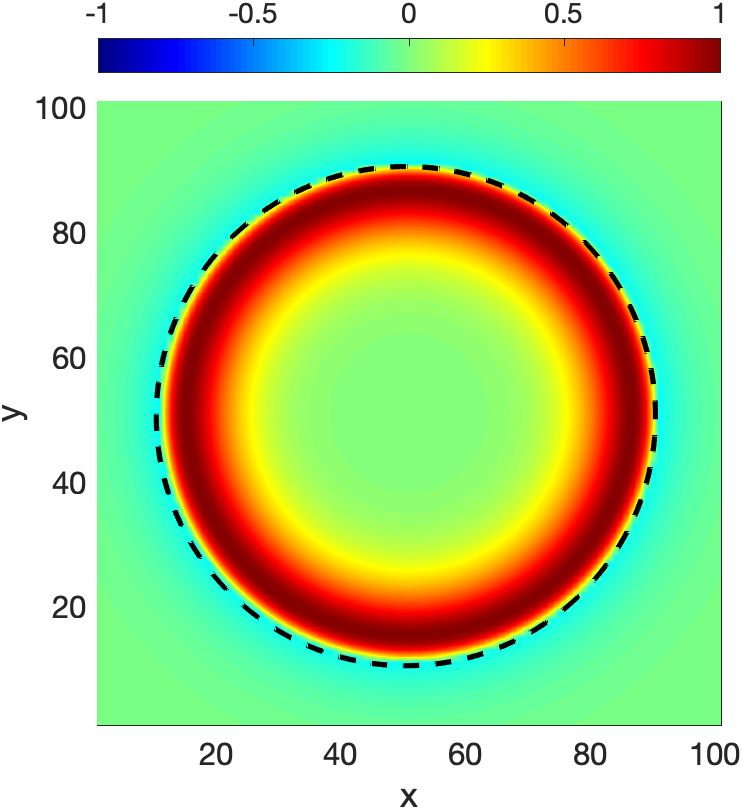}\\

\includegraphics[width=40mm]{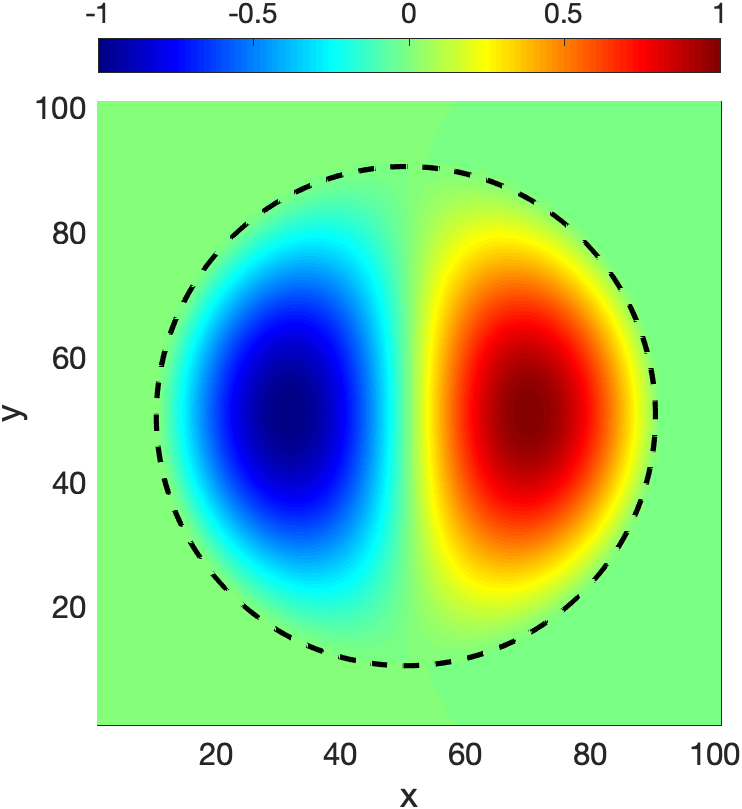}&
\includegraphics[width=40mm]{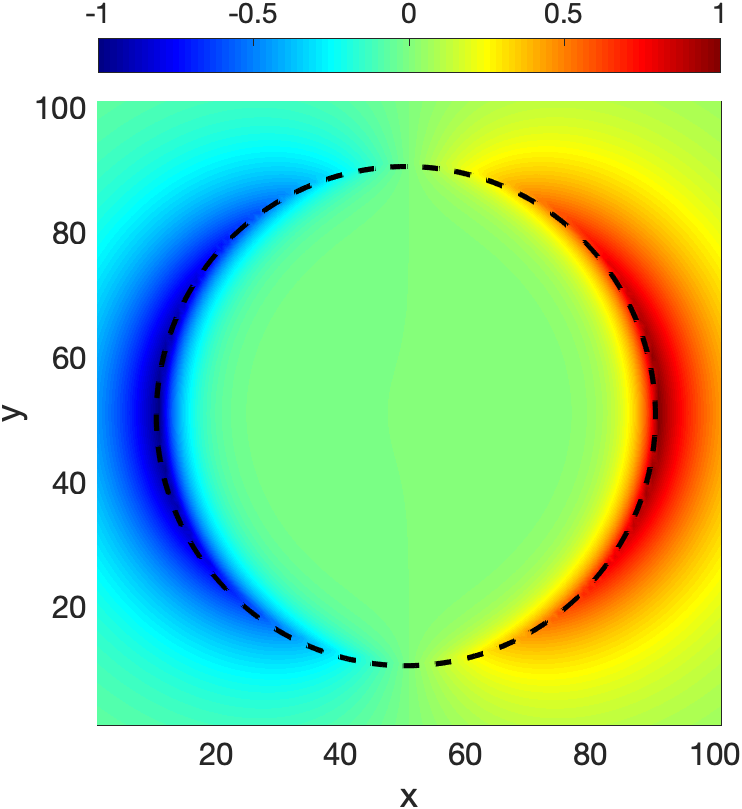}&
\includegraphics[width=40mm]{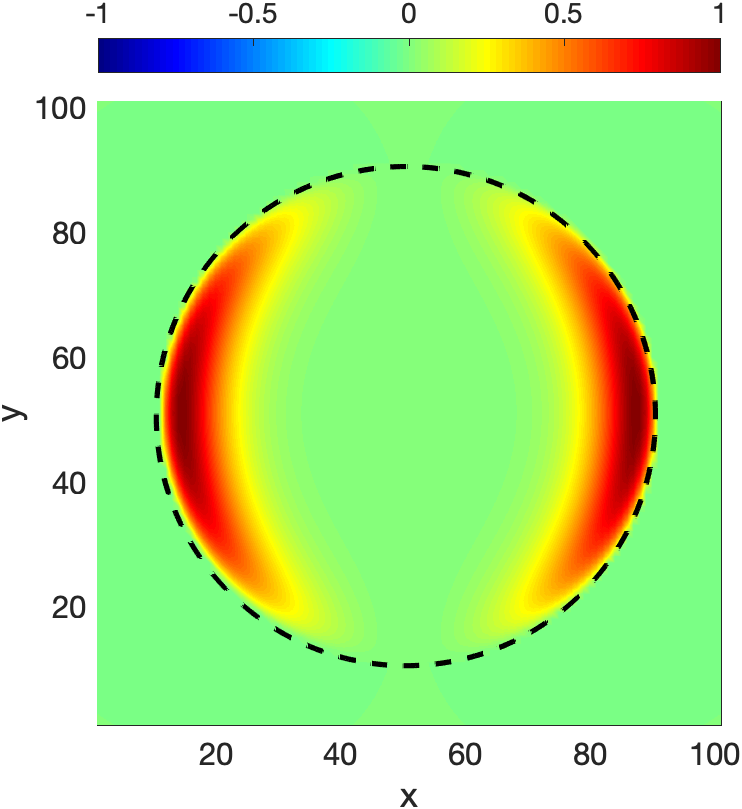}\\

\end{tabular}
\caption{This figure shows the sausage mode (first row) and the kink mode (second row) of the cylindrical magnetic flux tube, where the first column shows the slow body mode, the second column shows the slow surface mode and the third column shows the cross-correlation between the first and second columns.}
\label{fig: slow_surface_and_body_Corr}
\end{figure*}

\begin{figure*}[!t]
\centering
\begin{tabular}{ccc}
\includegraphics[width=40mm]{ZZ_sausage_slow_body.png}&
\includegraphics[width=40mm]{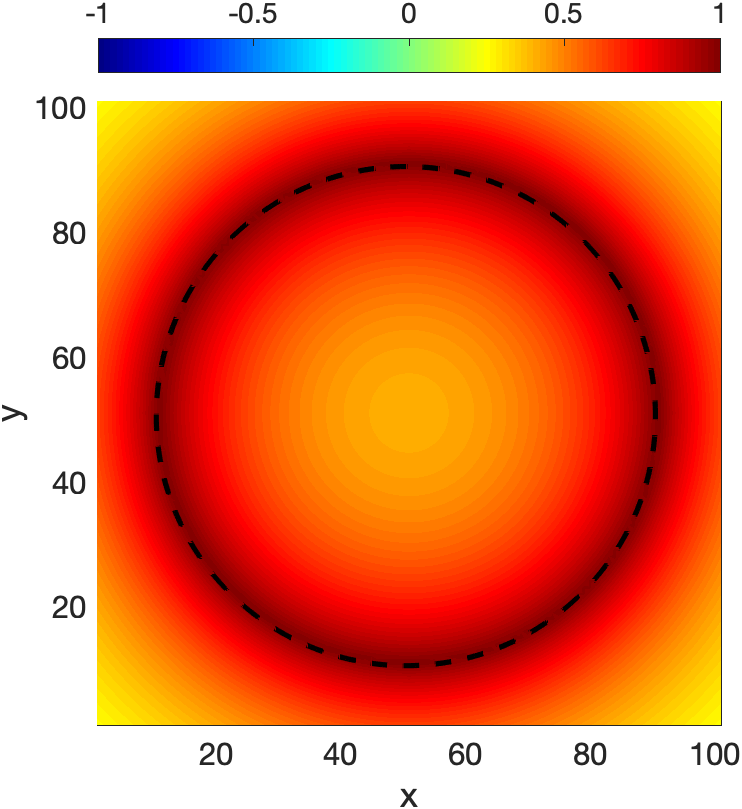}&
\includegraphics[width=40mm]{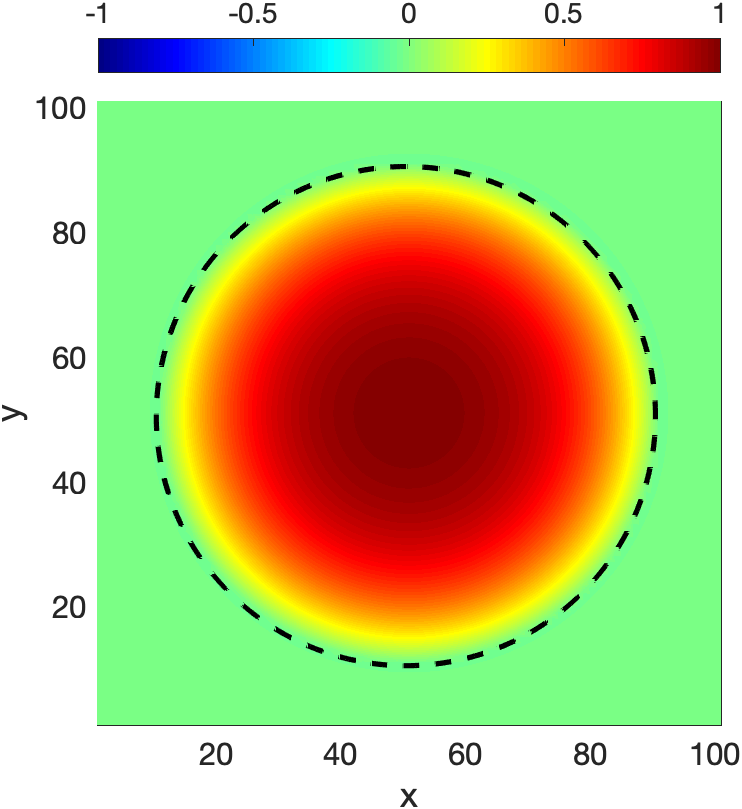}\\

\includegraphics[width=40mm]{ZZ_kink_slow_body.png}&
\includegraphics[width=40mm]{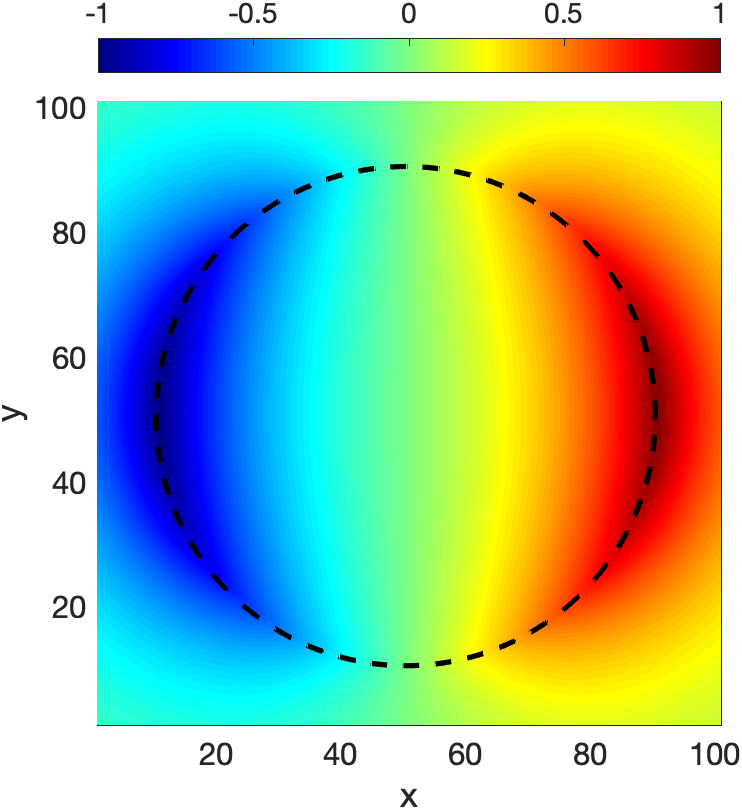}&
\includegraphics[width=40mm]{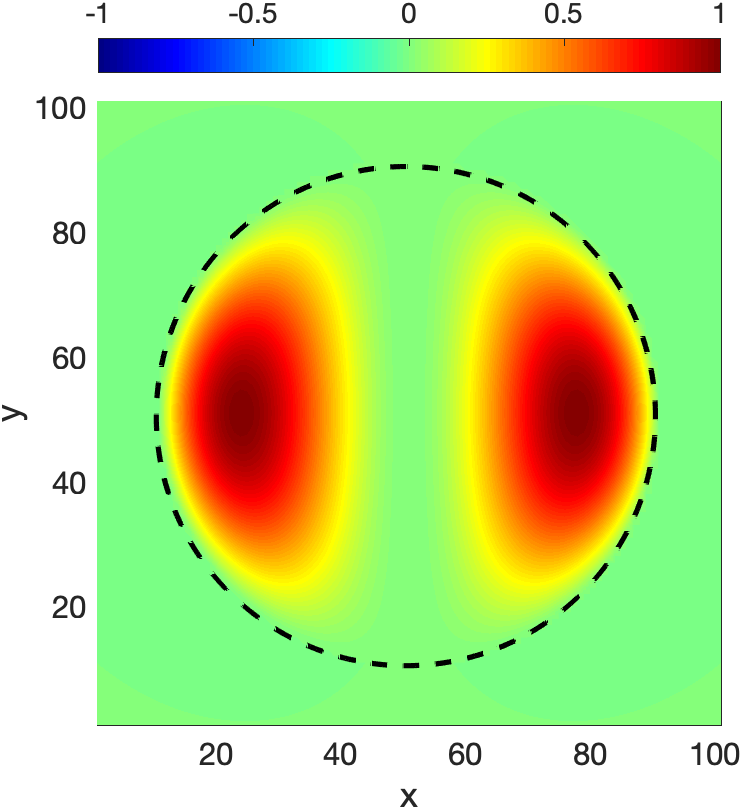}\\

\end{tabular}
\caption{This figure shows the sausage mode (first row) and the kink mode (second row) of the cylindrical magnetic flux tube, where the first column shows the slow body mode, the second column shows the fast surface mode and the third column shows the cross-correlation between the first and second columns.}
\label{fig: Fast_surface_and_body_Corr}
\end{figure*}

\begin{figure*}[!t]
\centering
\begin{tabular}{ccc}

\includegraphics[width=40mm]{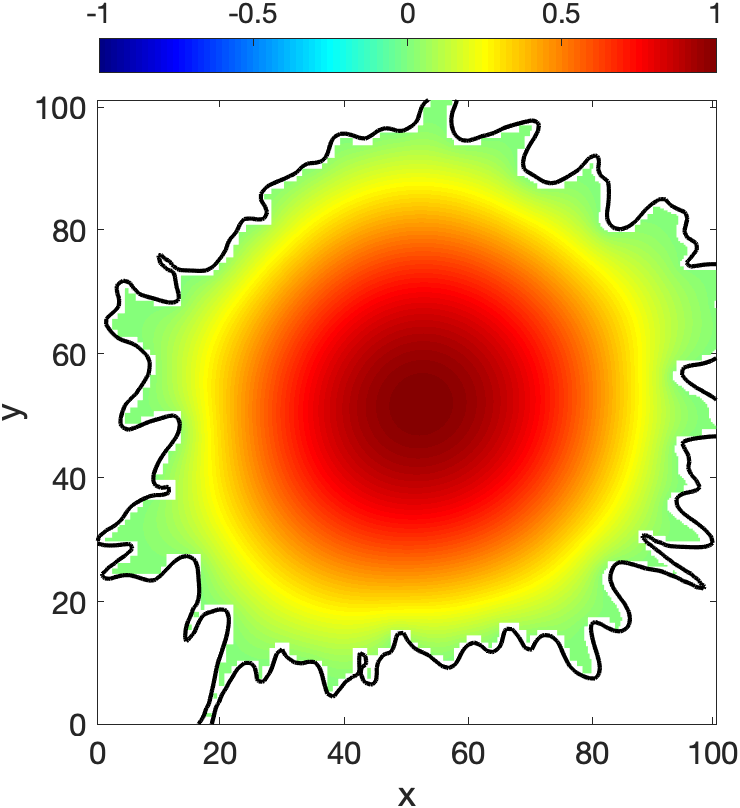}&
\includegraphics[width=40mm]{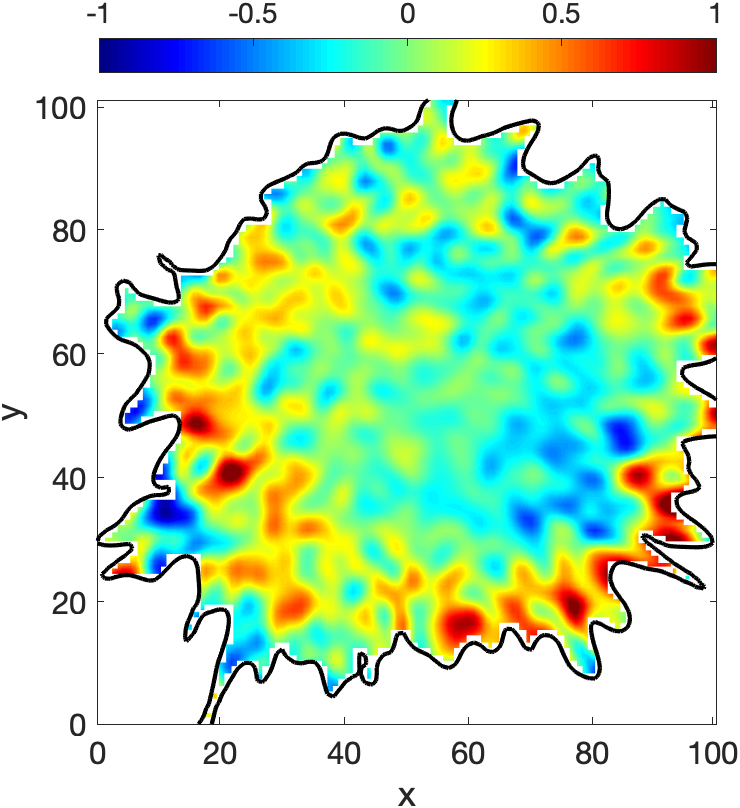}&
\includegraphics[width=40mm]{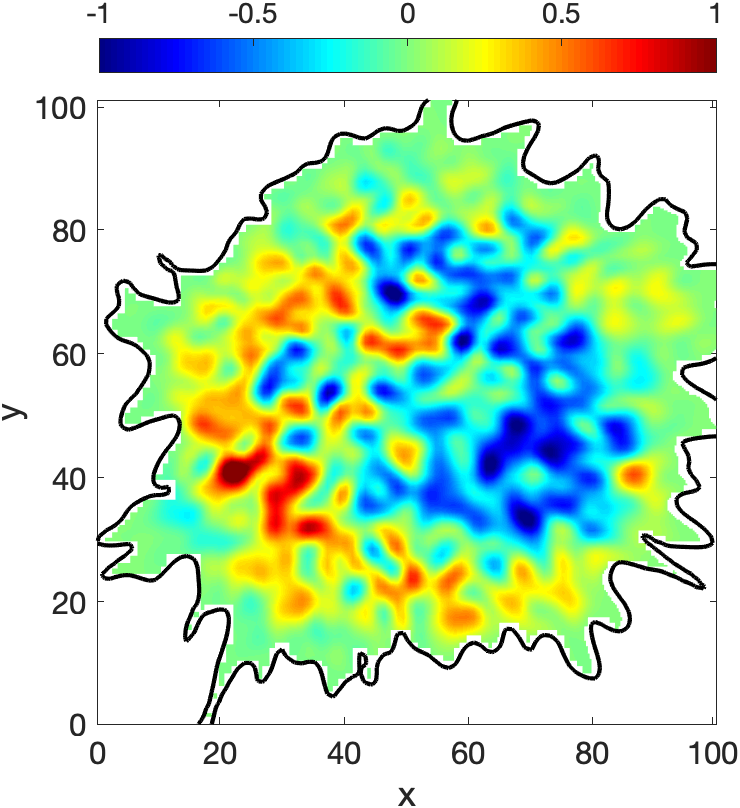}\\

\end{tabular}
\caption{This figure shows the fundamental slow body sausage mode as shown above in Figure \ref{fig:fund_sausage_Circular} (left panel), the spatial structure of POD 10 (middle panel) and the cross-correlation between the left panel and the middle panel (right panel)}.
\label{fig: susage_surface}
\end{figure*}

\begin{figure*}[!t]
\centering
\begin{tabular}{ccc}
\includegraphics[width=40mm]{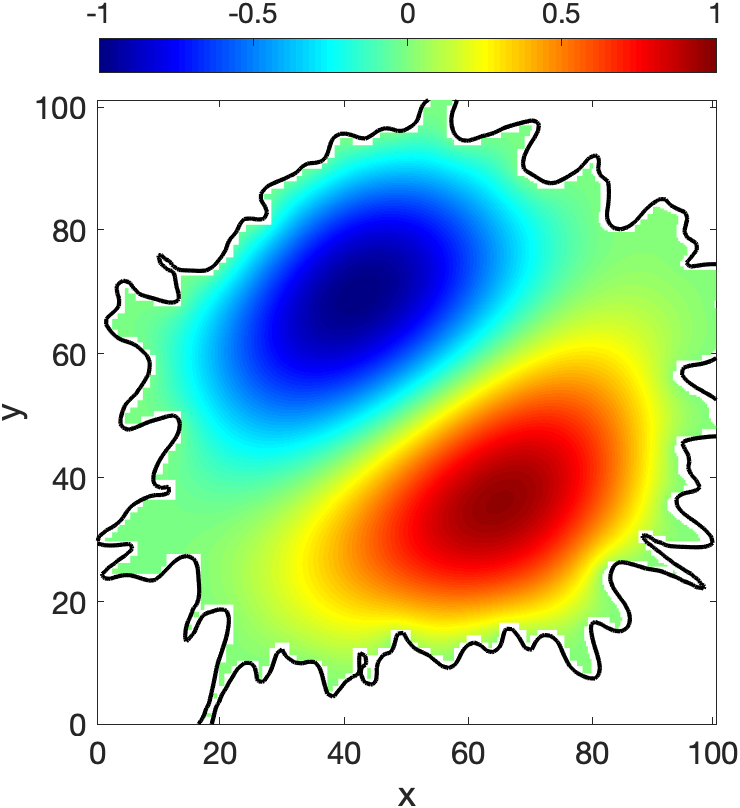}&
\includegraphics[width=40mm]{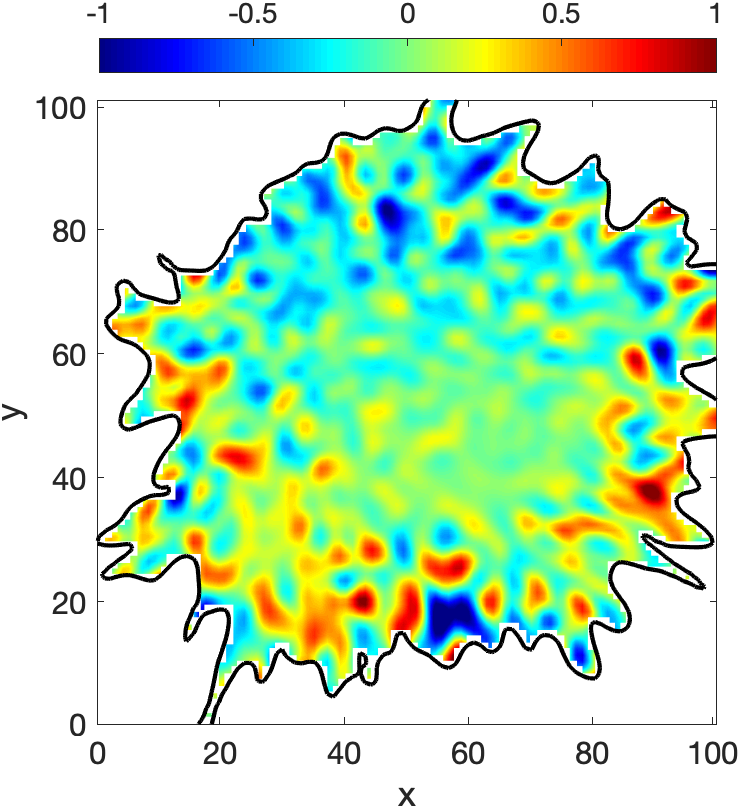}&
\includegraphics[width=40mm]{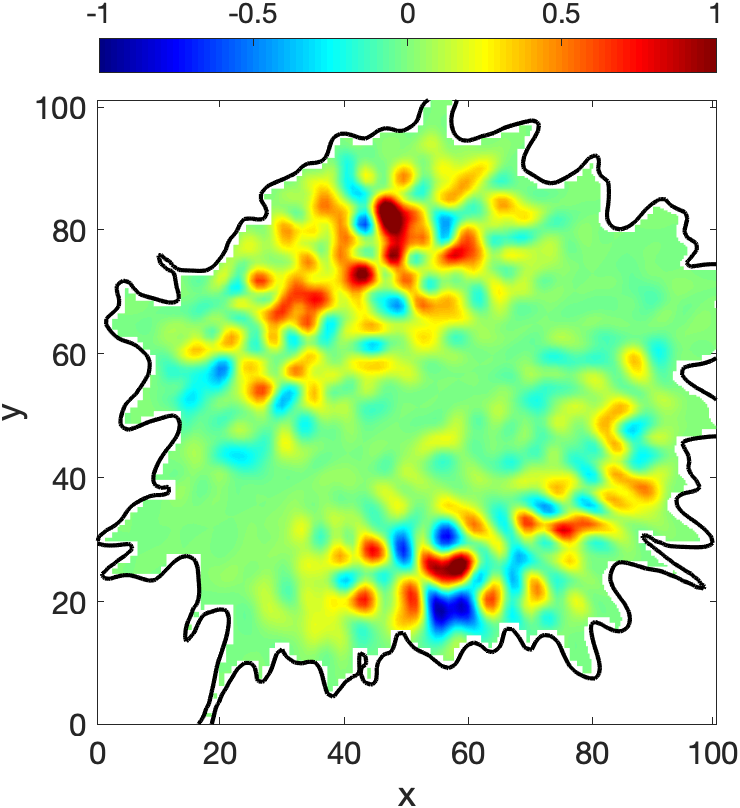}\\

\end{tabular}
\caption{This figure shows the fundamental slow body kink mode as shown above in Figure \ref{fig:fund_kink_Circular} (left panel), the spatial structure of POD 6 (middle panel) and the cross-correlation between the left panel and the middle panel (right panel)}.
\label{fig: kink_surface}
\end{figure*}

The POD and DMD techniques can also be applied to identify surface modes. In the POD analysis performed on the circular sunspot it was found that there were two modes that have the characteristics of surface waves: the POD 10 mode (see the middle panel of Figure \ref{fig: susage_surface}), which shows the azimuthal symmetry corresponding to the fundamental (slow or fast) surface sausage mode, and the POD 6 mode (see the middle panel of Figure \ref{fig: kink_surface}), which has a pattern close to the fundamental (slow or fast) surface kink mode. Our theoretical model is restricted  to the identification of slow body modes, i.e. modes corresponding to $v_z=0$ at the umbra/penumbra boundary (see section \ref{sec:irr}). Therefore, in the framework of this study, cross-correlation with possible surface modes detected with POD/DMD and their direct theoretical counterparts cannot be performed. Nevertheless, the cross-correlation between the theoretical slow body and slow surface modes in a magnetic cylinder, produces a distinctive closed ring for the sausage mode and a broken ring for the kink mode, see Figure \ref{fig: slow_surface_and_body_Corr}, with a clear in phase relationship. Moreover, the cross-correlation between the slow body mode and the fast surface mode also provides an in phase, but the spatial structure is very close to the slow body eigenmode shown in Figure \ref{fig: Fast_surface_and_body_Corr}. These distinctive signatures provide, at least, an indirect way of detecting slow and fast surface modes in the observational data.

The POD modes which appear most likely to be surface modes have been correlated with the fundamental slow body sausage and kink modes as shown in Figures \ref{fig: susage_surface} and \ref{fig: kink_surface}, respectively. From Figure \ref{fig: susage_surface} (middle panel) the red ring is present indicating a slow surface sausage mode but the blue regions of anti-phase inside the red ring and on the outer left edge cannot be explained by the theoretical model. The anti-phase regions are even more prominent in Figure \ref{fig: susage_surface} (right panel) and are also not predicted for the theoretical fast surface mode. The cross-correlation for the POD 6 mode shows stronger agreement with a fast surface kink mode as shown in Figure \ref{fig: kink_surface}, although there are still some small anti-phase regions which are not consistent with the theoretical model.

\begin{figure*}[!t]
\centering
\begin{tabular}{c}
\includegraphics[width=70mm]{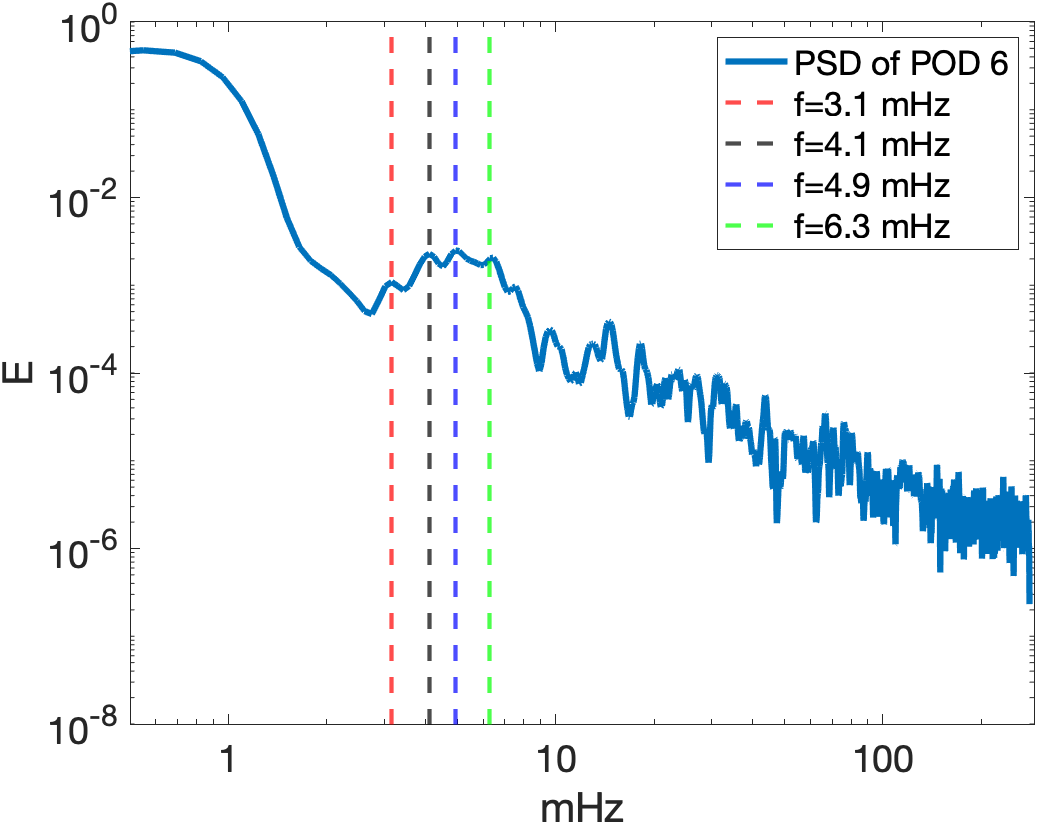}

\end{tabular}
\caption{The power spectrum density (PSD) of the time coefficients of POD 6 mode.}
\label{fig: PSD_surface}
\end{figure*}

\section{Summary and Conclusions} \label{sec:Conclusion}

The present study provided clear evidence of MHD sausage ($n=0$) and kink modes($n=1$) overtones and higher order ($n\geq 2$) fluting modes in sunspots. The results obtained are significant extension of the previous studies by \citet{2017ApJ...842...59J} and \citet{albidah2020RS} where MHD modes were recovered for the case of an approximately  circular sunspot. Firstly, the mode detection was carried out by means of the POD and DMD techniques. Our results were compared with their theoretical counterparts obtained assuming a cylindrical magnetic flux tube, as well as with the model of a magnetic flux
that corresponds to the actual irregular shape of the umbra boundary.

Secondly, the same techniques were also applied to a sunspot whose shape is close to an elliptical cross-section and compared the obtained results with the theoretical predictions of waves in an elliptical waveguide, as well as an irregularly shaped magnetic flux tube corresponding to the actual shape of the umbra boundary. The comparison between modes detected in observational data and in theoretical models was carried out by means of cross-correlation analysis calculated on a pixel-by-pixel basis. The correlation results demonstrate that the higher order MHD modes are more strongly affected by irregularities in the sunspot shape. 

The vertical wavenumber, $k_z$, and mode frequencies have been calculated by using the magneto-acoustic wave parameter ($m_i$) and Equation \ref{Eq:1} for sunspot with a circular cross-sectional shape and Equation \ref{Eq:2} for the sunspot with an elliptical cross-sectional shape (see Tables \ref{Table1} and \ref{Table2}). 

 The existence of these MHD waves were theoretically predicted almost 40 years ago \citep[see, e.g.][]{edwin1983wave}, so our study offers probably one of the first observational evidence for the existence of higher order modes in the chromosphere. Only few papers reported their observational presence
\citep[see, e.g.][]{yuan2015signature,Kang2019}.
These waves offer an unprecedented diagnostic tool for describing the dynamical state of the plasma and the structure of the magnetic field since they are carrying information about the medium in which they are propagating and seismological techniques can be applied to compare observationally determined quantities with theoretical predictions to infer values that cannot be (directly or indirectly) measured. Seismology using a single wave is a sort of “under-determined” system, because many variables are implicit and the variables that can be extracted are not independent (similar to a system of equations having $n\geq 2$ variables, but only $n-1$ equations are given). The observation of at least two or more modes in the same structure helps resolve this degeneracy.

Concurrent observations of different kinds of waves (including higher order modes presented in our study) could allow us to more fully understand the true nature of the dynamics and comprehensively describe the plasma state and structure of the magnetic field. Potentially, our results could help us better understand the nature and properties of modes in more realistically structured waveguides, where the sound and Alfv\'{e}n speeds are spatially varying, which would modify the eigenvalues and eigenfunctions, especially of the higher-order modes.

Higher order modes also give a more complete description of sub-surface driver. It is clear that in the present situation we are dealing with a broad band driver. However slow body modes are weakly dispersive and their phase speeds are confined to a narrow band between the tube speed and internal sound speed. This means that a helioseismological approach of exploiting detected $p$-modes, where modes in $\omega-k$ space correspond to distinct clear ridges, would certainly be a challenge. 

In addition, due to the presence of the higher order modes (as these are the most sensitive to the shape of the waveguide), we demonstrated that the using the exact cross-sectional shape of the waveguide is essential for the correct interpretation of waves.

The current study and techniques used for wave detection have important implications for the interpretation of observational data from next-generation ground-based observing facilities (in particular, the new 4m DKIST solar telescope).

\begin{acknowledgments}
ABA acknowledges the support by Majmaah University (Saudi Arabia) to carry out his PhD studies. VF, GV, IB and SSAS are grateful to The Royal Society, International Exchanges Scheme, collaboration with Chile (IES/R1/170301). VF and GV are grateful to Science and Technology Facilities Council (STFC) grant ST/V000977/1. This research has also received financial support from the European Union’s Horizon 2020 research and innovation program under grant agreement No. 824135 (SOLARNET). AA acknowledges the Deanship of Scientific Research (DSR), King Faisal University, Al-Hassa (KSA) for the financial support.

\end{acknowledgments}


\appendix
\begin{figure*}[!t]
\centering
\begin{tabular}{cc}
 \includegraphics[scale=0.38]{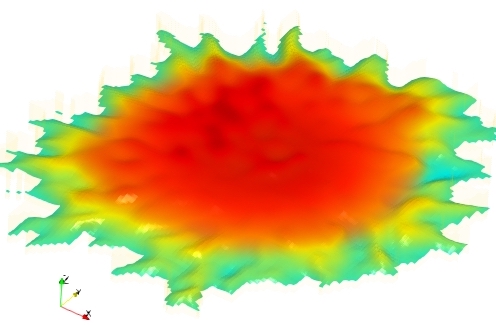}&
 \includegraphics[scale=0.38]{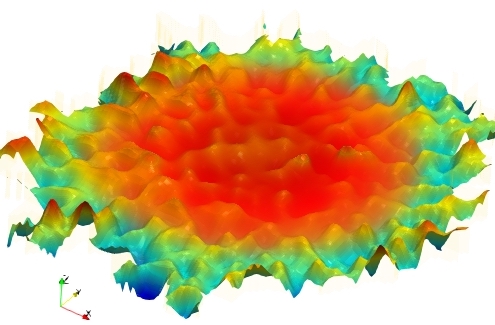}\\
 \includegraphics[scale=0.38]{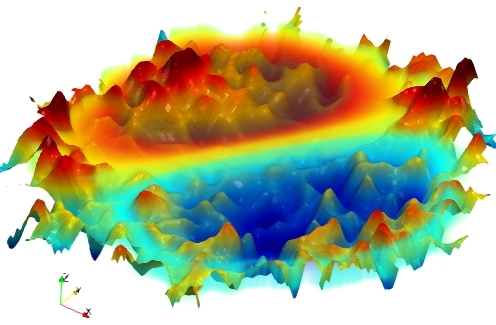}&
 \includegraphics[scale=0.38]{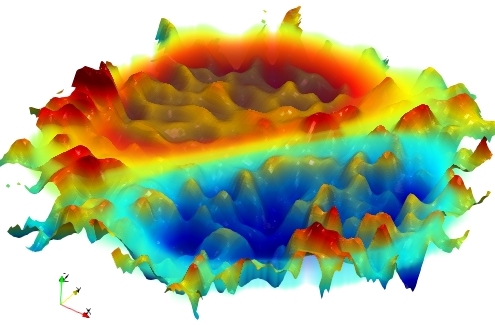}\\
 \includegraphics[scale=0.38]{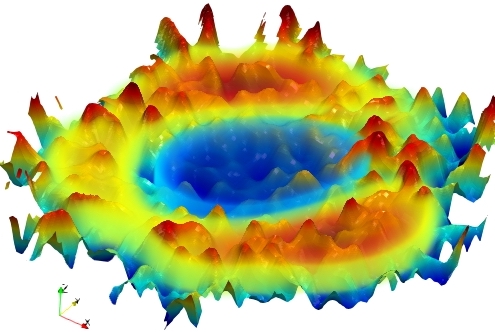}&
 \includegraphics[scale=0.31]{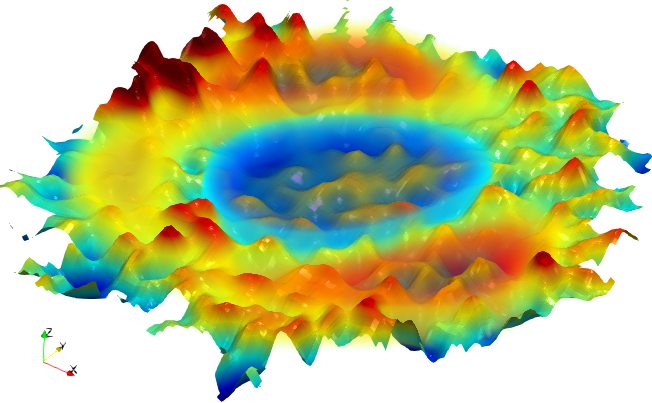}\\
 \includegraphics[scale=0.20]{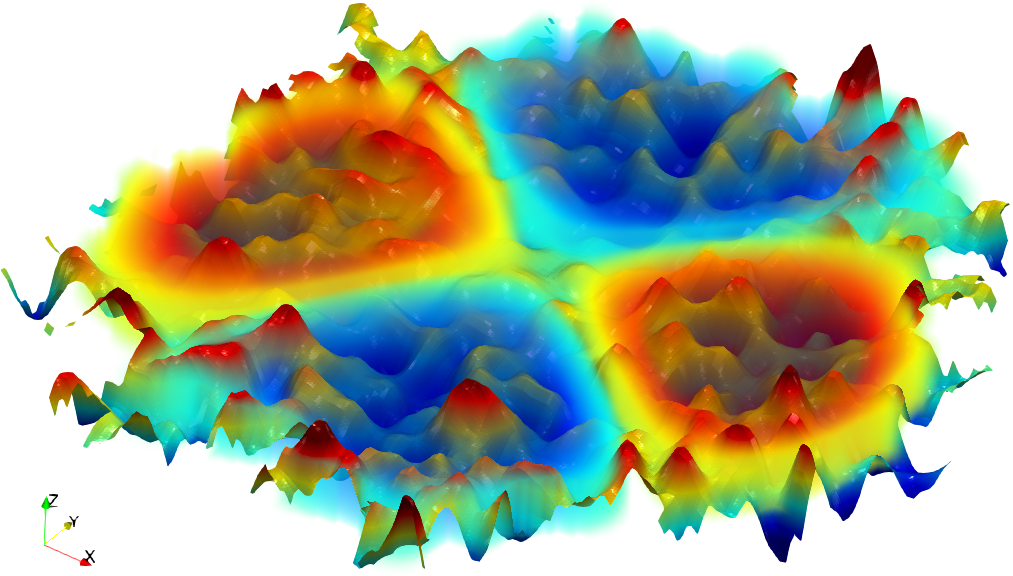}&
 \includegraphics[scale=0.38]{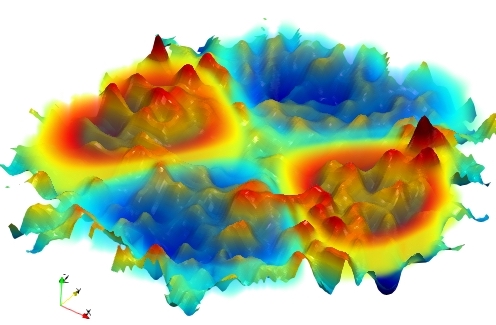}\\
 \includegraphics[scale=0.20]{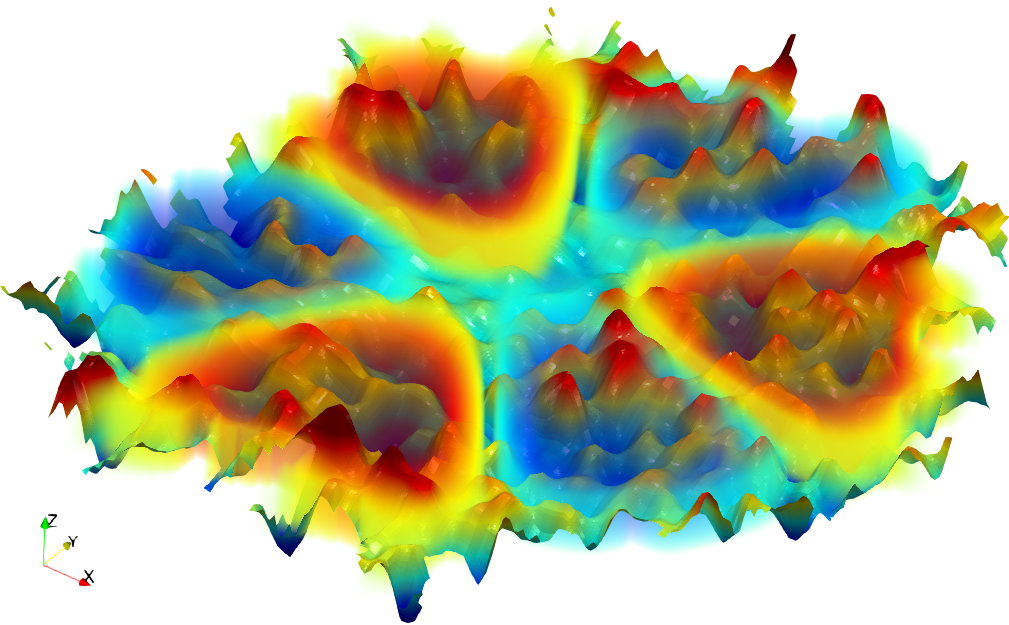}&
 \includegraphics[scale=0.38]{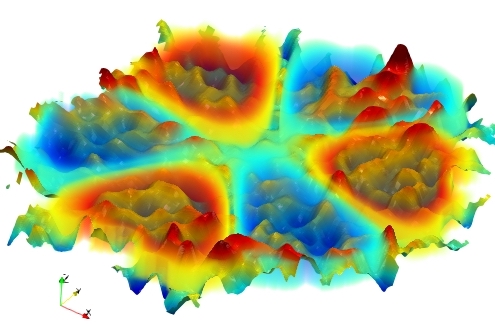}\\
\end{tabular}
\caption{This figure displays 3D surface plot of POD (the first column) and DMD (the second column) modes obtained for the case of circular sunspot as well as volume rendering of the theoretical MHD wave model which uses the same color code as the POD and DMD modes. Rows refer to particular identified MHD modes, that is the fundamental slow body sausage mode (first row), the fundamental slow body  kink mode (second row), the slow body sausage overtone (third row), the $n=2$ slow body fluting mode (fourth row) and the $n=3$ slow body fluting mode (last row).}
\label{fig:C_3D}
\end{figure*}


\begin{figure*}[!t]
\centering
 \begin{tabular}{cc}
 \includegraphics[scale=0.38]{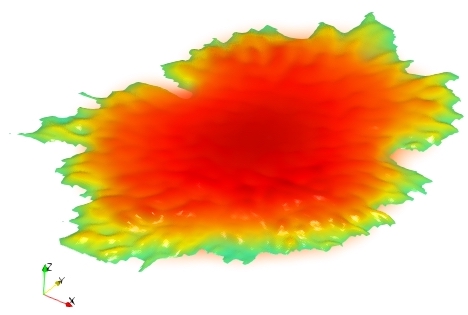}&
 \includegraphics[scale=0.38]{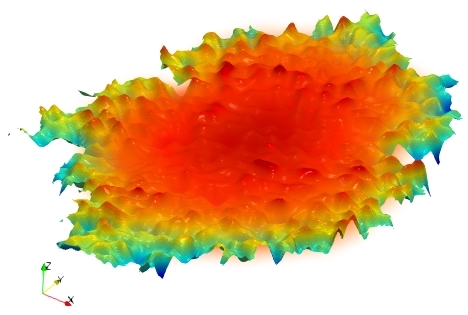}\\
 \includegraphics[scale=0.38]{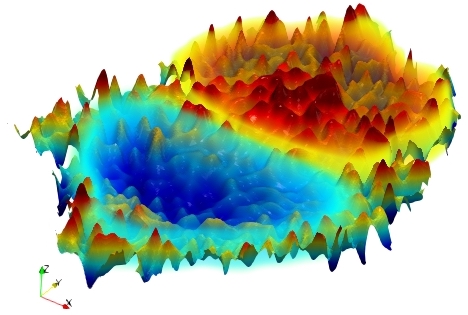}&
 \includegraphics[scale=0.38]{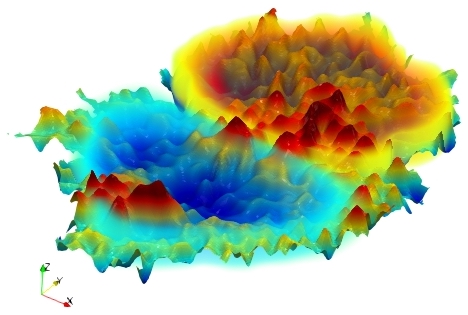}\\
 \includegraphics[scale=0.38]{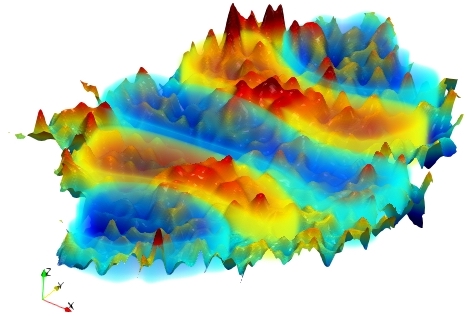}&
 \includegraphics[scale=0.20]{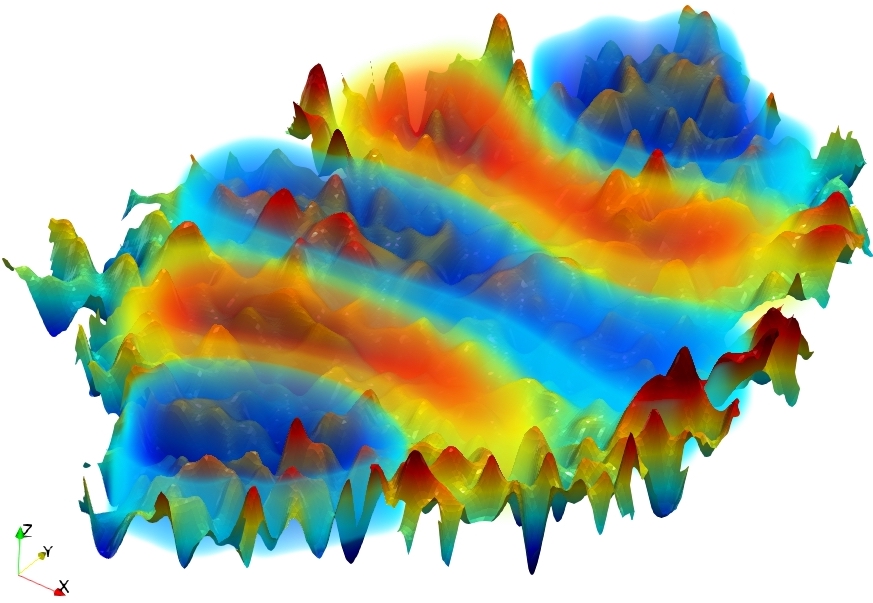}\\
 \includegraphics[scale=0.38]{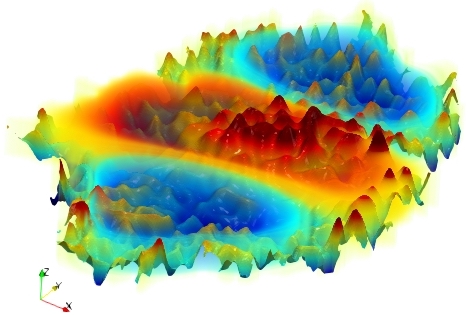}&
 \includegraphics[scale=0.38]{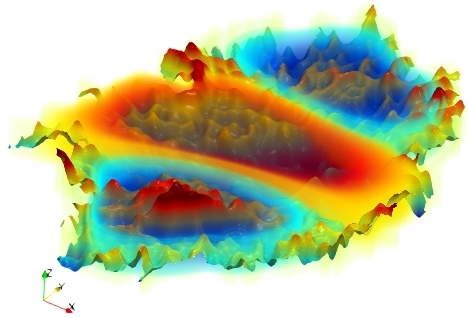}\\
 \includegraphics[scale=0.38]{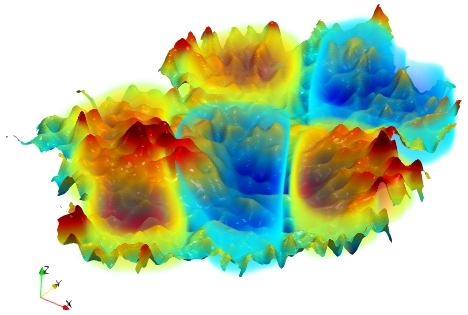}&
 \includegraphics[scale=0.20]{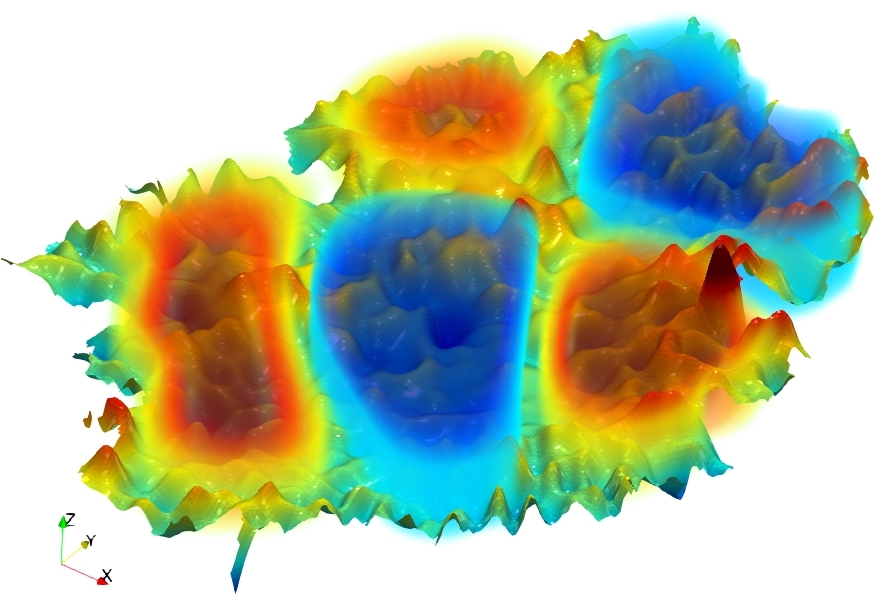}\\
\end{tabular}
\caption{The visualisation technique are same as Figure \ref{fig:C_3D}, but here rows refer to MHD wave mode recovered in the elliptical sunspot, that is the the fundamental slow body sausage (first row), the fundamental slow body kink mode (second row), the slow body overtone kink mode (third row), the $n=2$ slow body fluting mode (fourth row) and the $n=3$ slow body fluting mode (last row).
\label{fig:E_3D}}
\end{figure*}

\bibliographystyle{aasjournal}
\bibliography{ApJ}{}

\end{document}